%% file: main.tex
\documentclass[
  final,
	highlight,
  british,
  a4paper,
	format=acmsmall,
  nonacm=true,
	screen=true,
	natbib=true
]{acmart}
\input{preamble}

\omitproofs %

\title{Linear Logic, the $\pi$-calculus, and their Metatheory: A Recipe for Proofs as Processes}

\def\affIMADA{
  \department{Department of Mathematics and Computer Science}
  \institution{University of Southern Denmark}
  \streetaddress{Campusvej 55}
  \city{Odense}
  \postcode{5230}
  \country{Denmark}
}

\author[F.~Montesi]{Fabrizio Montesi}
\orcid{0000-0003-4666-901X}
\affiliation{\affIMADA}
\email{fmontesi@imada.sdu.dk}

\author[M.Peressotti]{Marco Peressotti}
\orcid{0000-0002-0243-0480}
\affiliation{\affIMADA}
\email{peressotti@imada.sdu.dk}

\begin{abstract}
\input{abstract}
\end{abstract}



\begin{document}

\maketitle

\input{introduction}

\input{mpill}

\input{pill}

\input{hopill}

\input{repill}

\input{conclusions}

\begin{acks}
The authors thank Lu\'\i s Cruz-Filipe and Davide Sangiorgi for useful discussions.

This work was partially sponsored by Villum Fonden, grant no.\ 29518, and by
Independent Research Fund Denmark, grant no.\ 0135-00219.
\end{acks}

\bibliography{biblio}

\clearpage
\FloatBarrier
\appendix

\input{hopill-full}

\end{document}

%% file: abstract.tex
The Curry-Howard correspondence between natural deduction and the $\lambda$-calculus has provided a canonical foundation for the study of typed functional languages.
Initiated by \citet{A94}, the Proofs as Processes agenda is to establish a similar foundation for the study of concurrent languages, by researching the connection between linear logic \citep{G87} and the $\pi$-calculus \citep{MPW92}.

To date, Proofs as Processes is still a partial success.
\citet{CP10} and \citet{W14} showed that linear propositions correspond to session types: structured communication protocols that prescribe the observable behaviour of processes \citep{H93}.
Further, \citet{CMS18}, \citet{MP18}, and \citet{KMP19} demonstrated that adopting devices from hypersequents \citep{A91} shapes proofs (typing derivations) such that they correspond to the expected syntactic structure of processes in the $\pi$-calculus.
What remains is reconstructing the expected metatheory of session types and the $\pi$-calculus.
In particular, the hallmark of session types is session fidelity: an operational correspondence between the observable behaviours of processes and their session types, formulated via labelled transition systems.
To date, this result still has to be reconstructed for Proofs as Processes, because it is unclear how a transition system for session types can arise from linear logic.

In this article, we present \pill, a new process calculus rooted in linear logic. The key novelty of \pill is that it comes with a carefully formulated design recipe for Proofs as Processes, which is based on a dialgebraic view of labelled transition systems and their homomorphisms \citep{CHM13}.
Thanks to our recipe, \pill offers the first reconstruction of the expected labelled transition systems of session types based on linear logic, which we use to establish session fidelity for Proofs as Processes.
In general, we use \pill to carry out the first thorough investigation of the metatheoretical properties enforced by linear logic on the observable behaviour of processes, uncovering connections with standard relations such as similarity and bisimilarity.
We also prove that \pill and our recipe offer a robust basis for the further exploration of Proofs as Processes, by considering different features and extensions: polymorphism, higher-order communications (code mobility), and recursion.
Our higher-order extension of \pill, called \hopill, operates as a higher-order linear logic that enforces linear usage of process variables.

%% file: introduction.tex
\section{Introduction}
\label{sec:introduction}

\begin{quote}
``What matters most in mobile interactive systems is not values, but connectivity and mobility of processes. With or without types, the unifying feature is \emph{behaviour}, and what it means to say that two different processes behave the same.''
\\
---Robin Milner, foreword for ``The $\pi$-calculus: A Theory of Mobile Processes'' \citep{SW01}.
\end{quote}

\paragraph{Background}
The $\pi$-calculus is the most influential theory of mobile processes, whereby concurrent processes can restructure networks by communicating over (channel) names \citep{MPW92}.
Its conception has been the origin of a prolific line of research on the metatheory of mobile processes, which studies the communication behaviours that can be expressed in the $\pi$-calculus and its variants \citep{SW01}.
Two concerns have been particularly relevant.
\begin{itemize}
\item \emph{Behavioural theory}, which investigates techniques for comparing processes.
\item \emph{Type systems}, which syntactically check process definitions to guarantee some desirable properties about their behaviours.
\end{itemize}

On the side of behavioural theory, the most accredited notion is bisimilarity.
Bisimilarity is widely
accepted as the finest desirable notion of equivalence between behaviours, and its solidity has been confirmed repeatedly over more than three decades of research in both concurrency theory and other fields \citep{S11}.
In fact, bisimilarity is so robust that researchers often use it to evaluate how ``good'' the design of a process calculus is: the operators of the calculus should preserve bisimilarity.
Having design principles like this is especially valuable because new process calculi are still developed regularly.

On the side of type systems, the situation is more unstable.
The benchmark for solid foundations in this area is the Propositions as Types correspondence between the $\lambda$-calculus and natural deduction uncovered by Curry and Howard, which to date is still a major source of inspiration for the development of functional programming languages and proof assistants \citep{W15}.
In \citeyear{A94}, \citeauthor{A94} launched the quest for finding a similar guiding light for the future development of typed process calculi. The core of this agenda is to develop a Proofs as Processes correspondence, where proofs in \citeauthor{G87}'s linear logic correspond to processes in the $\pi$-calculus.
While achieving Proofs as Processes in a way that is fully satisfactory on the side of processes proved to be elusive \citep{BS94},
linear logic has already been tremendously influential in the development of typed process calculi.
For example, it inspired the typing discipline of session types, where types prescribe the actions that a process will perform on a channel to participate in a protocol \citep{H93}. Session types grew to be one of the most popular approaches to typing processes to date \citep{Hetal16,Aetal16}, but the link to linear logic was purely inspirational and not rooted in a formal correspondence.

In \citeyear{CP10}, \citeauthor{CP10} kickstarted a renaissance of the exploration of Proofs as Processes, by discovering that propositions in \expandacronym{ILL} can be interpreted as session types.
Using sequents from \expandacronym{ILL} introduces a distinction between ``required'' and ``provided'' channels that is usually not present in the $\pi$-calculus.
\citet{W14} addressed the issue by revisiting \citeauthor{CP10}'s approach in \expandacronym{CLL} (\CLL), which removes the distinction.

The works by \citet{CP10} and \citet{W14} lacked a convincing interpretation of the parallel operator, the most fundamental structural operator of process calculi: no rule in the logic corresponds directly to constructing a parallel composition.
This shortcoming undermines the reconstruction of the expected labelled transition system (lts) semantics of the $\pi$-calculus and, consequently, its behavioural theory.
A series of works that link Proofs as Processes to choreographies (``Alice and Bob'' security protocol notation) provided a preliminary answer \citep{M13:phd,CMS18,CCMM18}: the parallel operator corresponds to composing hypersequents---collections of sequents, previously studied by \citep{A91}.
Inspired by this idea, \citet{MP18} introduced a device to classical linear logic for typing compositions of independent processes: if two processes $P$ and $Q$ implement respectively the types in the typing environments $\Gamma$ and $\Delta$, then we can type the parallel composition $P \pp Q$ with the \emph{hyperenvironment}\footnote{The term hyperenvironment was introduced later, in \citep{KMP19}.} $\Gamma \pp \Delta$; this indicates that all names in $\Gamma$ are used in parallel to those in $\Delta$.
The rule for composing hyperenvironments in parallel corresponds to the term constructor for composing processes in parallel, and supports proof transformations that can be used to construct an lts semantics \citep{MP18}.
\citet{KMP19} later used hyperenvironments to define a conservative extension of linear logic that supports an lts semantics for processes with non-blocking I/O, and proved that the behavioural theory of the resulting calculus can be explored using the standard toolbox of bisimilarity.

The results developed so far on Proofs as Processes are promising. However, they do not match what should be expected of a convincing foundation for the session-typed $\pi$-calculus yet.
The transitions of a well-typed process should guarantee \emph{session fidelity}: the property that all observable actions performed by the process are allowed precisely by its session types \citep{MY13,HYC16}.
By contrast, it is still unclear how this result could be reconstructed for transition systems built on linear logic \citep{MP18,KMP19}.
In general, there is still no clear recipe for building calculi based on linear logic that present session fidelity and the standard principles of behavioural theory found in the $\pi$-calculus.

This is unfortunate:
while previous studies of important extensions for the original $\pi$-calculus, such as polymorphism and higher-order processes, could rely on a solid core calculus equipped with a behavioural theory that carries clear principles for ``good'' design (e.g., that bisimilarity is a congruence), Proofs as Processes has not reached this position yet.
Indeed, while many calculi based on linear logic resemble each other, there is still no version that can claim to be a minimal and satisfactory foundation. For example, the behavioural theory in \citep{KMP19} assumes non-blocking I/O, which is not a basic feature of process calculi but rather an extension \citep{MS04}, and some of the transition rules therein are not structural:
instead of inspecting only the outer-most term constructor, they have side-conditions that traverse the whole syntactic structure of the process. These side-conditions are not present in standard presentations of the $\pi$-calculus, and go against the philosophy of the Structural Operational Semantics approach by \citet{P04}.
Furthermore, \citet{KMP19} sacrificed polymorphism, which was instead present in previous studies \citep{MP18,W14}.

In summary, while previous results suggest that transition systems based on hyperenvironments could play an important role in the relationship between linear logic and process calculi, it is still unclear whether they can form a solid foundation that Proofs as Processes can stand upon.

\paragraph{This article}
We present a design recipe for the development of session-typed process calculi based on linear logic, which addresses the shortcomings of previous approaches and ties up their loose ends.

Our recipe is unifying: it is carefully formulated to support a series of new metatheoretical results for Proofs as Processes that justify our lts semantics and explain the role of hyperenvironments, both from the point of view of behavioural theory and that of type systems. In particular, we achieve session fidelity and discover new connections between the metatheory of derivations and behavioural theory.

Our recipe is also robust: starting from a core calculus, we test the recipe with different features, including polymorphism and code mobility.
These extensions preserve our properties, and similarity and bisimilarity are congruences in all the related calculi.

The main contributions presented in this article are:
\begin{enumerate}
\item
We present our recipe by developing \pill: a new process calculus rooted in linear logic and hyperenvironments.
In \pill, proofs (typing derivations) are processes \citep{A94,BS94}, propositions are session types \citep{CP10,W14} and parallel composition of processes corresponds to the parallel composition of hyperenvironments \citep{CMS18,MP18,KMP19}.
Inspired by \citep{MP18}, the semantics of \pill is based on hyperenvironments and a labelled transition system (lts) for typing derivations that follows the Structural Operational Semantics approach by \citet{P04}.

The key novelty is our recipe.
By adopting a dialgebraic view of labelled transition systems and their homomorphisms \citep{CHM13}, which we apply for the first time in this context, we systematically extract from the lts of derivations an lts for untyped processes and an lts for typing environments.
The former models the expected process dynamics of a session-typed $\pi$-calculus. We show that well-typed processes enjoy erasure and preservation of typability under execution.

Our lts for typing environments is the first observable semantics of session types rooted in linear logic. Session type dynamics give an abstract view of the execution of protocols. Thanks to this transition system, we can finally prove session fidelity for Proofs as Processes.

\pill is also the first process calculus rooted in linear logic to support an lts semantics and behavioural polymorphism.

\item By investigating the metatheory of process dynamics in \pill, we perform the first thorough investigation of the properties enforced by first-order linear logic and hyperenvironments on the observable behaviour of processes.

Bisimilarity and similarity for \pill are defined as expected. We prove that they are, respectively, a congruence and a precongruence.

Our new lts of typing environments and similarity yield an elegant characterisation of session fidelity: well-typed processes are simulated by their types.
For well-typed processes, similarity actually characterises type inhabitancy: if a well-typed process is simulated by some typing environment, then the process necessarily inhabits it (this is derivable in the type system).
Furthermore, bisimilarity is a sound type equivalence check: for any two well-typed processes, if the two processes are bisimilar then they inhabit exactly the same types.

Parallelism has no side-effects.
If a process is ready to perform actions on different channels, then the order in which these actions are executed does not matter: they can be executed simultaneously or serially in any order.

We also uncover the first behavioural interpretation of parallel composition of hyperenvironments, by establishing a correspondence between the parallel operators for hyperenvironments and processes in terms of strong bisimilarity. Namely, if a process is typed by the parallel composition of two hyperenvironments, then it can be rewritten as a strongly-bisimilar parallel composition of two processes.
As we are going to see, this rewriting turns out to be a useful principle for proving other results, including readiness (a generalisation of progress) and the soundness of our proof theory wrt classical linear logic.

\item We develop \hopill, an extension of \pill to code mobility.
The proof theory of \hopill introduces rules for assuming that a sequent can be proven and providing a witness for such an assumption. On the process calculus side, these rules correspond to receiving and sending a process, respectively.
Differently from previous work on higher-order process calculi, \hopill introduces a resource interpretation of code, from which we gain control on how many times transmitted code can be used.

\citet{S93} established that that code mobility can be simulated by channel mobility in the $\pi$-calculus. The same holds for \hopill and \pill. Interestingly, our encoding from \hopill to \pill leverages the translation of proofs from \pill to \CLL, in particular the rewriting of processes within strong bisimilarity that makes parallelism syntactically manifest.

\item
We show that adding recursion and infinite protocols to \pill does not require extending its syntax or semantics. Rather, the addition of a typing rule for expanding types based on type equations is sufficient, because it introduces the feature of having self-calling servers.
This feature is powerful enough for modelling recursive procedures and the standard divergent process $\Omega$ from process calculi.

\end{enumerate}

We believe that our results demonstrate that \pill and our recipe form a solid candidate foundation for the Proofs as Processes agenda, by finally putting in harmony the two metatheoretical concerns mentioned at the beginning: behavioural theory and type systems.

\paragraph{Structure of the paper}
\Cref{sec:mpill} presents \pimll: a fragment of \pill that corresponds to the multiplicative fragment of linear logic.
\pimll suffices to present the key ideas of our development.
We extend \pimll to full linear logic in \cref{sec:pill}. The resulting calculus, called \pill, can express choices (associatives), replicated processes (exponentials), and polymorphism (first-order quantifiers).
In \cref{sec:hopill}, we present \hopill, the calculus that extends \pill to code mobility, and the translation from \hopill to \pill.
We then introduce how our theory can be extended to (potentially infinite) recursive protocols in \cref{sec:repill}.
We discuss related work in \cref{sec:related} and conclude in \cref{sec:conclusions}.

%% file: mpill.tex
\section{Multiplicative Fragment of \pill}
\label{sec:mpill}

We start our study from the multiplicative fragment of \pill, \pimll for short (following the nomenclature used for fragments of linear logic). This is the fragment of \pill that corresponds to the multiplicative fragment of linear logic. \pimll is minimalistic, yet it suffices to explain our development.

\subsection{Processes and Typing}
\paragraph{Processes}
Programs in \pimll are processes (\hl{$\phl{P}$, $\phl{Q}$, $\phl{R}$}). Processes communicate by using names (\hl{$\phl{x}$, $\phl{y}$, $\phl{z}$}), which represent endpoints of sessions.
Our typing discipline will enforce that sessions are binary, i.e., each session involves two endpoints.
We assume an infinite set of names, and that equality of names is decidable.
The process terms of \pimll are defined by the following grammar.%
\begin{bnftable}
  {\phl P, \phl Q \Coloneqq {}}%
  {\res{xy}{P}}
    \phl P, \phl Q \Coloneqq {}
            & \send{x}{y}{P}       & output $y$ on $x$ and continue as $P$\\
    \mid {} & \recv{x}{y}{P}       & input $y$ on $x$ and continue as $P$ \\
    \mid {} & \close{x}{P}         & output (empty message) on $x$ and continue as $P$\\
    \mid {} & \wait{x}{P}          & input (empty message) on $x$ and continue as $P$\\
    \mid {} & \res{xy}{P}          & name restriction, ``cut''\\
    \mid {} & P \pp Q              & parallel composition of processes $P$ and $Q$\\
    \mid {} & \nil                 & terminated process \\
\end{bnftable}%
In defining the terms for communication actions, we follow \citeauthor{W14}'s convention of indicating output with square brackets ``$[-]$'' and input with round parentheses ``$(-)$''.
Term $\send{x}{y}{P}$ allocates a fresh name $y$, outputs $y$ over $x$ and then proceeds as $P$. Dually, term $\recv{x}{y}{P}$ inputs a name $y$ over $x$ and then proceeds as $P$. Names that are sent or received are bound ($y$ in our syntax for output and input). Terms $\close{x}{P}$ and $\wait{x}{P}$ respectively model output and input with no content. A restriction term $\res{xy}{P}$ connects the endpoints $x$ and $y$ of $P$ to form a session, thus enabling communications from $x$ to $y$ and \viceversa. Restriction hides the endpoints $x$ and $y$ from the context. Term $P \pp Q$ is the parallel composition of two processes $P$ and $Q$. Term $\nil$ is the terminated process, which also acts as the unit of parallel composition. We often omit trailing $\nil$s in examples.

In both the communication terms $\send{x}{y}{P}$ and $\recv{x}{y}{P}$, the name $y$ is bound to the continuation $P$.
In the restriction term $\res{xy}{P}$, both $x$ and $y$ are bound to $P$. This gives rise to the expected notions of free names, bound names, and $\alpha$-renaming for processes \citep{SW01}. We write $\fn(P)$ and $\bn(P)$ for, respectively, the free and bound names of a process $P$.

\begin{example}\label{ex:latch-syntax}
The following process, $Latch_{xyz}$, models a simple linear latch that can concurrently receive signals on the endpoints $x$ and $y$. When they are both received, it signals the environment of the event through endpoint $z$.
\[
Latch_{xyz} \defeq \res{x_1x_2}{\res{y_1y_2}{(
  \wait{x}{\aclose{x_1}} \pp \wait y{\aclose {y_1}}
  \pp
  \wait{x_2}{\wait {y_2}{\aclose z}}
)}}
\]
\end{example}

\begin{remark}
\pimll is essentially a fragment of the internal $\pi$-calculus, the variant of the $\pi$-calculus where input and output are symmetric \citep{S96}. Having symmetric input and output primitives gives a more elegant behavioural theory without sacrificing expressivity: just as for the internal $\pi$-calculus, we shall see in \cref{ex:free-output} how the typical ``output of a free name'' primitive can be recovered by using forwarders.
\end{remark}

\begin{remark}
In the original $\pi$-calculus, two processes in parallel can communicate by using the same name $x$ on their input and output actions, and restriction has the form $\res{x}{P}$. Here we use the restriction term $\res{xy}{P}$, which was originally proposed by \citet{V12} for session-typed calculi and then later adopted in some presentations of process calculi typed with linear logic \citep{CLMSW16,KMP19}. We are going to come back to the reason behind this choice when we present the typing rules of \pimll.
\end{remark}

\paragraph{Types}
Types in \pimll (\hl{$\thl{A}$, $\thl{B}$, $\thl{C}$, \dots}) are the propositions of the multiplicative fragment of classical linear logic (\CLL) \citep{G87}.
A type specifies how an endpoint is used. We recall the syntax of propositions in multiplicative \CLL below, and describe how they are interpreted as types---the interpretation is the same as in \citep{CLMSW16}.
\begin{bnftable}[\thl]%
  {\thl{A},\thl{B} \Coloneqq {}}%
  {A \tensor B}%
  \thl{A},\thl{B} \Coloneqq {} & A \tensor B & send $A$, proceed as $B$\\
      \mid {} & A \parr B   & receive $A$, proceed as $B$\\
      \mid {} & \one        & empty output, unit for $\tensor$\\
      \mid {} & \bot        & empty input, unit for $\parr$
\end{bnftable}%
Recall that, in \CLL, each proposition has a dual. We write $\dual{A}$ for the dual of type $A$. We will use the notion of duality to check that two endpoints are used in compatible ways.
Duality is defined inductively as follows.
\begin{highlight}
\[
\thl{\dual{(A \tensor B)}} = \thl{\dual{A} \parr \dual{B}}
\qquad
\thl{\dual{(A \parr B)}} = \thl{\dual{A} \tensor \dual{B}}
\qquad
\thl{\dual{\one}} = \thl{\bot}
\qquad
\thl{\dual{\bot}} = \thl{\one}
\]
\end{highlight}
According to the behavioural interpretation of types, duality essentially checks that each send action has a corresponding receive action.
As usual, duality is an involution: $\dual{\left(\dual A\right)} = A$.

\paragraph{Environments}
An environment (\hl{$\thl{\Gamma}$, $\thl{\Delta}$, \dots}) associates names to types.
We write environments as lists; for example, $\Gamma = \cht{x_1}{A_1}, \ldots, \cht{x_n}{A_n}$ associates each $x_i$ to its respective type $A_i$, for $1 \leq i \leq n$. All names in an environment are distinct. Environments allow for exchange, that is, order in environments is ignored. Environments can be combined when they do not share names: assuming that $\Gamma$ and $\Delta$ do not share names, $\Gamma, \Delta$ is the environment that consists exactly of all the associations in $\Gamma$ and those in $\Delta$.
Environments carry the structure of a (partial) commutative monoid with ``$,$'' acting as sum and the empty environment $\emptyseq$ as unit: we equate environments according to the following rules, for all $\Gamma$, $\Delta$, and $\Xi$.
\begin{highlight}
\[
\thl{\Gamma, \emptyseq} = \thl{\Gamma}
\qquad
\thl{\Gamma , \Delta} = \thl{\Delta , \Gamma}
\qquad
\thl{(\Gamma , \Delta) , \Xi} = \thl{\Gamma , (\Delta , \Xi)}
\]
\end{highlight}

Environments specify how names are used, but they do not specify whether they are used independently. The next ingredient, hyperenvironments, deals exactly with this aspect.

\paragraph{Hyperenvironments}
A hyperenvironment (\hl{$\thl{\hyp{G}}$, $\thl{\hyp{H}}$, \dots}) is a collection of environments, which are composed by the parallel operator ``$\pp$''; for example, given some environments $\Gamma_1$, \ldots, $\Gamma_n$, then $\hyp G = \Gamma_1 \pp \cdots \pp \Gamma_n$ is a hyperenvironment.
All names in a hyperenvironment are distinct. Like environments, hyperenvironments can be combined when they do not share names: assuming that the names in $\hyp G$ and $\hyp H$ are all different, $\hyp G \pp \hyp H$ is the hyperenvironment containing exactly all the environments in $\hyp G$ and those in $\hyp H$.
We write $\emptyhyp$ for the empty hyperenvironment, \ie, the hyperenvironment containing no environments.
Hyperenvironments carry the structure of a (partial) commutative monoid with ``$\pp$'' acting as sum and $\emptyhyp$ as unit: we equate hyperenvironments according to the following rules, for all $\hyp G$, $\hyp H$, and $\hyp I$.
\begin{highlight}
\[
\thl{\hyp G \pp \emptyhyp} = \thl{\hyp G}
\qquad
\thl{\hyp G \pp \hyp H} = \thl{\hyp H \pp \hyp G}
\qquad
\thl{(\hyp G \pp \hyp H) \pp \hyp I} = \thl{\hyp G \pp (\hyp H \pp \hyp I)}
\]
\end{highlight}

As we are going to see, separation of (hyper)environments by ``$\pp$'' has meanings related to independence of derivability and process behaviour.

\paragraph{Judgements}
A judgement $\judge{}{P}{\hyp G}$ states that process $P$ uses its free names according to $\hyp G$.%
Judgements can be derived by using the inference rules displayed in \cref{fig:mpill-typing-rules}.

\begin{figure}[t]
  \begin{highlight}\small 
  \begin{spreadlines}{\typerulesskipamount}%
    \headertext{Structural rules}%
     \begin{gather*}
      \infer[\rlabel{\rname{Cut}}{rule:mpill-cut}]
      {\judge{}{\res{xy}{P}}{\hyp{G} \pp \Gamma, \Delta}}
      {\judge{}{P}{\hyp{G}\pp\Gamma, \cht{x}{A} \pp \Delta, \cht{y}{\dual{A}}}}
      \qquad
      \infer[\rlabel{\rname{Mix}}{rule:mpill-mix}]
      {\judge{}{P \pp Q}{\hyp{G} \pp \hyp{H}}}
      {\judge{}{P}{\hyp{G}}&\judge{}{Q}{\hyp{H}}}
      \qquad
      \infer[\rlabel{\rname{Mix$_0$}}{rule:mpill-mix-0}]
      {\judge{}{\nil}{\emptyhyp}}
      {}
    \end{gather*}%
    \headertext{Logical rules}%
    \begin{gather*}
      \infer[\rlabel{\rname{$\tensor$}}{rule:mpill-tensor}]
      {\judge{}{\send{x}{y}{P}}{\Gamma,\Delta,\cht{x}{A \tensor B}}}
      {\judge{}{P}{\Gamma,\cht{y}{A} \pp \Delta,\cht{x}{B}}}
      \qquad
      \infer[\rlabel{\rname{$\one$}}{rule:mpill-one}]
      {\judge{}{\close{x}{P}}{\cht{x}{\one}}}
      {\judge{}{P}{\emptyhyp}}
      \qquad
      \infer[\rlabel{\rname{$\parr$}}{rule:mpill-parr}]
      {\judge{}{\recv{x}{y}{P}}{\Gamma, \cht{x}{A \parr B}}}
      {\judge{}{P}{\Gamma, \cht{y}{A}, \cht{x}{B}}}
      \qquad
      \infer[\rlabel{\rname{$\bot$}}{rule:mpill-bot}]
      {\judge{}{\wait{x}{P}}{\Gamma, \cht{x}{\bot}}}
      {\judge{}{P}{\Gamma}}
    \end{gather*}%
    \footer%
  \end{spreadlines}%
  \end{highlight}%
  \caption{\pimll, typing rules.}
  \label{fig:mpill-typing-rules}
\end{figure}

The only axiom is \cref{rule:mpill-mix-0}, which types the inert process $\nil$ with the empty hyperenvironment.
\Cref{rule:mpill-mix} types the parallel composition of two processes, by composing their respective hyperenvironments. The information that the two premises are proven independently is recorded explicitly by the use of the parallel operator in the hyperenvironment of the conclusion.
This information allows us to reformulate rules of \CLL that require independent premises as rules that require a single premise with independent environments (denoted by the presence of ``$\pp$'').
For example, the standard cut rule of \CLL becomes \cref{rule:mpill-cut}. The rule types a restriction $\res{xy}{P}$ by checking that the endpoints $x$ and $y$ are used by parallel components in $P$ (the environments in the premise are separated by ``$\pp$'') in a dual way (as usual in \CLL).

Moving to the logical rules, \cref{rule:mpill-tensor} is the standard rule of \CLL, but reformulated to use hyperenvironments following the same intuition that we discussed for \cref{rule:mpill-cut}.
When typing an output $\send xyP$, we require that the names $x$ and $y$ are used independently (in parallel) in $P$. This is standard \citep{CP10,W14}, and it is going to be important when reasoning about progress.
\Cref{rule:mpill-one} types an empty output, requiring the continuation to have no free names.\footnote{For now, this means that the continuation of an empty output is necessarily term $\nil$, but this is not going to be the case when we extend \pill to recursive types in \cref{sec:repill}.}
\Cref{rule:mpill-parr,rule:mpill-bot} are exactly as in \CLL, and respectively type the input of a name and empty input.

We range over typing derivations with the letters \der D, \der E, and \der F.
Also, we write $\deduce{\judge{}{P}{\hyp G}}{\der D}$ whenever a derivation \der D has the judgement $\judge{}{P}{\hyp G}$ as conclusion.

In the remainder, we say that a process $P$ is well-typed if there exist some $\der D$ and $\hyp G$ such that $\deduce{\judge{}{P}{\hyp G}}{\der D}$.

\begin{example}\label{ex:latch-typing}
The process $Latch_{xyz}$ from \cref{ex:latch-syntax} has the typing $\judge{}{Latch_{xyz}}{\cht{x}{\bot}, \cht{y}{\bot}, \cht{z}{\one}}$, as shown by the following derivation. For precision, we report trailing $\nil$s.
\[
\footnotesize
\infer[\ref{rule:mpill-cut}]{
  \judge{}{
    \res{x_1x_2}{\res{y_1y_2}{(
      \wait{x}{\close{x_1}\nil} \pp \wait y{\close {y_1}\nil}
      \pp
      \wait{x_2}{\wait {y_2}{\close z\nil}})}}
  }{\cht{x}{\bot}, \cht{y}{\bot}, \cht{z}{\one}}
}{
  \infer[\ref{rule:mpill-cut}]{
    \judge{}{
      \res{y_1y_2}{(
      \wait{x}{\close{x_1}\nil} \pp \wait y{\close {y_1}\nil}
      \pp
      \wait{x_2}{\wait {y_2}{\close z\nil}})}
    }{\cht{x_1}{\one}, \cht{x}{\bot} \pp \cht{y}{\bot}, \cht{x_2}{\bot}, \cht{z}{\one}}
  }{
    \infer[\ref{rule:mpill-mix}]{
      \judge{}{
        \wait{x}{\close{x_1}\nil} \pp \wait y{\close {y_1}\nil}
        \pp
        \wait{x_2}{\wait {y_2}{\close z\nil}}
      }{\cht{x_1}{\one}, \cht{x}{\bot} \pp \cht{y_1}{\one}, \cht{y}{\bot} \pp \cht{x_2}{\bot}, \cht{y_2}{\bot}, \cht{z}{\one}}
    }{
      \infer[\ref{rule:mpill-bot}]{
        \judge{}{
          \wait{x}{\close{x_1}\nil}
        }{\cht{x_1}{\one}, \cht{x}{\bot}}
      }{
        \infer[\ref{rule:mpill-one}]{
          \judge{}{
            \close{x_1}\nil
          }{\cht{x_1}{\one}}
        }{
          \infer[\ref{rule:mpill-mix-0}]{
            \judge{}{
              \nil
            }{\emptyhyp}
          }{}
        }
      }
      &
      \infer[\ref{rule:mpill-mix}]{
        \judge{}{
          \wait y{\close {y_1}\nil}
          \pp
          \wait{x_2}{\wait {y_2}{\close z\nil}}
        }{\cht{y_1}{\one}, \cht{y}{\bot} \pp \cht{x_2}{\bot}, \cht{y_2}{\bot}, \cht{z}{\one}}
      }{
        \infer[\ref{rule:mpill-bot}]{
          \judge{}{
            \wait y{\close {y_1}\nil}
          }{\cht{y_1}{\one}, \cht{y}{\bot}}
        }{
          \infer[\ref{rule:mpill-one}]{
            \judge{}{
              \close {y_1}\nil
            }{\cht{y_1}{\one}}
          }{
            \infer[\ref{rule:mpill-mix-0}]{
              \judge{}{
                \nil
              }{\emptyhyp}
            }{}
          }
        }
        &
        \infer[\ref{rule:mpill-bot}]{
          \judge{}{
            \wait{x_2}{\wait {y_2}{\close z\nil}}
          }{\cht{x_2}{\bot}, \cht{y_2}{\bot}, \cht{z}{\one}}
        }{
          \infer[\ref{rule:mpill-bot}]{
            \judge{}{
              \wait {y_2}{\close z\nil}
            }{\cht{y_2}{\bot}, \cht{z}{\one}}
          }{
            \infer[\ref{rule:mpill-one}]{
              \judge{}{
                \close z\nil
              }{\cht{z}{\one}}
            }{
              \infer[\ref{rule:mpill-mix-0}]{
                \judge{}{
                  \nil
                }{\emptyhyp}
              }{}
            } %
          }
        }
      }
    }
  }
}
\]
\end{example}

\subsection{Operational Semantics of Derivations}
\label{sec:mpill-sos}
We define a labelled transition system (\lts) for derivations by following the approach originally proposed by \citet{MP18}, which applies the Structural Operational Semantics (SOS) style by \citet{P04}. Specifically, we view:
\begin{itemize} %
  \item inference rules as operations of a (sorted) signature;
  \item derivations as terms generated by this signature;
  \item (labelled) transformations of derivations as (labelled) transitions;
  \item and a specification of rules for deriving transformations of derivations as an SOS specification.
\end{itemize}
In \cref{sec:mpill-proc-hyp}, we will use the lts for derivations to obtain corresponding semantics for processes and types.

As an example, consider a derivation that ends with an application of \cref{rule:mpill-bot}.
\begin{highlight}%
\[
\infer[\ref{rule:mpill-bot}]
      {\judge{}{\wait{x}{P}}{\Gamma, \cht{x}{\bot}}}
      {\deduce
        {\judge{}{P}{\Gamma}}
        {\der D}}
\]
\end{highlight}%
We can view \cref{rule:mpill-bot} as the outermost operation used in the derivation: the subderivation $\der D$ of the premise $\judge{}{P}{\Gamma}$ is the only argument of the operation and $x$ is a parameter (operations are on derivations).
Notice that this operation corresponds to the term constructor $\wait{x}{(-)}$ in the syntax of processes, which in this example takes the continuation $P$ (corresponding to the derivation $\der D$) as argument.
In the $\pi$-calculus, such terms define observable actions that come with corresponding transition rules, where the target (a.k.a. \emph{derivative}) is the operation argument and the transition label represents the applied operation.
Similarly, in our setting, we can define the transition axiom below---for readability, we \cboxed{\pboxcolor}{\mbox{box}} proofs.%
\begin{drules}
\plto
  {\infer[\ref{rule:mpill-bot}]
      {\judge{}{\wait{x}{P}}{\Gamma, \cht{x}{\bot}}}
      {\deduce
        {\judge{}{P}{\Gamma}}
        {\der{D}}}}
  {\lwait{x}}
  {\deduce
    {\judge{}{P}{\Gamma}}
    {\der{D}}}
\end{drules}
The label $\lwait{x}$ represents unambiguously the rule application that we consume in the transition (for convenience, we use as labels the same syntactic form of the corresponding action prefix).

Following the same methodology outlined for \cref{rule:mpill-bot}, we obtain the following three axioms for the semantics of \crefv{rule:mpill-one,rule:mpill-tensor,rule:mpill-parr}.
\begin{drules}
\plto
  {\infer[\ref{rule:mpill-one}]
    {\judge{}{\close{x}{P}}{\cht{x}{\one}}}
    {\deduce
      {\judge{}{P}{\emptyhyp}}
      {\der{D}}}}
  {\lclose{x}}
  {\deduce
    {\judge{}{P}{\emptyhyp}}
    {\der{D}}}
\\
\plto
  {\infer[\ref{rule:mpill-tensor}]
    {\judge{}{\send{x}{x'}{P}}{\Gamma,\Delta,\cht{x}{A \tensor B}}}
    {\deduce
      {\judge{}{P}{\Gamma,\cht{x'}{A} \pp \Delta,\cht{x}{B}}}
      {\der{D}}}}
  {\lsend{x}{x'}}
  {\deduce
    {\judge{}{P}{\Gamma,\cht{x'}{A} \pp \Delta,\cht{x}{B}}}
    {\der{D}}}
\\
\plto
  {\infer[\ref{rule:mpill-parr}]
    {\judge{}{\recv{x}{x'}{P}}{\Gamma, \cht{x}{A \parr B}}}
    {\deduce
      {\judge{}{P}{\Gamma, \cht{x'}{A}, \cht{x}{B}}}
      {\der{D}}}}
  {\lrecv{x}{x'}}
  {\deduce
    {\judge{}{P}{\Gamma, \cht{x'}{A}, \cht{x}{B}}}
    {\der{D}}}
\end{drules}
The labels for the transitions discussed so far correspond to the action prefixes in \pimll and form the set 
$\ActSet = \{ \lclose{x}, \lwait{x}, \lsend{x}{y}, \lrecv{x}{y} \mid x,y \text{ names} \}$. 

For rule \cref{rule:mpill-mix}, we define three rules: two for executions that involve only either the left or the right component (\cref{rule:mpill-mix-1,rule:mpill-mix-2}) and one where the two components interact (\cref{rule:mpill-mix-sync}).
\begin{drules}
\def\regh{\vphantom{\hyp{G}'\pp}}
\infer[\rlabel{\rname{Par$_1$}}{rule:mpill-mix-1}]
  {\plto
    {\infer[\ref{rule:mpill-mix}]
      {\judge{}{P \pp Q}{\hyp{G} \pp \hyp{H}}\regh}
      {\deduce
        {\judge{}{P}{\hyp{G}}\regh}
        {\der{D}}
      &\deduce
        {\judge{}{Q}{\hyp{H}}\regh}
        {\der{E}}}}
    {l}
    {\infer[\ref{rule:mpill-mix}]
      {\judge{}{P' \pp Q}{\hyp{G}' \pp \hyp{H}}\regh}
      {\deduce
        {\judge{}{P'}{\hyp{G}'}\regh}
        {\der{D}'}
      &\deduce
        {\judge{}{Q}{\hyp{H}}\regh}
        {\der{E}}}}}
  {\plto
    {\deduce
      {\judge{}{P}{\hyp{G}}\regh}
      {\der{D}}}
    {l}
    {\deduce
      {\judge{}{P'}{\hyp{G}'}\regh}
      {\der{D}'}}
  &\phl{\bn(l) \cap \fn(Q) = \emptyset}}
\\
\def\regh{\vphantom{\pp\hyp{H}'}}
\infer[\rlabel{\rname{Par$_2$}}{rule:mpill-mix-2}]
  {\plto
    {\infer[\ref{rule:mpill-mix}]
      {\judge{}{P \pp Q}{\hyp{G} \pp \hyp{H}}\regh}
      {\deduce
        {\judge{}{P}{\hyp{G}}\regh}
        {\der{D}}
      &\deduce
        {\judge{}{Q}{\hyp{H}}\regh}
        {\der{E}}}}
    {l}
    {\infer[\ref{rule:mpill-mix}]
      {\judge{}{P \pp Q'}{\hyp{G} \pp \hyp{H}'}\regh}
      {\deduce
        {\judge{}{P}{\hyp{G}}\regh}
        {\der{D}}
      &\deduce
        {\judge{}{Q'}{\hyp{H}'}\regh}
        {\der{E}'}}}}
  {\plto
    {\deduce
      {\judge{}{Q}{\hyp{H}}\regh}
      {\der{E}}}
    {l}
    {\deduce
      {\judge{}{Q'}{\hyp{H}'}\regh}
      {\der{E}'}}
  &\phl{\bn(l) \cap \fn(P) = \emptyset}}
\\
\def\regh{\vphantom{\hyp{G}'\pp\hyp{H}'}}
\infer[\rlabel{\rname{Syn}}{rule:mpill-mix-sync}]
  {\plto
    {\infer[\ref{rule:mpill-mix}]
      {\judge{}{P \pp Q}{\hyp{G} \pp \hyp{H}}\regh}
      {\deduce
        {\judge{}{P}{\hyp{G}}\regh}
        {\der{D}}
      &\deduce
        {\judge{}{Q}{\hyp{H}}\regh}
        {\der{E}}}}
    {\lsync{l}{l'}}
    {\infer[\ref{rule:mpill-mix}]
      {\judge{}{P' \pp Q'}{\hyp{G}' \pp \hyp{H}'}\regh}
      {\deduce
        {\judge{}{P'}{\hyp{G}'}\regh}
        {\der{D}'}
      &\deduce
        {\judge{}{Q'}{\hyp{H}'}\regh}
        {\der{E}'}}}}
  {\plto
    {\deduce
      {\judge{}{P}{\hyp{G}}\regh}
      {\der{D}}}
    {l}
    {\deduce
      {\judge{}{P'}{\hyp{G}'}\regh}
      {\der{D}'}}
  &\plto
    {\deduce
      {\judge{}{Q}{\hyp{H}}\regh}
      {\der{E}}}
    {l}
    {\deduce
      {\judge{}{Q'}{\hyp{H}'}\regh}
      {\der{E}'}}
  & \phl{\bn(l) \cap \bn(l') = \emptyset}}
\end{drules}%
\Cref{rule:mpill-mix-1,rule:mpill-mix-2} allows for a parallel component to perform an action independently of the other, provided the usual hygienic condition on the bound names of the action from one component being different from the free names in the other parallel component.
This condition arises from the requirement of having distinct names in hyperenvironments. Pleasantly, this condition is standard in the internal $\pi$-calculus (\cf \citep{S93}), so our development can be seen as a logical justification of it.
\Cref{rule:mpill-mix-sync} synchronises two actions $l$ and $l'$ performed by parallel components, yielding a transition labelled by the parallel composition $\lsync{l}{l'}$. When writing $\lsync{l}{l'}$, we require that $l$ and $l'$ are not pairs themselves (\ie, $l,l' \in \ActSet$), as interactions in \pimll have two parties.
Also, the order in such pairs of labels is ignored: for all $l$ and $l'$, $\lsync{l}{l'}$ and $\lsync{l'}{l}$ are considered the same.
The condition on disjointness of bound names in \cref{rule:mpill-mix-sync} arises from the well-formedness of the resulting hyperenvironments.

As in \citep{V12}, communication in \pimll takes place when two compatible actions can be performed over endpoints connected by a restriction term.
This is obtained by transition rules that simplify applications of \cref{rule:mpill-cut}. The rule below simplifies a cut on units, resulting in an internal transition with the standard label $\tau$.
\begin{drules}
\infer[\rlabel{\rname{\ref*{rule:mpill-one}\ref*{rule:mpill-bot}}}{rule:mpill-cut-one}]
  {\plto
    {\infer[\ref{rule:mpill-cut}]
      {\judge{}{\res{xy}{P}}{\hyp{G} \pp \Gamma}}
      {\deduce
        {\judge{}{P}{\hyp{G}\pp \cht{x}{\one} \pp \Gamma, \cht{y}{\bot}}}
        {\der{D}}}}
    {\tau}
    {\deduce
      {\judge{}{P'}{\hyp{G}\pp \Gamma}}
      {\der{D'}}}}
  {\plto
    {\deduce
      {\judge{}{P}{\hyp{G}\pp \cht{x}{\one} \pp \Gamma, \cht{y}{\bot}}}
      {\der{D}}}
    {\lsync
      {\lclose{x}}
      {\lwait{y}}}
    {\deduce
      {\judge{}{P'}{\hyp{G}\pp \Gamma}}
      {\der{D'}}}}
\intertext{Similarly, the following rule simplifies a cut between a $\tensor$ and a $\parr$.}
\infer[\rlabel{\rname{\ref*{rule:mpill-tensor}\ref*{rule:mpill-parr}}}{rule:mpill-cut-tensor}]
  {\plto
    {\infer[\ref{rule:mpill-cut}]
      {\judge{}{\res{xy}{P}}{\hyp{G} \pp \Gamma,\Delta, \Xi}}
      {\deduce
        {\judge{}{P}{\hyp{G}\pp\Gamma,\Delta, \cht{x}{A \tensor B} \pp \Xi, \cht{y}{\dual{A} \parr \dual{B}}}}
        {\der{D}}}}
    {\tau}
    {\infer[\ref{rule:mpill-cut}]
      {\judge{}{\res{xy}{\res{x'y'}{P'}}}{\hyp{G} \pp \Gamma, \Delta, \Xi}}
      {\infer[\ref{rule:mpill-cut}]
        {\judge{}{\res{x'y'}{P'}}{\hyp{G} \pp \Gamma,\cht{x}{B} \pp \Delta, \Xi,\cht{y}{\dual{B}}}}
        {\deduce
          {\judge{}{P'}{\hyp{G} \pp \Gamma, \cht{x}{B} \pp \Delta,\cht{x'}{A} \pp \Xi, \cht{y}{\dual{B}}, \cht{y'}{\dual{A}}}}
          {\der{D}'}}}}}
  {\plto
    {\deduce
      {\judge{}{P}{\hyp{G}\pp\Gamma,\Delta, \cht{x}{A \tensor B} \pp \Xi, \cht{y}{\dual{A} \parr \dual{B}}}}
      {\der{D}}}
    {\lsync
      {\lsend{x}{x'}}
      {\lrecv{y}{y'}}}
    {\deduce
      {\judge{}{P'}{\hyp{G} \pp \Gamma, \cht{x}{B} \pp \Delta,\cht{x'}{A} \pp \Xi, \cht{y}{\dual{B}}, \cht{y'}{\dual{A}}}}
      {\der{D}'}}}
\end{drules}%
Both rules for simplifying an application of \cref{rule:mpill-cut} assume a transition that deconstructs dual propositions associated to the names connected by the cut. These transformations do not interact with the context nor have any effect on the types of the conclusions, as expected by transitions labelled with $\tau$ in typed process calculi.
These rules are the last ones to introduce labels for \pimll.
Formally, the set $\LblSet$ of labels for \pimll is 
$\{\tau, l, \lsync{l}{l'} \mid l,l' \in \ActSet \}$.

Lastly, the following rule performs the standard propagation of unrestricted actions, as found in the $\pi$-calculus.
\begin{drules}
\infer[\rlabel{\rname{Res}}{rule:mpill-cut-res}]
  {\plto
    {\infer[\ref{rule:mpill-cut}]
      {\judge{}{\res{xy}{P}}{\hyp{G} \pp \Gamma,\Delta}}
      {\deduce
        {\judge{}{P}{\hyp{G} \pp \Gamma,\cht{x}{A} \pp \Delta,\cht{y}{\dual{A}}}}
        {\der{D}}}}
    {l}
    {\infer[\ref{rule:mpill-cut}]
      {\judge{}{\res{xy}{P'}}{\hyp{G}' \pp \Gamma',\Delta'}}
      {\deduce
        {\judge{}{P'}{\hyp{G}' \pp \Gamma',\cht{x}{A} \pp \Delta',\cht{y}{\dual{A}}}}
        {\der{D'}}}}}
  {\plto
    {\deduce
      {\judge{}{P}{\hyp{G} \pp \Gamma,\cht{x}{A} \pp \Delta,\cht{y}{\dual{A}}}}
      {\der{D}}}
    {l}
    {\deduce
      {\judge{}{P'}{\hyp{G}' \pp \Gamma',\cht{x}{A} \pp \Delta',\cht{y}{\dual{A}}}}
      {\der{D'}}}
  &\phl{x,y \notin \cn(l)}}
\end{drules}

The rule below ensures that $\alpha$-convertible derivations support the same transformations.
\begin{drules}
\infer[\rlabel{\rname{$\aleq$}}{rule:mpill-alpha}]
  {\plto
    {\deduce
      {\judge{}{P}{\hyp{G}}}
      {\der{D}}}
    {l}
    {\deduce
      {\judge{}{Q'}{\hyp{G'}}}
      {\der{E'}}}}
  {\phl{P \aleq Q}
  &\plto
    {\deduce
      {\judge{}{Q}{\hyp{G}}}
      {\der{E}}}
    {l}
    {\deduce
      {\judge{}{Q'}{\hyp{G'}}}
      {\der{E'}}}}
\end{drules}

\begin{definition}%
\label{def:mpill-lts-derivations}
The \lts of derivations for \pimll, denoted $lts_d$, is the triple $(\DerSet,\LblSet,\lto{})$, where:
\begin{itemize}
\item The set $\DerSet$ is the set of typing derivations for \pimll.
\item The set $\LblSet$ is the set of transition labels: $\LblSet = \{\tau, l, \lsync{l}{l'} \mid l,l' \in \ActSet \}$.
\item The relation ${\lto{}} \subseteq \DerSet \times \LblSet \times \DerSet $ is the least relation closed under the SOS rules stated in this subsection.
\end{itemize}
\end{definition}

\begin{example}\label{ex:mmpills}
Consider the following derivation $\der D$:
\begin{highlight}
\[
\der{D} = 
  \pbox{
    \infer[\ref{rule:mpill-cut}]{
      \judge{}{\res{xy}{\left(\send{x}{x'}{Q} \pp \recv{y}{y'}{\wait{z}{R}}\right)}}{\Gamma, \Gamma', \Delta, \cht{z}{\bot}}
    }{
      \infer[\ref{rule:mpill-mix}]{
        \judge{}{\send{x}{x'}{Q} \pp \recv{y}{y'}{\wait{z}{R}}}{\Gamma, \Gamma', \cht{x}{A\tensor B} \pp \Delta, \cht{y}{\dual{A} \parr \dual{B}}, \cht{z}{\bot}}
      }{
        \deduce{\judge{}{\send{x}{x'}{Q}}{\Gamma, \Gamma', \cht{x}{A} \tensor B}}{\der E}
        &
        \deduce{\judge{}{\recv{y}{y'}{\wait{z}{R}}}{\Delta, \cht{y}{\dual{A} \parr \dual{B}}, \cht{z}{\bot}}}{\der F}
      }
    }
  }
\]
where
\[
\der E = \pbox{
    \infer[\ref{rule:mpill-tensor}]{
      \judge{}{\send{x}{x'}{Q}}{\Gamma, \Gamma', \cht{x}{A \tensor B}}
    }{
      \deduce{\judge{}{Q}{\Gamma, \cht{x}{A} \pp \Gamma', \cht{x'}{B}}}{\der E'}
    }
}
\qquad
\der F =
  \pbox{
    \infer[\ref{rule:mpill-parr}]{
      \judge{}{\recv{y}{y'}{\wait{z}{R}}}{\Delta, \cht{y}{\dual{A} \parr \dual{B}}, \cht{z}{\bot}}
    }{
      \infer[\ref{rule:mpill-bot}]{
        \judge{}{\wait{z}{R}}{\Delta, \cht{y}{\dual{A}}, \cht{y'}{\dual{B}}, \cht{z}{\bot}}
      }{
        \deduce{\judge{}{R}{\Delta, \cht{y}{\dual{A}}, \cht{y'}{\dual{B}}}}{\der F'}
      }
    }
  }
\]
for some $\der E'$ and $\der F'$.
\end{highlight}
Then, we have the following transitions.
\def\vlto#1#2{\mspace{6mu}\left\downarrow\vbox to 2.4ex{\vfill\hbox to 84ex{$\mspace{3mu}\scriptstyle #1$\qquad#2\hfill}\vfill}\right.}
\begin{highlight}
\begin{spreadlines}{3pt}
\begin{equation}\label[ltstrace]{eq:mmpills-trans}
\begin{alignedat}{2}
&\pbox{\der D}& \\
&\vlto{\tau}{(by \cref{rule:mpill-cut-tensor,rule:mpill-mix-sync}, and the axioms for \ref{rule:mpill-tensor} and \ref{rule:mpill-parr})} & \\
&\pbox{
  \infer[\ref{rule:mpill-cut}]{
    \judge{}{\res{xy}{\res{x'y'}{\left(Q \pp \wait{z}{R}\right)}}}{\Gamma, \Gamma', \Delta, \cht{z}{\bot}}
  }{
    \infer[\ref{rule:mpill-cut}]{
      \judge{}{\res{x'y'}{\left(Q \pp \wait{z}{R}\right)}}{\Gamma, \cht{x}{A} \pp \Gamma', \Delta, \cht{y}{\dual{A}}, \cht{z}{\bot}}
    }{
      \infer[\ref{rule:mpill-mix}]{
        \judge{}{Q \pp \wait{z}{R}}{\Gamma, \cht{x}{A} \pp \Gamma', \cht{x'}{B} \pp \Delta, \cht{y}{\dual{A}}, \cht{y'}{\dual{B}}, \cht{z}{\bot}}
      }{
        \deduce{\judge{}{Q}{\Gamma, \cht{x}{A} \pp \Gamma', \cht{x'}{B}}}{\der E'}
        &
        \infer[\ref{rule:mpill-bot}]{
          \judge{}{\wait{z}{R}}{\Delta, \cht{y}{\dual{A}}, \cht{y'}{\dual{B}}, \cht{z}{\bot}}
        }{
          \deduce{\judge{}{R}{\Delta, \cht{y}{\dual{A}}, \cht{y'}{\dual{B}}}}{\der F'}
        }
      }
    }
  }}
& \\
&\vlto{\lwait{z}}{(by \cref{rule:mpill-cut-res,rule:mpill-mix-2}, and the axiom for \ref{rule:mpill-bot})} & \\
&\pbox{
  \infer[\ref{rule:mpill-cut}]{
    \judge{}{\res{xy}{\res{x'y'}{\left(Q \pp R\right)}}}{\Gamma, \Gamma', \Delta}
  }{
    \infer[\ref{rule:mpill-cut}]{
      \judge{}{\res{x'y'}{\left(Q \pp R\right)}}{\Gamma, \cht{x}{A} \pp \Gamma', \Delta, \phl y:\dual A}
    }{
      \infer[\ref{rule:mpill-mix}]{
        \judge{}{Q \pp R}{\Gamma, \cht{x}{A} \pp \Gamma', \cht{x'}{B} \pp \Delta, \cht{y}{\dual{A}}, \cht{y'}{\dual{B}}}
      }{
        \deduce{\judge{}{Q}{\Gamma, \cht{x}{A} \pp \Gamma', \cht{x'}{B}}}{\der E'}
        &
        \deduce{\judge{}{R}{\Delta, \cht{y}{\dual{A}}, \cht{y'}{\dual{B}}}}{\der F'}
      }
    }
  }}
&
\end{alignedat}
\end{equation}
\end{spreadlines}
\end{highlight}
\end{example}

\subsection{From Transitions for Derivations to Processes and Environments}
\label{sec:mpill-proc-hyp}

Process calculi, like the $\pi$-calculus, typically come with a semantics for processes that does not depend on typing.
The point of session types is then to guarantee that observable actions performed by a (well-typed) process match what has been promised in its types. This property is known as \emph{session fidelity} \citep{MY13,HYC16}.
In this subsection, we show a recipe for obtaining all these ingredients and their properties directly from our semantics of derivations.

\subsubsection{Semantics of Processes}

Observe that the transitions in \cref{eq:mmpills-trans} make sense even if we completely ignore types, and they resemble the expected ones in an untyped calculus. Specifically, the following transitions for processes are taken by reading off the red parts in the conclusions of the derivations in the transitions from \cref{eq:mmpills-trans}.
\begin{highlight}
\[
\phl{\res{xy}{\left(\send{x}{x'}{Q} \pp \recv{y}{y'}{\wait{z}{R}}\right)}}
\lto{\tau}
\phl{\res{xy}{\res{x'y'}{\left(Q \pp \wait{z}{R}\right)}}}
\lto{\lwait{z}}
\phl{\res{xy}{\res{x'y'}{\left(Q \pp R\right)}}}
\]
\end{highlight}

We formalise the intuitive idea of ``erasing typing from derivations'' as a function $\proc$ that goes from the set of derivation $\DerSet$ to the set of process terms $\ProcSet$.
This function is defined by recursion on the structure of derivations: it just homomorphically maps the application of a rule to the term constructor introduced in its conclusion, as shown below in the case of \cref{rule:mpill-bot}.
\begin{highlight}
\[
  \proc\begin{pmatrix}
    \infer[\ref{rule:mpill-bot}]
          {\judge{}{\wait{x}{P}}{\Gamma, \cht{x}{\bot}}}
          {\mbox{$\deduce
            {\judge{}{P}{\Gamma}}
            {\der D'}$}}
  \end{pmatrix}
  \defeq 
  \phl{\wait{x}{\nohl{\proc\begin{pmatrix}\deduce
              {\judge{}{P}{\Gamma}}
              {\der D'}\end{pmatrix}}}}
\]
\end{highlight}
Because the structure of a well-typed \pimll term drives the structure of its derivation, this definition is equivalent to taking a derivation to the process term in its conclusion:
\begin{highlight}
\[
  \proc\begin{pmatrix}\deduce
              {\judge{}{P}{\hyp{G}}}
              {\der D}\end{pmatrix}
  = \phl{P}
  \text{.}
\]
\end{highlight}

In defining a semantics for process terms we want to ensure that it agrees with the semantics of derivations for well-types terms. That is, for any derivation $\der{D} \in \DerSet$ and $l \in \LblSet$:
\begin{itemize}
\item
if $\der{D} \lto{l} \der{D'}$ then $\proc(\der{D}) \lto{l} \proc(\der{D'})$, and
\item
if $\proc(\der{D}) \lto{l} P'$ then $\der{D} \lto{l} \der{D'}$ for some $\proc(\der{D'}) = P'$.
\end{itemize}
We call this property of a semantics for processes \emph{erasure}.

As we will see in \cref{sec:mpill-metatheory}, erasure plays a crucial role in the metatheory of \pimll and it is reasonable to expect the same from extensions of the language that want to preserve its metatheory. Because many of these extensions introduce changes to the syntax of processes, typing judgements, and semantics (\cf \cref{sec:pill,sec:hopill}), we need a uniform definition that abstracts from these aspects.
To this end, we observe that the characterisation of erasure given above is equivalent to the requirement that the function $\proc\colon \DerSet \to \ProcSet$ carries a homomorphism of labelled transition systems: from the \lts of derivations, $lts_d$, to the \lts of processes, say $lts_p$. The function $\proc$ carries such homomorphism if it makes the diagram below commute where 
$\pw(-)$ and $\lb(-)$ are the functors over the category $\Set$ of sets and functions defined aside ($X$ stands for a set and $f\colon X \to Y$ for a function from $X$ to $Y$).\footnote{We adopt a dialgebraic interpretation of labelled transition systems and their homomorphisms \cite{CHM13} instead of the more common coalgebraic interpretation ($\DerSet \to \pw(\DerSet)^{\LblSet}$) because it simplifies the treatment of higher-order behaviours in \cref{sec:hopill}.}
\begin{equation}
\label{eq:mpill-erasure}
\begin{tikzpicture}[
    auto, 
    xscale=3.2, 
    yscale=1.7,
    baseline=(current bounding box.center)]
  \node (d0) at (0,1) {$\lb(\DerSet)$};
  \node (d1) at (0,0) {$\pw(\DerSet)$};
  \draw[->] (d0) to node[swap] {$lts_d$} (d1);

  \node (p0) at (1,1) {$\lb(\ProcSet)$};
  \node (p1) at (1,0) {$\pw(\ProcSet)$};
  \draw[->] (p0) to node[] {$lts_p$} (p1);
  
  \draw[->] (d0) to node[] {$\lb(\proc)$} (p0);
  \draw[->] (d1) to node[swap] {$\pw(\proc)$} (p1);
\end{tikzpicture}
\qquad\qquad
\begin{aligned}
\pw(X) = {} & \{ X' \mid X' \subseteq X\}\\
\pw(f) = {} & \lambda X'.\{ f(x) \mid x \in X'\}\\
\lb(X) = {}& X \times \LblSet\\
\lb(f) = {}& \lambda (x,l).(f(x),l)
\end{aligned}
\end{equation}
The presentation of labelled transition systems used in the diagram above is equivalent to the one based on triples we used before. Inf act, one has the following bijection (we use a generic set of labels $L$ instead of $\LblSet$ in the definition of $\lb(-)$):
\begin{itemize}
\item a triple $(S,L,\lto{})$ corresponds to the function mapping $(s,l) \in S \times L$ to $\{ s' \mid s \lto{l} s' \} \in \pw(S)$;
\item a function $lts\colon S \times L \to \pw(S)$ corresponds to the triple $(S,L,\lto{})$ such that $s \lto{l} s'$ iff $s' \in lts(s,l)$.
\end{itemize}
In the sequel, we will use there two representations intercheangeably. 

\begin{definition}[Erasure]
\label{def:erasure}
Let $\lb(-)$ be a functor over $\Set$ and fix an type erasure function $\proc\colon \DerSet \to \ProcSet$. 
A labelled transition system $lts_1\colon \lb(\ProcSet) \to \pw(\ProcSet)$ enjoys \emph{erasure} \wrt $lts_2\colon \lb(\DerSet) \to \pw(\DerSet)$ if the function $\proc$ is a homomorphism from $lts_2$ to $lts_1$.
\end{definition}

\begin{figure}
  \begin{highlight}\small
  \begin{spreadlines}{\termltsskipamount}
    \headertext{Actions}
    \begin{gather*}
        \phl{\send{x}{x'}{P}}
        \lto{\lsend{x}{x'}}
        \phl{P}
      \qquad
        \phl{\recv{x}{x'}{P}}
        \lto{\lrecv{x}{x'}}
        \phl{P}
      \qquad
        \phl{\close{x}{P}}
        \lto{\lclose{x}}
        \phl{P}
      \qquad
        \phl{\wait{x}{P}}
        \lto{\lwait{x}}
        \phl{P}
    \end{gather*}%
    \headertext{Structure}
    \begin{gather*}
      \infer[\rlabel{\ref*{rule:mpill-alpha}}{rule:mpill-proc-alpha}]
        {\phl{P} \lto{l} \phl{Q'}}
        {\phl{P \aleq Q}
        &\phl{Q} \lto{l} \phl{Q'}}
      \quad
      \infer[\rlabel{\ref*{rule:mpill-mix-1}}{rule:mpill-proc-mix-1}]
        {\phl{P \pp Q} \lto{l} \phl{P' \pp Q}}
        {\phl{P} \lto{l} \phl{P'}
        &\phl{\bn(l) \cap \fn(Q) = \emptyset}}
      \quad
      \infer[\rlabel{\ref*{rule:mpill-mix-2}}{rule:mpill-proc-mix-2}]
        {\phl{P \pp Q} \lto{l} \phl{P \pp Q'}}
        {\phl{Q} \lto{l} \phl{Q'}
        &\phl{\bn(l) \cap \fn(P) = \emptyset}}
      \\
      \infer[\rlabel{\ref*{rule:mpill-mix-sync}}{rule:mpill-proc-mix-sync}]
        {\phl{P \pp Q} \lto{\lsync{l}{l'}} \phl{P' \pp Q'}}
        {\phl{P} \lto{l} \phl{P'}
        &\phl{Q} \lto{l'} \phl{Q'}
        &\phl{\bn(l) \cap \bn(l') = \emptyset}}
      \qquad
      \infer[\rlabel{\ref*{rule:mpill-cut-res}}{rule:mpill-proc-cut-res}]
        {\phl{\res{xy}{P}} \lto{l} \phl{\res{xy}{P'}}}
        {\phl{P} \lto{l} \phl{P'}
        &\phl{x,y \notin \cn(l)}}
    \end{gather*}%
    \headertext{Communication}
    \begin{gather*}
      \infer[\rlabel{\ref*{rule:mpill-cut-one}}{rule:mpill-proc-cut-one}]
        {\phl{\res{xy}{P}} \lto{l} \phl{P'}}
        {\phl{P} \lto{\lsync{\lclose{x}}{\lwait{y}}} \phl{P'}}
      \qquad
      \infer[\rlabel{\ref*{rule:mpill-cut-tensor}}{rule:mpill-proc-cut-tensor}]
        {\phl{\res{xy}{P}} \lto{l} \phl{\res{xy}{\res{x'y'}{P'}}}}
        {\phl{P} \lto{\lsync{\lsend{x}{x'}}{\lrecv{y}{y'}}} \phl{P'}}
    \end{gather*}
    \footer%
  \end{spreadlines}
  \end{highlight}
  \caption{\pimll, process transitions.}
  \label{fig:mpill-sos-processes}
\end{figure}

The principle of ``erasing types'' can be applied also to the SOS specification for $lts_d$ to derive an SOS specification for processes that is independent from typing. For every rule
\[
\infer[]
  {\der{D} \lto{l} \der{D'}}
  {\der{D}_1 \lto{l_1} \der{D}'_1
  &\dots
  &\der{D}_n \lto{l_n} \der{D}'_n
  &\text{condition on names}}
\]
in the SOS specification for derivations, we introduce a corresponding rule
\[
\infer[]
  {\proc(\der{D}) \lto{l} \proc(\der{D'})}
  {\proc(\der{D}_1) \lto{l_1} \proc(\der{D}'_1)
  &\dots
  &\proc(\der{D}_n) \lto{l_n} \proc(\der{D}'_n)
  &\text{condition on names}}
\]
to the SOS specification for processes, obtained by point-wise application of $\proc$.
The resulting SOS specification is displayed in \cref{fig:mpill-sos-processes}.

\begin{definition}%
\label{def:mpill-lts-processes}
The \lts of processes for \pimll, denoted $lts_p$, is the triple $(\ProcSet,\LblSet,\lto{})$ where
\begin{itemize}
\item $\ProcSet$ is the set of process terms for \pimll,
\item $\LblSet$ is the set of labels for \pimll (see \cref{def:mpill-lts-derivations}),
\item ${\lto{}} \subseteq \ProcSet \times \LblSet \times \ProcSet $ is the least relation closed under the SOS rules in \cref{fig:mpill-sos-processes}.
\end{itemize}
\end{definition}

The \lts of processes enjoys erasure (wrt $lts_d$). The proof of this result will become clear later, when we present the metatheory of \pimll.

\begin{theorem}[Erasure]
\label{thm:mpill-erasure}
$lts_p$ enjoys erasure wrt $lts_d$.
\end{theorem}

\begin{remark}
  Erasure constrains only the semantics of well-typed processes and allows us to attribute any meaning to ill-typed processes like $\wait{x}{\close{x}{\nil}}$. 
  By selecting a semantics for processes that is based on the SOS specification of derivations we obtain some level of ``uniformity'' and a familiar semantics also for ill-typed processes. This decision allows us to generalise some of the metatheory of \pill (\eg, \cref{thm:mpill-congruence}) to all processes and not just well-typed ones. Indeed, these results dependos only on properties of the SOS specification for processes and not on their typing.
\end{remark}

From erasure, we get the standard result of typability preservation as a corollary.
\begin{corollary}[Typability Preservation]
Let $P$ be well-typed. Then, $P \lto{l} P'$ implies that $P'$ is well-typed.
\end{corollary}

\begin{example}\label{ex:latch-proc-transitions}
Recall the process $Latch_{xyz}$ from \cref{ex:latch-syntax}.
\[
Latch_{xyz} \defeq \res{x_1x_2}{\res{y_1y_2}{(
  \wait{x}{\aclose{x_1}} \pp \wait y{\aclose {y_1}}
  \pp
  \wait{x_2}{\wait {y_2}{\aclose z}}
)}}
\]
The following transitions are valid, showing the concurrency informally described in \cref{ex:latch-syntax}.
\begin{align*}
Latch_{xyz} & \lto{\lwait x} \lto{\lwait y} \lto{\tau} \lto{\tau} \lto{\lclose z} \nil
&
Latch_{xyz} & \lto{\lwait y} \lto{\lwait x} \lto{\tau} \lto{\tau} \lto{\lclose z} \nil
\\
Latch_{xyz} & \lto{\lwait x} \lto{\tau} \lto{\lwait y} \lto{\tau} \lto{\lclose z} \nil
&
Latch_{xyz} & \lto{\lwait y} \lto{\tau} \lto{\lwait x} \lto{\tau} \lto{\lclose z} \nil
\end{align*}
\end{example}

\subsubsection{Semantics of Typing Environments}

Recall again the derivation transitions in \cref{eq:mmpills-trans}, and observe that $\tau$-transitions (transitions labelled with $\tau$) do not alter environments, while transitions with observable actions do: in the second transition the typing of name $z$ disappears.
In the area of session types, the safety of these modifications to types by observable actions is known as session fidelity.

Previous work not based on linear logic formalised session fidelity as an operational correspondence between processes and their environments \citep{MY13,HYC16}.
So far, this result has not been reproduced in the research line of Proofs as Processes, because it was unclear how the \lts of environments could be justified by the proof theory of linear logic.

As it sometimes happens in works inspired by the Curry-Howard correspondence, the answer is obvious as soon as it is shown. Just like we can read off the red part of derivation transitions to obtain process transitions, we can read off the blue part to obtain transitions of (hyper)environments.
The transitions below are obtained by reading the blue part of \cref{eq:mmpills-trans}.
\begin{highlight}
\[
\thl{\Gamma, \Gamma', \Delta, \cht{z}{\bot}}
\lto{\tau}
\thl{\Gamma, \Gamma', \Delta, \cht{z}{\bot}}
\lto{\lwait{z}}
\thl{\Gamma, \Gamma', \Delta}
\]
\end{highlight}

The hyperenvironment of a well-typed process represents its interface to the context. Hence, transitions of hyperenvironments give us an immediate way to formulate what ``going wrong'' means for a well-typed process: to perform an observable action that cannot be matched by a transition of its hyperenvironment.

To obtain a semantics for hyperenvironments, we follow the same approach shown for the semantics of process terms.
Let $\env$ be the function from the set of derivations $\DerSet$ to the set of hyperenvironments $\EnvSet$ that maps a derivation to the hyperenvironment in its conclusion:
\begin{highlight}
\[
  \env\begin{pmatrix}\deduce
              {\judge{}{P}{\hyp{G}}}
              {\der D}\end{pmatrix}
  = \thl{\hyp{G}}
  \text{.}
\]
\end{highlight}

We use the semantics of hyperenvironments to define a notion of safety that captures session fidelity.
Specifically, the notion of session fidelity wrt a semantics for environments $lts_e = (\EnvSet,\LblSet,\lto{})$ requires that, given $\judge{}{P}{\hyp{G}}$, $P \lto{l} P'$ for any $l$ and $P'$ implies $\hyp{G} \lto{l} \hyp{G'}$ for some $\hyp{G'}$ such that $\judge{}{P'}{\hyp{G'}}$.
Like erasure, session fidelity can be generalised using \lts-homomorphisms: the key observation is that the span of functions $(\proc,\env)$ identifies all $P$ and $\hyp{G}$ such that $\judge{}{P}{\hyp{G}}$ as well as every witness of this judgement (every $\der{D}$ such that $\proc(\der{D}) = P$ and $\env(\der{D}) = \hyp{G}$). 

Consider for a moment the \lts of derivations (by erasure, this is equivalent to considering well-typed processes). Then, the notion of session fidelity above corresponds to the requirement that if $\der{D} \lto{l} \der{D'}$ then $\env(\der{D}) \lto{l} \env(\der{D'})$ or, in terms of homomorphisms, that the function $\env\colon\DerSet \to \EnvSet$ carries a lax homomorphism of labelled transition systems from $lts_d$ to $lts_e$, as shown by the diagram below ($\lb(-)$ and $\pw(-)$ are the functors over $\Set$ defined above).
\[\begin{tikzpicture}[
    auto, 
    xscale=3.2, 
    yscale=1.7,
    baseline=(current bounding box.center)]
  \node (d0) at (0,1) {$\lb(\DerSet)$};
  \node (d1) at (0,0) {$\pw(\DerSet)$};
  \draw[->] (d0) to node[swap] {$lts_d$} (d1);
  
  \node (e0) at (1,1) {$\lb(\EnvSet)$};
  \node (e1) at (1,0) {$\pw(\EnvSet)$};
  \draw[->] (e0) to node[] {$lts_e$} (e1);
  
  \draw[->] (d0) to node[] {$\lb(\env)$} (e0);
  \draw[->] (d1) to node[swap] {$\pw(\env)$} (e1);
  
  \node at ($(d0)!.5!(e1)$) {$\subseteq$};
\end{tikzpicture}\]
Then, to capture session fidelity for processes we only need the second condition of erasure, which states that if $\proc(\der{D}) \lto{l} P' $ then $\der{D} \lto{l} \der{D'}$ for some $\der{D'}$ such that $\proc(\der{D'}) = P'$. This condition is equivalent to the requirement that $\proc$ carries an oplax homomorphism.

\begin{definition}[Session fidelity]
\label{def:safety}
Let $\lb(-)$ be a functor over $\Set$ and fix the erasure functions $\proc\colon \DerSet \to \ProcSet$ and $\env\colon \DerSet \to \EnvSet$. 
A labelled transition system $lts_1\colon \lb(\ProcSet) \to \pw(\ProcSet)$ enjoys \emph{session fidelity} \wrt $lts_2\colon \lb(\DerSet) \to \pw(\DerSet)$ iff the diagram below holds for some $lts_3$.
\[\begin{tikzpicture}[
    auto, 
    xscale=3.2, 
    yscale=1.7,
    baseline=(current bounding box.center)]
  \node (d0) at (0,1) {$\lb(\DerSet)$};
  \node (d1) at (0,0) {$\pw(\DerSet)$};
  \draw[->] (d0) to node[swap] {$lts_3$} (d1);

  \node (p0) at (-1,1) {$\lb(\ProcSet)$};
  \node (p1) at (-1,0) {$\pw(\ProcSet)$};
  \draw[->] (p0) to node[swap] {$lts_1$} (p1);
  
  \node (e0) at (1,1) {$\lb(\EnvSet)$};
  \node (e1) at (1,0) {$\pw(\EnvSet)$};
  \draw[->] (e0) to node[] {$lts_2$} (e1);
  
  \draw[->] (d0) to node[swap] {$\lb(\proc)$} (p0);
  \draw[->] (d1) to node[] {$\pw(\proc)$} (p1);
  
  \draw[->] (d0) to node[] {$\lb(\env)$} (e0);
  \draw[->] (d1) to node[swap] {$\pw(\env)$} (e1);
  
  \node at ($(d0)!.5!(e1)$) {$\subseteq$};
  \node at ($(p0)!.5!(d1)$) {$\subseteq$};
\end{tikzpicture}\]
\end{definition}

\begin{figure}
\begin{highlight}
\begin{spreadlines}{\envltsskipamount}
\header
\begin{gather*}
{\thl{\Gamma, \Delta, \cht{x}{A \tensor B}}
\lto{\lsend{x}{x'}}
\thl{\Gamma, \cht{x}{B} \pp \Delta, \cht{x'}{A}}}
\qquad
{\thl{\Gamma, \cht{x}{A \parr B}}
\lto{\lrecv{x}{x'}}
\thl{\Gamma, \cht{x}{B}, \cht{x'}{A}}}
\\
{\thl{\cht{x}{\one}}
\lto{\lclose{x}}
\thl{\emptyhyp}}
\qquad
{\thl{\Gamma, \cht{x}{\bot}}
\lto{\lwait{x}}
\thl{\Gamma}}
\qquad
{\thl{\hyp{G}}
\lto{\tau}
\thl{\hyp{G}}}
\\
\infer
  {\thl{\hyp{G} \pp \hyp{H}} \lto{l} \thl{\hyp{G'} \pp \hyp{H}}}
  {\thl{\hyp{G}} \lto{l} \thl{\hyp{G'}}}
\qquad
\infer
  {\thl{\hyp{G} \pp \hyp{H}} \lto{l} \thl{\hyp{G} \pp \hyp{H'}}}
  {\thl{\hyp{H}} \lto{l} \thl{\hyp{H'}}}
\qquad
\infer
  {\thl{\hyp{G} \pp \hyp{H}} \lto{l} \thl{\hyp{G'} \pp \hyp{H'}}}
  {\thl{\hyp{G}} \lto{l} \thl{\hyp{G'}}
  &\thl{\hyp{H}} \lto{l} \thl{\hyp{H'}}}
\end{gather*}
\footer
\end{spreadlines}
\end{highlight}
  \caption{\pimll, typing environment transitions.}
  \label{fig:mpill-sos-environments}
\end{figure}

To obtain a semantics for (hyper)environments, we apply the same principle of ``erasing processes'' to the SOS specification of $lts_d$, which gives us the specification in \cref{fig:mpill-sos-environments}.

\begin{definition}
\label{def:mpill-lts-environments}
The \lts of environments for \pimll, denoted $lts_e$, is the triple $(\EnvSet,\LblSet,\lto{})$ where
\begin{itemize}
\item $\EnvSet$ is the set of hyperenvironments for \pimll ($\hyp{G}$, $\hyp{H}$, \dots),
\item $\LblSet$ is the set of labels for \pimll (see \cref{def:mpill-lts-derivations}),
\item ${\lto{}} \subseteq \EnvSet \times \LblSet \times \EnvSet $ is the least relation closed under the SOS rules in \cref{fig:mpill-sos-environments}.
\end{itemize}
\end{definition}

Well-typed processes never go wrong, as stated by the following result. The proof of this result depends on some properties about the behavioural theory of \pimll, which are presented in the next subsection and make session fidelity straightforward.

\begin{theorem}[session fidelity]
\label{thm:mpill-safety}
$lts_p$ enjoys session fidelity wrt $lts_e$.
\end{theorem}

\begin{example}\label{ex:latch-safety}
Recall from \cref{ex:latch-typing} that $\judge{}{Latch_{xyz}}{\cht{x}{\bot}, \cht{y}{\bot}, \cht{z}{\one}}$.
The process $Latch_{xyz}$ can perform a transition with label $\lwait x$ or another with label $\lwait y$. In either case, its typing hyperenvironment can mimic it. We show the first case.
\begin{align*}
Latch_{xyz} & \lto{\lwait x} \res{x_1x_2}{\res{y_1y_2}{(
  \aclose{x_1} \pp \wait y{\aclose {y_1}}
  \pp
  \wait{x_2}{\wait {y_2}{\aclose z}}
)}}
\\
\cht{x}{\bot}, \cht{y}{\bot}, \cht{z}{\one} & \lto{\lwait x} \cht{y}{\bot}, \cht{z}{\one}
\end{align*}
The derivative of the process is typed by the derivative of the hyperenvironment, as expected by session fidelity. The former can now perform either a transition with label $\lwait y$ or another with label $\tau$ (by resolving its internal communication on $x_1$ and $x_2$). In the second case we obtain the following, and the derivatives remain related by typing.
\begin{align*}
\res{x_1x_2}{\res{y_1y_2}{(
  \aclose{x_1} \pp \wait y{\aclose {y_1}}
  \pp
  \wait{x_2}{\wait {y_2}{\aclose z}}
)}} & \lto\tau \res{y_1y_2}{(
  \wait y{\aclose {y_1}}
  \pp
  \wait {y_2}{\aclose z}
)}
\\
\cht{y}{\bot}, \cht{z}{\one} & \lto{\tau} \cht{y}{\bot}, \cht{z}{\one}
\end{align*}
\end{example}

\section{Metatheory}
\label{sec:mpill-metatheory}

We now move to the metatheoretical study of \pimll.

We start from behavioural theory, showing that the transition system of \pimll gives rise to the expected notions of similarity and bisimilarity.
Then, we leverage these notions to prove that \pimll enjoys erasure and session fidelity, by establishing appropriate (bi)simulations between processes, derivations, and environments.

Afterwards, we focus on studying the metatheoretical implications of hyperenvironments. We find that well-typed processes enjoy serialisation and non-interference for independent actions, and that independence is guaranteed by separation (parallel composition) in hyperenvironments.

Interestingly, our investigation uncovers that standard bisimilarity laws underpin the definition of a deterministic function akin to previous formulations of a transformation called ``disentanglement'' \citep{KMP18}, which rewrites a process into a parallel composition of independent components. Formulating disentanglement based on bisimilarity laws leads to important benefits, including: it reveals that it does not alter behaviour (strong bisimilarity), and it simplifies the proof of readiness for \pimll (a property that generalises progress and deadlock-freedom) compared to previous works.

We end this subsection by showing that \pimll is in a strict relationship with classical linear logic.

\subsection{Behavioural Theory}
\label{sec:mpill-behavioural-theory}

The standard definitions of simulations and bisimulations for the $\pi$-calculus can be adopted for \pimll and its family of semantics.

Similarly to erasure and session fidelity, we state simulations and bisimulations in terms of homomorphisms of labelled transition systems abstracting from the specific $\lb(-)$ and set of labels of \pimll---our definitions are straightforward consequences of \cite{CHM13,H06}.
\begin{definition}[Strong Bisimilarity]
  \label{def:strong-bisimilarity}
  Let $\lb(-)$ be a functor over the category of sets and functions, $lts_1\colon \lb(S_1)\to \pw(S_1)$ and $lts_2\colon \lb(S_2)\to \pw(S_2)$ two labelled transition systems, $\crel{R}$ a relation between their statespaces $S_1$ and $S_2$, and $fst\colon {\crel{R}} \to S_1$ and $snd\colon {\crel{R}} \to S_2$ the associated canonical projections.
  \begin{itemize}
  \item $\crel{R}$ is a \emph{strong forward simulation} from $lts_1$ to $lts_2$ if the diagram below holds for some $lts_r$.
  \[\begin{tikzpicture}[
      auto, 
      xscale=3.2, 
      yscale=1.2,
      baseline=(current bounding box.center)]
    \node (d0) at (0,1) {$\lb(\crel{R})$};
    \node (d1) at (0,0) {$\pw(\crel{R})$};
    \draw[->] (d0) to node[swap] {$lts_r$} (d1);
  
    \node (p0) at (-1,1) {$\lb(S_1)$};
    \node (p1) at (-1,0) {$\pw(S_1)$};
    \draw[->] (p0) to node[swap] {$lts_1$} (p1);
    
    \node (e0) at (1,1) {$\lb(S_2)$};
    \node (e1) at (1,0) {$\pw(S_2t)$};
    \draw[->] (e0) to node[] {$lts_2$} (e1);
    
    \draw[->] (d0) to node[swap] {$\lb(fst)$} (p0);
    \draw[->] (d1) to node[] {$\pw(fst)$} (p1);
    
    \draw[->] (d0) to node[] {$\lb(snd)$} (e0);
    \draw[->] (d1) to node[swap] {$\pw(snd)$} (e1);
    
    \node at ($(d0)!.5!(e1)$) {$\subseteq$};
    \node at ($(p0)!.5!(d1)$) {$\subseteq$};
  \end{tikzpicture}\]
  Strong forward similarity is the largest relation $\ssimby$ that is a strong forward simulation.
  \item $\crel{R}$ is a \emph{strong backward simulation} from $lts_1$ to $lts_2$ if its symmetric $\crel{R}^{-1}$ is a strong forward simulation from $lts_2$ to $lts_1$ \ie if the diagram below holds for some $lts_r$.
  \[\begin{tikzpicture}[
      auto, 
      xscale=3.2, 
      yscale=1.2,
      baseline=(current bounding box.center)]
    \node (d0) at (0,1) {$\lb(\crel{R})$};
    \node (d1) at (0,0) {$\pw(\crel{R})$};
    \draw[->] (d0) to node[swap] {$lts_r$} (d1);
  
    \node (p0) at (-1,1) {$\lb(S_1)$};
    \node (p1) at (-1,0) {$\pw(S_1)$};
    \draw[->] (p0) to node[swap] {$lts_1$} (p1);
    
    \node (e0) at (1,1) {$\lb(S_2)$};
    \node (e1) at (1,0) {$\pw(S_2t)$};
    \draw[->] (e0) to node[] {$lts_2$} (e1);
    
    \draw[->] (d0) to node[swap] {$\lb(fst)$} (p0);
    \draw[->] (d1) to node[] {$\pw(fst)$} (p1);
    
    \draw[->] (d0) to node[] {$\lb(snd)$} (e0);
    \draw[->] (d1) to node[swap] {$\pw(snd)$} (e1);
    
    \node at ($(d0)!.5!(e1)$) {$\supseteq$};
    \node at ($(p0)!.5!(d1)$) {$\supseteq$};
  \end{tikzpicture}\]
  Strong backward similarity is the largest relation $\ssims$ that is a strong backward simulation.
  \item $\crel{R}$ is a \emph{strong bisimulation} for $lts_1$ and $lts_2$ if it is both a strong forward and backward simulation from $lts_1$ to $lts_2$ \ie if the diagram below commutes for some $lts_r$.  
  \[\begin{tikzpicture}[
      auto, 
      xscale=3.2, 
      yscale=1.2,
      baseline=(current bounding box.center)]
    \node (d0) at (0,1) {$\lb(\crel{R})$};
    \node (d1) at (0,0) {$\pw(\crel{R})$};
    \draw[->] (d0) to node[swap] {$lts_r$} (d1);
  
    \node (p0) at (-1,1) {$\lb(S_1)$};
    \node (p1) at (-1,0) {$\pw(S_1)$};
    \draw[->] (p0) to node[swap] {$lts_1$} (p1);
    
    \node (e0) at (1,1) {$\lb(S_2)$};
    \node (e1) at (1,0) {$\pw(S_2t)$};
    \draw[->] (e0) to node[] {$lts_2$} (e1);
    
    \draw[->] (d0) to node[swap] {$\lb(fst)$} (p0);
    \draw[->] (d1) to node[] {$\pw(fst)$} (p1);
    
    \draw[->] (d0) to node[] {$\lb(snd)$} (e0);
    \draw[->] (d1) to node[swap] {$\pw(snd)$} (e1);
  \end{tikzpicture}\]
  Strong bisimilarity is the largest relation $\sbis$ that is a strong  bisimulation.
  \end{itemize}
\end{definition}

To instantiate \cref{def:strong-bisimilarity} on \pimll we simply have to use $\lb(-)$ defined in \eqref{eq:mpill-erasure} \ie 
\[
  \lb(X) = X \times \LblSet \qquad \lb(f) = \lambda(x,l).(f(x),l)
\]
where $X$ is a set and $f\colon X \to Y$ a function.
When we unfold the definition of homomorphism, we obtain the familiar phrasing of simulation and bisimulations for the internal $\pi$-calculus (here $lts_1 = (S_1,\LblSet,\lto{}_1)$ and $lts_2 = (S_2,\LblSet,\lto{}_2)$).
\begin{itemize}
\item $\crel{R}$ is a strong forward simulation from $lts_1$ to $lts_2$ if $s_1 \crel{R} s_2$ implies that if $s_1 \lto{l}_1 s_1'$ then  $s_2 \lto{l}_2 s_2'$ for some $s_2'$ such that $s_1' \crel{R} s_2'$.
\item $\crel{R}$ is a strong backward simulation from $lts_1$ to $lts_2$ if $s_1 \crel{R} s_2$ implies that if $s_2 \lto{l}_2 s_2'$ then  $s_1 \lto{l}_1 s_1'$ for some $s_1'$ such that $s_1' \crel{R} s_2'$.
\item $\crel{R}$ is a strong bisimulation for $lts_1$ and $lts_2$ if $s_1 \crel{R} s_2$ implies that 
\begin{itemize}
\item  if $s_1 \lto{l}_1 s_1'$ then  $s_2 \lto{l}_2 s_2'$ for some $s_2'$ such that $s_1' \crel{R} s_2'$; and that 
\item if $s_2 \lto{l}_2 s_2'$ then  $s_1 \lto{l}_1 s_1'$ for some $s_1'$ such that $s_1' \crel{R} s_2'$.
\end{itemize}
\end{itemize}

\begin{fact}
\label{fact:bisimilarity-laws}
As in standard process algebras, parallel composition and $\nil$ obey the
laws of abelian monoids under (strong) bisimilarity. Formally, for any $P$, $Q$, and $R$:
\begin{equation*}
  P  \pp \nil \sbis P
  \qquad
  P \pp Q \sbis Q \pp P
  \qquad
  P \pp (Q \pp R) \sbis (P \pp Q) \pp R
\end{equation*}
Restriction distributes over parallel composition and restriction, provided that they do not depend on the restricted channel. For $x,y \notin (\fn(Q))$:
\begin{equation*}
  \res{xy}{(P \pp Q)} \sbis \res{xy}{(P)} \pp Q
  \quad
  \res{xy}{\res{x'y'}{P}} \sbis \res{x'y'}{\res{xy}{P}}
\end{equation*}
\end{fact}

Strong behavioural relations can discriminate processes whose behaviour differs only by $\tau$-labelled transitions. The standard approach for calculi in the $\pi$-calculus family is to define (bi)\-simulations in terms of ``saturated'' transitions \cite{M89,SW01}. 
Formally, the \emph{saturation} of an \lts $(S,L,\lto{})$ with $\tau \in L$ is the \lts $(S,L,\slto{})$ where $\slto{}$ is the smallest relation such that: $s \slto{\tau} s$ for all $s \in S$; and if $s_1 \slto{\tau} s_2$, $s_2 \lto{l} s_3$, and $s_3 \slto{\tau} s_4$, then $s_1 \slto{l} s_4$.

\begin{definition}[Bisimilarity]
  \label{def:bisimilarity}
  Let $lts_1$ and $lts_2$ be two labelled transition systems over the same set of labels.
  A relation $\crel{R}$ between their statespaces is a 
  simulation (resp.~bisimulation) from $lst_1$ to $lts_2$ whenever
  it is a strong simulation (resp.~bisimulation) from the saturation of $lst_1$ to the saturation of $lts_2$.
  Similarity (resp.~bisimilarity) is the largest relation $\simby$ (resp.~$\sbis$) that is a simulation (resp.~bisimulation).
\end{definition}

It follows from the inclusion of \cref{rule:mpill-alpha} in the SOS specifications of \pimll that ${\aleq} \subsetneq {\sbis}$. Also, from the definition of saturation, it follows that ${\sbis} \subsetneq {\ssimby} \subsetneq {\simby}$ and ${\sbis} \subsetneq {\bis} \subsetneq {\simby}$ \cite{BMP15}.

The behavioural relations that we have presented are congruences for the \lts of processes, so they allow for local reasoning.

\begin{definition}[Congruence]
  \label{def:congruence}
  An equivalence (resp.~preorder) relation $\asymp$ over processes is a congruence (resp.~precongruence) if it is closed under all syntactic constructs of the language.
\end{definition}
For \pimll, $\asymp$ is a (pre)congruence if $P \asymp Q$ implies that:
\begin{enumerate}[nosep, noitemsep]
    \item ${P \pp R} \asymp {Q \pp R}$ for any $R$ (and the symmetric);
  \item $\prefixed{\pi}{P} \asymp \prefixed{\pi}{Q}$ for any prefix $\pi$;
  \item $\res{xy}{P} \asymp \res{xy}{Q}$ for any $x$, $y$.
\end{enumerate}

\begin{theorem}[Congruence]
\label{thm:mpill-congruence}
On the \lts of processes, $\ssimby$ and $\simby$ are precongruences and $\sbis$ and $\bis$ are congruences.
\end{theorem}

Similar results hold for the \lts of derivations and the \lts of hyperenvironments: replacing a subderivation $\der{E}$ in $\der{D}$ with $\der{E'} \asymp \der{E}$ yields a derivation $\der{D'} \asymp \der{D}$. This claim can be proved with minor adaptations to the proof of \cref{thm:mpill-congruence}.

The standard tools and techniques of behavioural theory are central to the developments in this work. As a first instance, we observe that erasure and session fidelity (\cref{def:erasure,def:safety}) can be stated in terms of strong (bi)simulations. This observation holds also for extensions of the language thanks to our formulation in terms of homomorphisms. 

\begin{lemma}[Erasure, behaviourally]
\label{thm:erasure-correspondence}
Let $lts_p' = (\ProcSet,\LblSet,\lto{})$ and $lts_d' = (\DerSet,\LblSet,\lto{})$. The following statements are equivalent.
\begin{itemize}
\item $lts_p'$ enjoys erasure wrt $lts_d'$.
\item $\{ (\proc(\der{D}),\der{D}) \mid \der{D} \in \DerSet \}$ is a strong bisimulation for $lts_p'$ and $lts_d'$.
\end{itemize}
\end{lemma}

\begin{lemma}[session fidelity, behaviourally]
\label{thm:safety-correspondence}
Let $lts_p' = (\ProcSet,\LblSet,\lto{})$ and $lts_d' = (\EnvSet,\LblSet,\lto{})$. The following statements are equivalent.
\begin{itemize}
\item $lts_p'$ enjoys session fidelity wrt $lts_e'$.
\item $\{ (\proc(\der{D}),\env(\der{D})) \mid \der{D} \in \DerSet \}$ is a strong forward simulation from $lts_p'$ to $lts_e'$.
\end{itemize}
\end{lemma}

The behavioural characterisations of erasure and session fidelity give us a proof strategy for obtaining these properties.
\begin{theorem}
\label{thm:mpill-erasure-beh}
The relation $\{ (\proc(\der{D}),\der{D}) \mid \der{D} \in \DerSet \}$ is a strong bisimulation for $lts_p$ and $lts_d$.
\end{theorem}

\begin{theorem}
\label{thm:mpill-safety-beh}
The relation $\{ (\proc(\der{D}),\env(\der{D})) \mid \der{D} \in \DerSet \}$ is a strong forward simulation from $lts_p$ to $lts_e$.
\end{theorem}

The symmetric direction of \cref{thm:mpill-safety-beh} holds, granted some mild assumptions: if a hyperenvironment simulates a process and does not contain more names, then it can be used to type that process.
\begin{lemma}
\label{thm:mpill-typing-if-sim}
Let $P$ be well-typed and $\hyp{G}$ such that $\fn(P) = \cn(\hyp{G})$. 
If $P \simby \hyp{G}$, then $\judge{}{P}{\hyp{G}}$.
\end{lemma}
In general, the fact that a process is simulated by a hyperenvironment does not imply that it can also be typed by the same environmnet if the process is ill-typed. Consider the process $P = \close{x}{\res{xy}{\nil}}$,
we have that $P \simby \cht{x}{\one}$ and that $P$ is ill-typed.

Combining the last two results we have that on well-typed processes similarity identifies all types inhabited by a given process.

\begin{theorem}
\label{thm:mpill-typing-iff-sim}
For $P$ well-typed and $\hyp{G}$ such that $\fn(P) = \cn(\hyp{G})$, 
$P \simby \hyp{G}$ iff $\judge{}{P}{\hyp{G}}$.
\end{theorem}

Because composing simulations with bisimulations yields simulations, it follows that bisimilarity of well-typed processes implies type equivalence. This result positions any procedure for checking bisimilarity as sound procedure for checking type equivalence.

\begin{corollary}
For $P$ and $Q$ well-typed, if $P \bis Q$ and $\judge{}{Q}{\hyp{G}}$, then $\judge{}{P}{\hyp{G}}$.
\end{corollary}

\subsection{Metatheory of Parallelism}
\label{sec:mpill-metatheory-parallelism}

In the following, we  write $l_1 \disjoint l_2$ to denote that the labels labels $l_1$ and $l_2$ do not share free names, i.e., $\fn(l_1) \cap \fn(l_2) = \emptyset$.

\paragraph{Diamond Property}

We say that an \lts enjoys the diamond property when concurrent actions on distinct names can be interleaved in any order without affecting the trace final state.
\begin{definition}[Diamond Property]
  \label{def:diamond}
  An \lts $(S,\LblSet,\lto{})$ enjoys the \emph{diamond property} provided that for any $s \in S$ and $l_1 \disjoint l_2$ if $s \lto{l_1} s_1$ and $s \lto{l_2} s_2$, then there exists $s_3$ such that $s_1 \lto{l_2} s_3$ and $s_1 \lto{l_1} s_3$.
\end{definition}
Saturation preserves the diamon property.
\begin{lemma}
  \label{thm:diamond-saturation}
  If $(S,\LblSet,\lto{})$ enjoys the diamond property, then its saturation $(S,\LblSet,\slto{})$ enjoys the diamond property.
\end{lemma}

\begin{theorem}
  \label{thm:mpill-diamond}
  The \lts of derivations for \pimll enjoys the diamond property.
\end{theorem}
By erasure, well-typed processes and their transitions enjoy the diamanod property.
\begin{corollary}
\label{thm:mpill-processes-diamond}
  Let $P$ be well-typed and $l_1 \disjoint l_2$.
  \begin{itemize}
  \item If $P \lto{l_1} P_1$ and $P \lto{l_2} P_2$, then there exists $P_3$ such that $P_1 \lto{l_2} P_3$ and $P_2 \lto{l_1} P_3$.
  \item If $P \slto{l_1} P_1$ and $P \slto{l_2} P_2$, then there exists $P_3$ such that $P_1 \slto{l_2} P_3$ and $P_2 \slto{l_1} P_3$.
  \end{itemize}
\end{corollary}
The lts of processes does not enjoy the diamon property because there are processes that violate the property \eg $\close{x}{\close{y}{\nil}} \pp \close{x}{\close{z}{\nil}}$. Any such process is necessarily ill-typed.

A consequence of the diamond property is that $\tau$-labelled transitions do not affect the possible interactions with the environment: any interaction available before an internal step will still be available afterwards.
\begin{lemma}\label{thm:tau-bis}
  Let $(S,L\lto{})$ enjoy the diamond property. 
  If $s \lto{\tau} s'$, then $s \bis s'$.
\end{lemma}

\paragraph{Serialisation}

We say that an \lts enjoys serialisation if whenever a state can perform two actions in parallel (i.e., it is the source of a transition with a label like $\lsync{l_1}{l_2}$) it can also be perform the same actions sequentially.
\begin{definition}[Serialisation]
  \label{def:serialisation}
  An \lts $(S,\LblSet,\lto{})$ enjoys \emph{serialisation} provided that for
  any $s \in S$ and $l_1 \disjoint l_2$ 
  if $s \lto{\lsync{l_1}{l_2}} s'$, then $s \lto{l_1} \lto{l_2} s'$.
\end{definition}

Saturation preserves serialisation.
\begin{lemma}
  \label{thm:serialisation-saturation}
  If $(S,\LblSet,\lto{})$ enjoys serialisation, then its saturation $(S,\LblSet,\slto{})$ enjoys serialisation.
\end{lemma}

\begin{theorem}
  \label{thm:mpill-serialisation}
  The \lts of derivations for \pimll enjoys serialisation.
\end{theorem}

It follows by erasure that well typed processes and their transitions enjoy the diamanod property.
\begin{corollary}
\label{thm:mpill-processes-serialisation}
  Let $P$ be well-typed. %
  \begin{itemize}
  \item If $P \lto{\lsync{l_1}{l_2}} P'$, then $P \lto{l_1} \lto{l_2} P'$.
  \item If $P \slto{\lsync{l_1}{l_2}} P'$, then $P \slto{l_1} \slto{l_2} P'$.
  \end{itemize}
\end{corollary}

\paragraph{Non-interference}

We say that an \lts enjoys non-interference if when a state is ready to perform two actions on different names, then these can be parallelised, i.e., they do not interfere.
\begin{definition}[Non-interference]
  \label{def:non-interference}
  An \lts $(S,\LblSet,\lto{})$ enjoys \emph{non-interference} provided that for any $s \in S$ and $l_1,l_2 \in \ActSet$ if $l_1 \disjoint l_2$, $s \lto{l_1}$, and $s \lto{l_2}$, then $s \lto{\lsync{l_1}{l_2}}$.
\end{definition}
Saturation preserves non-interference.
\begin{lemma}
  \label{thm:non-interference-saturation}
  If $(S,\LblSet,\lto{})$ enjoys non-interference, then its saturation $(S,\LblSet,\slto{})$ enjoys non-interference.
\end{lemma}

\begin{theorem}
  \label{thm:mpill-non-interference}
  The \lts of derivations for \pimll enjoys the non-interference.
\end{theorem}

It follows by erasure that well typed processes and their transitions enjoy non-interference.
\begin{corollary}
\label{thm:mpill-processes-non-interference}
  Let $P$ be well-typed. %
  \begin{itemize}
  \item If $P \lto{l_1}$ and $P \lto{l_2}$ if and only if $P \lto{\lsync{l_1}{l_2}}$.
  \item If  $P \slto{l_1}$ and $P \slto{l_2}$ if and only if $P \slto{\lsync{l_1}{l_2}}$.
  \end{itemize}
\end{corollary}

\paragraph{Disentanglement}

\citet[Lemma~2.6]{KMP19} observed that applications of \cref{rule:mpill-mix} can be moved until either they reach the top level or become attached to cuts or the formation of tensors.
They use this ``disentanglement'' result to show that if a hyperenvironment is derivable in their system, then any of its environments is derivable in \CLL using a proof closely related to some sub-derivation rooted under a top-level mix in the disentangled derivation.

Differently here, we show that disentanglement can be seen as a rewriting relation derived from standard laws of strong bisimilarity (\cf \cref{fact:bisimilarity-laws}). Our definition does not rely on types: it can be applied to any process, even ill-typed ones. Also, it is a deterministic procedure for computing disentangled processes (hence derivations).
Thanks to its semantic foundation and constructive definition, our notion of disentanglement has applications that go beyond relating the type theory of our calculi to \CLL. Notably, we use it later in this subsection to prove that \pimll enjoys readiness (a generalisation of progress) and in \cref{sec:hopill} to encode higher-order communications into first-order communications.

\begin{definition}[Disentanglement]
The disentanglement $\disen{P}$ of a well-typed process $P$ is the process obtained by recursively rewriting a processes using the laws
\begin{itemize}
\item $P \pp \nil \sbis P$;
\item $\res{xy}{(P \pp Q)} \sbis \res{xy}P \pp Q$ if $x,y \notin \fn(Q)$;
\item $\res{xy}{(P \pp Q)} \sbis P \pp \res{xy} Q$ if $x,y \notin \fn(P)$;
\end{itemize}
read from left to right until it is no longer possible.
\end{definition}

\begin{example}\label{ex:latch-dis}
Consider the following process, which is a subterm of the latch in
\cref{ex:latch-syntax}.
\[
\res{y_1y_2}{(
  \wait{x}{\aclose{x_1}} \pp \wait y{\aclose {y_1}}
  \pp
  \wait{x_2}{\wait {y_2}{\aclose z}}
)}
\]
The following process is the disentanglement of the term above.
\[
  \wait{x}{\aclose{x_1}} \pp \res{y_1y_2}{( \wait y{\aclose {y_1}}
  \pp
  \wait{x_2}{\wait {y_2}{\aclose z}}
)}\text{.}
\]
\end{example}

This definition is equivalent to the procedure defined below by recursion on the structure of processes (for restriction we need to apply several applications of the binary law above at once).
\begin{lemma}
\label{thm:mpill-disentanglement}
The disentanglement $\disen{P}$ of a process $P$ is the process recursively defined as follows:
\begin{gather*}
\disen{\prefixed{\pi}{P}} = \prefixed{\pi}{\disen{P}}
\qquad
\disen{\nil} = \nil 
\qquad
\disen{P \pp Q} = \begin{cases}
  \disen{P} & \text{if } Q = \nil\\
  \disen{P} \pp \disen{Q}
  & \text{otherwise}
\end{cases}
\\
\disen{\res{xy}{P}} =  \res{xy}{(P_i \pp P_j)} \pp (\disen{P}\setminus P_i,P_j)
\text{ where } \disen{P} = P_1 \pp \dots \pp P_{n}\text{, } 
x \in \fn(P_i) \text{, } 
y \in \fn(P_j)
\end{gather*}
\end{lemma}

Disentanglement preserves the semantics of processes.
\begin{lemma}
  \label{thm:disentanglement-bisim}
  $P \sbis \disen{P}$.
\end{lemma}

When applied to well-typed processes, disentanglement preserves typing. Furthermore, each parallel component of the disentangled process corresponds to and is typed by a parallel component of the hyperenvironment that types the initial process.
\begin{lemma}\label{thm:disentanglement-shape}
Let $\judge{}{P}{\Gamma_1 \pp \cdots \pp \Gamma_{n}}$. Then, $\disen{P} = P_1 \pp \cdots \pp P_{n}$ s.t.~$\judge{}{P_i}{\Gamma_i}$ for each $i \in [1,n]$.
\end{lemma}

\paragraph{Readiness}

A well-typed process whose hyperenvironment consists of a single environment is ready to perform a saturated transition on at least one of its free names. In the following, we write $l_x$ for a label with $x$ free ($\lclose{x}$, $\lsend{x}{y}$, \dots).
\begin{lemma}\label{thm:mpill-readiness-single}
If $\judge{}{P}{\Gamma}$, then there exist $x \in \cn(\Gamma)$ and $l$ such that $x \in fv(l)$ and $P \slto{l}$.
\end{lemma}

It thus follow from disentanglement, that any well-typed process is always ready on at least one name for each component of its hyperenvironment.
\begin{theorem}[Readiness]
\label{thm:mpill-readiness}
Let $\judge{}{P}{\Gamma_1 \pp \cdots \pp \Gamma_n}$. For every $i \in [1,n]$, there exist $x\in\cn(\Gamma_i)$ and $l$ such that $x \in \fn(l)$ and $P \slto{l}$.
\end{theorem}

\begin{example}
Consider the typing environemnt $\Gamma = \cht{x}{\bot}, \cht{y}{\bot}, \cht{z}{\one}$. From readiness we know that in any process $P$ typed by $\Gamma$ must be ready on at least one name in $\Gamma$ and by session fidelity we know that this name cannot be $z$ (its type is $\one$). 
Process $Latch_{xyz}$ from \cref{ex:latch-typing} is typed by $\Gamma$ and indeed it is ready on both $x$ and $y$.
Other examples of processes typed by $\Gamma$ are $P=\wait{x}{\wait{y}{\close{z}{nil}}}$ which is ready only on $x$ and
$Q=\wait{y}{\wait{x}{\close{z}{nil}}}$ which is ready only on $y$.
Any other process $R$ typed by $\Gamma$ can only differ from $Latch_{xyz}$, $P$ or $Q$ on its internal behaviour (the type of $x$, $y$, and $z$ prescribe thir use and any other action must be on a bound name). 
It follows that $Q$ is necessarily bisimilar to one of the three processes above.
\end{example}

\subsection{Relation to Linear Logic and Classical Processes}
\label{sec:mpill-mcp}

The theory of \pill is strongly connected to \CLL and its corresponding process calculus of Classical Processes (\CP) \citep{W14}. In particular, this allows for lifting existing proof search techniques for \CLL to \pill. Here we present the connection between \pimll and \MCLL and \MCP, the multiplicative fragments of \CLL and \CP. We will later extend this connection to the full theories of \pill, \CLL, and \CP in \cref{sec:pill}.

\def\cprname#1{{#1\textsuperscript{\tiny\CP}}}
\begin{figure}[t]
  \begin{highlight}\small 
    \header%
     \begin{spreadlines}{\typerulesskipamount}%
     \begin{gather*}
      \infer[\rlabel{\cprname{\rname{Cut}}}{rule:mcp-cut}]
      {\cpjudge{\res{xy}{(P \pp Q)}}{\Gamma, \Delta}}
      {\cpjudge{P}{\Gamma, \cht{x}{A}}
      &\cpjudge{Q}{\Delta, \cht{y}{\dual{A}}}}
      \qquad
      \infer[\rlabel{\cprname{\rname{$\tensor$}}}{rule:mcp-tensor}]
      {\cpjudge{\send{x}{y}{(P\pp Q)}}{\Gamma,\Delta,\cht{x}{A \tensor B}}}
      {\cpjudge{P}{\Gamma,\cht{y}{A}}
      &\cpjudge{Q}{\Delta,\cht{x}{B}}}
      \qquad
      \infer[\rlabel{\cprname{\rname{$\one$}}}{rule:mcp-one}]
      {\cpjudge{\aclose{x}}{\cht{x}{\one}}}
      {\phantom{\judge{}{P}{\emptyhyp}}}
      \\
      \infer[\rlabel{\cprname{\rname{$\parr$}}}{rule:mcp-parr}]
      {\cpjudge{\recv{x}{y}{P}}{\Gamma, \cht{x}{A \parr B}}}
      {\cpjudge{P}{\Gamma, \cht{y}{A}, \cht{x}{B}}}
      \qquad
      \infer[\rlabel{\cprname{\rname{$\bot$}}}{rule:mcp-bot}]
      {\cpjudge{\wait{x}{P}}{\Gamma, \cht{x}{\bot}}}
      {\cpjudge{P}{\Gamma}}
    \end{gather*}%
    \end{spreadlines}%
    \footer%
  \end{highlight}%
  \caption{Multiplicative \CP, typing rules.}
  \label{fig:mcp-typing-rules}
\end{figure}

\subsubsection{Multiplicative Classical Processes (\MCP)}
Process terms in the multiplicative fragment of \CP are defined by the grammar below (we follow the presentation of \CP by \citet{CLMSW16}, which is based on endpoints like \pill).
\begin{bnftable}
  {\phl P, \phl Q, \phl R \Coloneqq {}}%
  {\res{xy}{(P \pp Q)}}
    \phl P, \phl Q \Coloneqq {}
            & \send{x}{y}{(P \pp Q)} & output $y$ on $x$\\
    \mid {} & \recv{x}{y}{P}         & input $y$ on $x$\\
    \mid {} & \aclose{x}             & output (empty message) on $x$\\
    \mid {} & \wait{x}{P}            & input (empty message) on $x$\\
    \mid {} & \res{xy}{(P \pp Q)}     & name restriction, ``cut''
\end{bnftable}%
The typing discipline of \MCP corresponds to \MCLL and is shown in \cref{fig:mcp-typing-rules}.
\Cref{rule:mcp-cut} corresponds to the standard cut rule of \CLL.
The cut rule of \CLL requires two separate premises, which in \CP corresponds to two processes. As a result, \cref{rule:mcp-cut} is intensional: it must inspect the syntactic structure of the process under the restriction to check that there are two parallel components.
In other words, the rule requires treating restriction and parallel as a single combined operator.
This is in contrast to \pimll where hyperenvironments allow us to treat restriction and parallel as two separate operators, as usual in process calculi.
Similarly to cut, the original rule $\tensor$ from \CLL has two separate premises and the corresponding rule in \CP, \ref{rule:mcp-tensor}, is also intensional (whereas \cref{rule:mpill-tensor} in \pimll is not).
From the viewpoint of process calculi, the fact that \cref{rule:mpill-tensor} has a single premise is key to reconstructing the typical output term of the internal $\pi$-calculus, which has a single continuation \citep{S96}.
The remaining rules in \cref{fig:mcp-typing-rules} all have an equivalent rule (or literally the same rule) in \pimll.

Derivations in \MCP can be interpreted as derivations in \pimll where the \crefv{rule:mcp-cut,rule:mcp-tensor,rule:mcp-one} are derivable in terms of \crefv{rule:mpill-cut,rule:mpill-tensor,rule:mpill-mix,rule:mpill-one,rule:mpill-mix-0}. In terms of processes, this corresponds to regarding the term constructors $\res{xy}{(- \pp -)}$, $\send{x}{y}{(- \pp -)}$, and $\aclose{x}$ of multiplicative \CP as syntactic sugar for \pimll.
\begin{proposition}
If $\cpjudge{P}{\Gamma}$ in \MCP then $\judge{}{P}{\Gamma}$ in \pimll.
\end{proposition}

\begin{remark}
\citet{W14} proposed a different rule for typing parallel composition, inspired by the \rname{Mix} rule by \citet{G87}.
\begin{highlight}
\[\infer[\cprname{\rname{Mix}}]
  {\cpjudge{P \pp Q}{\Gamma,\Delta}}
  {\cpjudge{P}{\Gamma}
  &\cpjudge{Q}{\Delta}}
\]
\end{highlight}
Unfortunately, combining the environments from the premises by using ``,'' makes the rule forget that the names in $\Gamma$ are implemented in parallel to the names in $\Delta$. This prevents $P$ and $Q$ from ever communicating in \CP, since we cannot be sure that connecting them with a restriction term will not introduce a cycle.
\end{remark}

\subsubsection{From \pimll to \MCP, and Back}

There are well-typed \pimll processes that are not valid processes in \CP, for example $\res{xy}{(P \pp Q \pp R)}$. However, if a \pimll process can be typed using a single environment, then its disentanglement can be typed in \CP with the same environment.

\begin{proposition}
\label{thm:mpill-complete-mcp-single-env}
If $\judge{}{P}{\Gamma}$ in \pimll then $\cpjudge{\disen{P}}{\Gamma}$ in \MCP.
\end{proposition}

It follows that the parallel components of the disentanglement of any well-typed \pill process are well-typed processes.

\begin{lemma}\label{lemma:disen-cp}
Let $\judge{}{P}{\Gamma_1 \pp \cdots \pp \Gamma_n}$. Then, $\disen{P} = P_1 \pp \cdots \pp P_n$ and $\cpjudge{P_i}{\Gamma_i}$ in \CP for all $i \in [1,n]$.
\end{lemma}

The result in \cref{lemma:disen-cp} inspires the exploration of the more general question:
\begin{quote}
Given any \pimll process $P$, can we systematically construct a \CP process $P'$ that is well-typed if and only if $P$ is? 
\end{quote}

\citet{KMP18,KMP19} provide a partial answer to this question at the level of typing environments in their calculi: for any $\hyp{G}$, there is $\Gamma$ such that $\hyp{G}$ is inhabited iff $\Gamma$ is inhabited in \CP. We extend the development presented therein to processes.

In \CLL, it is well known that any environment $\Gamma$ can be internalised as the type $\bigparr \Gamma$, meaning that $\Gamma$ is derivable iff $\bigparr \Gamma$ is. This type can be defined as follows.
\begin{highlight}
\[
  \thl{\bigparr \emptyseq} \defeq \thl{\bot}
  \qquad 
  \thl{\bigparr \Gamma,A} \defeq \thl{A \parr \bigparr \Gamma}
\]
\end{highlight}
\citet{KMP18} observed that like ``$,$'' can be internalised as ``$\parr$'', so can ``$\pp$'' be internalised as ``$\tensor$''. Formally, a hyperenvironment $\hyp{G}$ is internalised as the type $\bigtensor \hyp{G}$, defined as follows.
\begin{highlight}
\[
  \thl{\bigtensor \emptyhyp} \defeq \thl{\one}
  \qquad 
  \thl{\bigtensor \hyp{G} \pp \Gamma} \defeq \thl{\bigparr \Gamma \tensor \bigtensor \hyp{G}}
\]
\end{highlight}
Thus,
$\hyp{G}$ is inhabited iff $\cht{x}{\bigtensor \hyp{G}}$ is inhabited \cite[Theorem~4.11]{KMP18}. 

The same result holds for \pimll. Additionally, we extend this recipe to processes by providing an encoding of \pimll into \CP.
At the core of our encoding are disentanglement and multi-hole contexts, inspired by the internalisation of ``$,$'' and ``$\pp$'' recalled above.
We call these contexts \emph{packing contexts}, because of their effect on process interfaces and the role that they play in the encoding of higher-order into first-order processes that we will present \cref{sec:hopill}.
Below, we write $\hole{}$ to denote a hole in the syntactic tree of a process that can be replaced with a process term \citep{SW01}.

\begin{definition}[Packing context]
The packing context for $\Gamma$ and $y \notin \cn(\Gamma)$ is the $1$-holed context $\pack[\parr]{y}{\Gamma}$ defined below.
\[
\pack[\parr]{y}{\emptyseq} \defeq \wait{y}{\hole{}}
\qquad
\pack[\parr]{y}{\Gamma,\cht{z}{A}} \defeq \recv{y}{z}{\pack[\parr]{y}{\Gamma}}
\]
The packing context for $\hyp{G} = \Gamma_1 \pp \dots \pp \Gamma_n$ and $x \notin \cn(\hyp{G})$ is the $n$-holed context $\pack[\tensor]{x}{\hyp{G}}$, which is defined as follows.
\[
\pack[\tensor]{x}{\emptyhyp} \defeq \close{x}{\nil}
\qquad
\pack[\tensor]{x}{\hyp G \pp \Gamma} \defeq \send{x}{y}{(\pack[\tensor]{x}{\hyp{G}} \pp \pack[\parr]{y}{\Gamma})}
\]
\end{definition}
The choice of $x$ and $y$ in the definition above is immaterial as long as they do not occur in the (hyper)environment associated to the packing context.
In the remainder we will often omit them, and simply call $\pack[\tensor]{x}{\hyp{G}}$ and $\pack[\parr]{y}{\Gamma}$ packing contexts for $\hyp{G}$ and for $\Gamma$, respectively.

We write $\pack[\parr]{x}{\Gamma}[P]$ for the process obtained by replacing the hole $\hole{}$ in $\pack[\parr]{x}{\Gamma}$ with $P$, and $\pack[\tensor]{x}{\hyp{G}}[P_1, \ldots, P_n]$ for the process obtained by replacing each hole in $\pack[\tensor]{x}{\hyp{G}}$ with the respective process $P_i$ (from left to right).

As the name suggest, a packing context ``packs'' the interface of a process and exposes it over the given channel. 
\begin{lemma}
Let $P$, $P_1$, \dots, $P_n$ be processes.
\begin{itemize}
\item If $\judge{}{P}{\Gamma}$ and $x \notin \cn(\Gamma)$, then $\judge{}{\pack[\parr]{x}{\Gamma}[P]}{\cht{x}{\bigparr{\Gamma}}}$.
\item If $\judge{}{P_1 \pp \dots \pp P_n}{\hyp G}$  and $x \notin \cn(\hyp{G})$, then $\judge{}{\pack[\tensor]{x}{\hyp G}[P_1, \dots, P_n]}{\cht{x}{\bigtensor\hyp{G}}}$.
\end{itemize}
\end{lemma}

To pack an arbitrary well-typed process in \pimll, we simply disentangle its parallel components and plug them into the packing context for its typing hyperenvironment.
\begin{definition}[Process Packing]
Let $\judge{}{P}{\hyp{G}}$ and $x \notin \fn(P)$. 
The packing of $P$, denoted $\pack{x}{\judge{}{P}{\hyp{G}}}$, is the process $\pack[\tensor]{x}{\hyp{G}}[P_1, \dots, P_n]$ where $\disen{P} = P_1 \pp \dots \pp P_n$.
\end{definition}

Process packing allows us to demonstrate that the multiplicative fragment of \CP is a ``complete'' theory for \pimll, in the sense of the following theorem.

\begin{theorem}
\label{thm:mcp-sound-mpill}
If $\judge{}{P}{\hyp G}$ in \pill, then $\cpjudge{\pack{x}{\judge{}{P}{\hyp G}}}{\cht{x}{\bigtensor \hyp G}}$ in \MCP.
\end{theorem}

The converse of \cref{thm:mcp-sound-mpill} holds as well. To prove this claim, we define a transformation for \CP derivations of packed processes. By \cref{thm:mpill-complete-mcp-single-env} we can transform any derivation of $\cpjudge{\pack{x}{\judge{}{P}{\hyp G}}}{\cht{x}{\bigtensor \hyp G}}$ into a derivation $\der{D}$ of $\judge{}{\pack{x}{\judge{}{P}{\hyp G}}}{\cht{x}{\bigtensor \hyp G}}$ in the theory of \pimll. The derivation $\der{D}$ has the following form (for $\hyp{G} = \Gamma_1 \pp \dots \pp \Gamma_n$).
\[
  \infer=[\ref{rule:mpill-tensor}]
    {\judge{}{\pack[\tensor]{x}{\hyp{G}}[P_1,\dots,P_n]}{\bigtensor\hyp{G}}}
    {\infer=[\ref{rule:mpill-parr}]
      {\judge{}{\pack[\parr]{x_1}{\Gamma_1}[P_1]}{\bigparr\Gamma_1}}
      {\deduce
        {\judge{}{P_1}{\cht{x_1}{\Gamma_1}}}
        {\der{E}_1}}
    & \dots 
    &\infer=[\ref{rule:mpill-parr}]
          {\judge{}{\pack[\parr]{x_n}{\Gamma_n}[P_n]}{\bigparr\Gamma_n}}
          {\deduce
            {\judge{}{P_n}{\cht{x_n}{\Gamma_n}}}
            {\der{E}_n}}}
\]
By repeatedly applying \cref{rule:mpill-mix} to the premises $\der E_1$, \ldots, $\der E_n$, we obtain the following derivation of $\judge{}{P_1 \pp \dots \pp P_n}{\hyp{G}}$ in \pimll.
\[
  \infer=[\ref{rule:mpill-mix}]
    {\judge{}{P_1 \pp \dots \pp P_n}{\hyp{G}}}
    {\deduce
      {\judge{}{P_1}{\cht{x_1}{\Gamma_1}}}
      {\der{E}_1}
    & \dots 
    &\deduce
      {\judge{}{P_n}{\cht{x_n}{\Gamma_n}}}
      {\der{E}_n}}
\]
Let $\psi$ denote this transformation of derivations, which essentially replaces the applications of \cref{rule:mpill-parr,rule:mpill-tensor} for typing a packing context with applications of \cref{rule:mpill-mix}. We use it to prove the following lemma.

\begin{lemma}\label{lemma:pack-cp-pimll}
If $\cpjudge{\pack{x}{\judge{}{P}{\hyp G}}}{\cht{x}{\bigtensor \hyp G}}$ in \MCP, then $\judge{}{P}{\hyp G}$ in \pill.
\end{lemma}

\Cref{lemma:pack-cp-pimll} is quite inconvenient if our aim is to reuse proof search methods for \CLL in the context of \pimll, because we have to guess $P$ and pack it: what if we wanted to find out whether a generic $\hyp G$ is derivable or not in \pimll?
To reach a more general procedure, we first observe 
that, when we consider the cut-free fragment of \CP, any process that can be typed by the internalisation of a hyperenvironment is necessarily a packing.

\begin{lemma}
Let $\hyp{G} = \Gamma_1 \pp \dots \pp \Gamma_n$.
If $\cpjudge{P}{\cht{x}{\bigtensor \hyp{G}}}$ and $P$ is cut-free, then there are $Q_1$, \dots, $Q_n$ such that
$P = \pack[\tensor]{x}{\hyp{G}}[Q_1,\dots,Q_n]$ and 
$\cpjudge{Q_i}{\cht{x_i}{\Gamma_i}}$ for $i \in [1,n]$.
\end{lemma}

It follows that given a cut-free derivation $\der{D}$ for $\cpjudge{P}{\cht{x}{\bigtensor \hyp G}}$, we can exploit the transformation $\psi$ defined above to obtain the derivation $\psi(\der{D})$ such that $\env(\psi(\der{D})) = \hyp{G}$.
In other words, any proof search strategy for \CLL can be the base for a proof search strategy for \pimll: we simply search for a proof in \CLL that concludes the internalisation of the hyperenvironment of interest, and then apply $\psi$ to the resulting derivation.

\begin{corollary}\label{thm:mcll-search}
If $\der{D}$ is a cut-free derivation of $\bigtensor \hyp{G}$ in the multiplicative fragment of \CLL, then 
$\psi(\der{D})$ is a cut-free derivation of $\hyp{G}$ in \pimll.
\end{corollary}

Since cut is admissible in \CLL, \cref{cor:cll-search} implies in general that proof search in \pimll can be reduced to proof search in \CLL.

%% file: pill.tex
\section{\pill}
\label{sec:pill}

In this section, we present \pill, a calculus in correspondence with classical linear logic with first-order quantifiers (\CLLZeroOne, herein simply \CLL). Compared to \pimll/\MCLL (\cref{sec:mpill}), \pill introduces terms in correspondence with additives, exponentials, and first-order quantifiers.

\subsection{Processes and Typing}

\paragraph{Processes} Processes in \pill are given by the following grammar. We note on the right-hand side which terms have already been discussed in \pimll.
\def\bnfpimll{\small(\pimll)}
\begin{bnftable2}
  {\phl P, \phl Q \Coloneqq {}}%
  {\choice{x}{P}{Q}}
  {(\bnfpimll)}
    \phl P, \phl Q \Coloneqq {}
            & \send{x}{y}{P}       & output $y$ on $x$ and continue as $P$ &\bnfpimll\\
    \mid {} & \recv{x}{y}{P}       & input $y$ on $x$ and continue as $P$ &\bnfpimll\\
    \mid {} & \close{x}{P}            & output (empty message) on $x$ and continue as $P$ &\bnfpimll\\
    \mid {} & \wait{x}{P}          & input (empty message) on $x$ and continue as $P$ &\bnfpimll\\
    \mid {} & \inl{x}{P}           & select left on $x$ and continue as $P$&\\
    \mid {} & \inr{x}{P}           & select right on $x$ and continue as $P$&\\
    \mid {} & \choice{x}{P}{Q}     & offer a binary choice between $P$ (left) or Q (right) on $x$&\\
    \mid {} & \sendtype{x}{A}{P}   & output type $A$ on $x$ and continue as $P$&\\
    \mid {} & \recvtype{x}{X}{P}   & input a type as $X$ on $x$ and continue as $P$&\\
    \mid {} & \server{x}{y}{P}     & offer a service&\\
    \mid {} & \clientuse{x}{y}{P}  & consume a service&\\
    \mid {} & \clientdup{x}{x_1}{x_2}{P} & duplicate a service&\\
    \mid {} & \clientdisp{x}{P}    & dispose of a service&\\
    \mid {} & \res{xy}{P}          & name restriction, ``cut''&\bnfpimll\\
    \mid {} & P \pp Q              & parallel composition of processes $P$ and $Q$&\bnfpimll\\
    \mid {} & \nil                 & terminated process&\bnfpimll\\
    \mid {} & \forward{x}{y}       & link $x$ and $y$&
\end{bnftable2}%
We describe only the new terms (the others have already been discussed in \cref{sec:mpill}).

Terms $\inl{x}{P}$ and $\inr{x}{P}$ respectively send on $x$ the selection of the left or right branch of a (binary) offer available on the other end of the channel before proceeding as $P$. Dually, term $\choice{x}{P}{Q}$ offers on $x$ a choice between proceeding as $P$ (left branch) or $Q$ (right branch).
Terms $\sendtype{x}{A}{P}$ and $\recvtype{x}{X}{P}$ enable polymorphism: term $\sendtype{x}{A}{P}$ sends type $A$ over $x$ and proceeds as $P$; term $\recvtype{x}{X}{P}$ receives a type over $x$ abstracted by the type variable $X$ and then proceeds as $P$, where $X$ is bound in $P$.
Term $\server{x}{y}{P}$ is a server that offers on $x$ a replicable process $P$, where $y$ is bound in $P$.
A server can be used by clients any number of times. Accordingly, we have three client terms to interact with a server.
The client term $\clientuse{x}{y}{P}$ requests exactly one process by the server on $x$, and then proceeds by communicating with the process on channel $y$.
The client term $\clientdisp{x}{P}$ disposes of the server on $x$---the server is used zero times.
The client term $\clientdup{x}{x_1}{x_2}{P}$ requests that the server on $x$ is duplicated, and that the two resulting servers be available on the new channels $x_1$ and $x_2$, respectively.
Term $\forward{x}{y}$ links the endpoints $x$ and $y$ and has the effect of merging their sessions; it can be intuitively though of as a forwarding proxy.

In the remainder, we use $\pi$ to range over term prefixes: $\asend{x}{x'}$, $\arecv{x}{x'}$, $\aclose{x}$, $\await{x}$, $\ainl{x}$, $\ainr{x}$, $\aserveruse{x}{y}$, $\aclientuse{x}{y}$, $\aclientdup{x}{x_1}{x_2}$, and $\aclientdisp{x}$.

Free and bound names of processes and prefixes are defined as expected, as well as $\alpha$-conversion.
In particular, all names that appear inside of round parentheses ``$(-)$'' or square brackets ``$[-]$'' in a term prefix are bound to the continuation of the prefix. As in \pimll, a restriction $\res{xy}{P}$ binds $x$ and $y$ to $P$. All other names are free.

\begin{example}[Free output]
\label{ex:free-output}
Forwarders allow for recovering the usual output primitive for sending free names found in the original $\pi$-calculus, as syntactic sugar \citep{ALM16}.
\[
\sendfree{x}{y}{P} \defeq \send{x}{z}{(\forward{y}{z} \pp P)}
\]
Similar considerations apply to polyadic communications (terms that send multiple names) \cite{SW01}.
\end{example}

\begin{example}[Bit Operations \citep{ALM16,KMP19}]
\label{ex:and-server}
We report how to write a server that computes the logical AND of two bits in calculi based on linear logic.
We use selections to model sending bits.
The example uses the following syntactic sugar.
\begin{gather*}
\send x0P \defeq \send{x}{x'}{\inl{x'}{\close {x'}P}}
\qquad
\send x1P \defeq \send{x}{x'}{\inr{x'}{\close {x'}P}}
\\
\choicebit xPQ \defeq \choice x{\wait xP}{\wait xQ}
\end{gather*}
With these abbreviations, we can write a server that offers a service for computing logical AND.
\begin{equation*}
Server_y \defeq 
  \server{y}{y'}
  {\recv{y'}{p}
    {\recv{y'}{q}
      {\choicebit*{p}
        {\choicebit{q}
          {\send{y'}{0}{\close{y'}{\nil}}}
          {\send{y'}{0}{\close{y'}{\nil}}}}
        {\choicebit{q}
          {\send{y'}{0}{\close{y'}{\nil}}}
          {\send{y'}{1}{\close{y'}{\nil}}}}}}}
\end{equation*}
We now define a compatible client, $Client^{b_1b_2}_{xz}$, which sends bits $b_1$ and $b_2$ (0 or 1) to a server that accepts two bits on $x$ (the client abstracts from the concrete operation that the server computes).
The client uses the result to decide whether to select left or right on another channel $z$.
\begin{equation*}
Client^{b_1b_2}_{xz} =
\clientuse x{x'}{
\send{x'}{b_1}{
\send{x'}{b_2}{\choicebit{x'}{
\wait {x'}{\inl z{\close z\nil}}
}{
\wait {x'}{\inr z{\close z\nil}}
}
}}}
\end{equation*}
\end{example}

\begin{example}[Polymorphic API Gateway]
In the software paradigm of microservices \citep{DGLMMMS17}, an API gateway (Application Programming Interface gateway) is a polymorphic proxy server that offers a single endpoint through which clients can access other servers \citep{MW16}.
We can model this behaviour as the following process.
\[
Gateway_{x} \defeq
  \recvtype{x}{X_l}
    {\recvtype{x}{X_r}
      {\recv{x}{x_l}
        {\recv{x}{x_r}
          {\server{x}{x'}
            {\choice*{x'}
              {\clientdisp{x_r}{\clientuse{x_l}{x_l'}{\forward{x_l'}{x'}}}}
              {\clientdisp{x_l}{\clientuse{x_r}{x_r'}{\forward{x_r'}{x'}}}}}}}}}
\]
Process $Gateway_x$ models a gateway for two other servers.
First, it receives on $x$ the types of the two servers that clients will be able to use, abstracted by $X_l$ and $X_r$, along with the endpoints over which these servers are available.
Then, it waits to receive on $x$ client invocations, which subsequently decide whether they want to use the first or the second API. The gateway then sets up a connection to the right server by using appropriate client requests and a forwarder.

Let us see an example of how $Gateway_x$ can be used. Let $S_l$ and $S_r$ be servers that offer services at the endpoints $y_l$ and $y_r$, respectively. We can aggregate them using $Gateway_{x}$ to offer both services on $x$ as follows, where $A_l$ and $A_r$ stand for the respective types of the behaviours offered on $y_l$ and $y_r$.
\[
S_{l+r} \defeq 
\res{xy}{( Gateway_{x} \pp
  \sendtype{y}{A_l}
    {\sendtype{y}{A_r}
      {\send{y}{y_l}{( S_l \pp \send{y}{y_r}{( S_r \pp
        \forward{y}{z})})}}})}
\]
To use one of the services behind the gateway, a client simply needs to first send a left or right selection. For example, $\clientuse{x}{x'}{\inl{x}{P}}$ is a client for $S_{l+r}$ that uses the service offered by $S_l$ through the gateway.
\[
  \res{xz}{(\clientuse{x}{x'}{\inl{x}{P}} \pp S_{l+r})}
\]
\end{example}

\paragraph{Typing}
Types in \pill extend those of \pimll to include also the additive and exponential propositions of \CLL.
\begin{bnftable2}[\thl]%
  {\thl{A},\thl{B} \Coloneqq {}}%
  {A \tensor B}%
  {(\bnfpimll)}
  \thl{A},\thl{B} \Coloneqq {} & A \tensor B & send $A$, proceed as $B$&\bnfpimll\\
    \mid {} & A \parr B   & receive $A$, proceed as $B$&\bnfpimll\\
    \mid {} & \one        & empty output, unit for $\tensor$&\bnfpimll\\
    \mid {} & \bot        & empty input, unit for $\parr$&\bnfpimll\\
    \mid {} & {A \oplus B} & select $A$ or $B$&\\
    \mid {} & {A \with B} & offer $A$ or $B$&\\
    \mid {} & {X} & type variable&\\
    \mid {} & {\dual X} & dual of type variable&\\
    \mid {} & {\exists X.A} & existential, type output&\\
    \mid {} & {\forall X.A} & universal, type input&\\
    \mid {} & {\query A} & client request&\\
    \mid {} & {\bang A} & server accept&\\
\end{bnftable2}%
Duality is extended to additives and exponentials exactly as in \CLL.
\begin{highlight}
\begin{gather*}
\thl{\dual{(A \tensor B)}} = \thl{\dual{A} \parr \dual{B}}
\qquad
\thl{\dual{(A \parr B)}} = \thl{\dual{A} \tensor \dual{B}}
\qquad
\thl{\dual{\one}} = \thl{\bot}
\qquad
\thl{\dual{\bot}} = \thl{\one}
\\
\thl{\dual{(A \oplus B)}} = \thl{\dual{A} \with \dual{B}}
\qquad
\thl{\dual{(A \with B)}} = \thl{\dual{A} \oplus \dual{B}}
\qquad
\thl{\dual{(\query A)}} = \thl{\bang\dual{A}}
\qquad
\thl{\dual{(\bang A)}} = \thl{\query\dual{A}}
\\
\thl{\dual{(X)}} = \thl{\dual{X}}
\qquad
\thl{\dual{(\dual X)}} = \thl{X}
\qquad
\thl{\dual{(\exists X.A)}} = \thl{\forall X.\dual{A}}
\qquad
\thl{\dual{(\forall X.A)}} = \thl{\exists X.\dual{A}}
\end{gather*}
\end{highlight}

Environments and hyperenvironments are defined as in \pimll (but contain types in the extended syntax). Typing judgements have the same form, $\judge{}{P}{\hyp G}$, as well.

\begin{figure}[t]
  \begin{highlight}\small 
  \begin{spreadlines}{\typerulesskipamount}%
    \headertext{Structural rules}%
     \begin{gather*}
      \infer[\rlabel{\rname{Ax}}{rule:pill-axiom}]
      {\judge{}{\forward{x}{y}}{\cht{x}{A^\bot}, \cht{y}{A}}}
      {}
      \quad
      \infer[\rrelabel{rule:mpill-cut}{rule:pill-cut}]
      {\judge{}{\res{xy}{P}}{\hyp{G} \pp \Gamma, \Delta}}
      {\judge{}{P}{\hyp{G}\pp\Gamma, \cht{x}{A} \pp \Delta, \cht{y}{\dual{A}}}}
      \quad
      \infer[\rrelabel{rule:mpill-mix}{rule:pill-mix}]
      {\judge{}{P \pp Q}{\hyp{G} \pp \hyp{H}}}
      {\judge{}{P}{\hyp{G}}&\judge{}{Q}{\hyp{H}}}
      \quad
      \infer[\rrelabel{rule:mpill-mix-0}{rule:pill-mix-0}]
      {\judge{}{\nil}{\emptyhyp}}
      {}
    \end{gather*}%
    \headertext{Logical rules}%
    \begin{gather*}
      \infer[\rrelabel{rule:mpill-tensor}{rule:pill-tensor}]
        {\judge{}{\send{x}{y}{P}}{\Gamma,\Delta,\cht{x}{A \tensor B}}}
        {\judge{}{P}{\Gamma,\cht{y}{A} \pp \Delta,\cht{x}{B}}}
      \qquad
      \infer[\rrelabel{rule:mpill-one}{rule:pill-one}]
        {\judge{}{\close{x}{P}}{\cht{x}{\one}}}
        {\judge{}{P}{\emptyhyp}}
      \qquad
      \infer[\rrelabel{rule:mpill-parr}{rule:pill-parr}]
        {\judge{}{\recv{x}{y}{P}}{\Gamma, \cht{x}{A \parr B}}}
        {\judge{}{P}{\Gamma, \cht{y}{A}, \cht{x}{B}}}
      \qquad
      \infer[\rrelabel{rule:mpill-bot}{rule:pill-bot}]
        {\judge{}{\wait{x}{P}}{\Gamma, \cht{x}{\bot}}}
        {\judge{}{P}{\Gamma}}
      \\
      \infer[\rlabel{\rname{$\oplus_1$}}{rule:pill-oplus_1}]
        {\judge{}{\inl{x}{P}}{\Gamma, \cht{x}{A \oplus B}}}
        {\judge{}{P}{\Gamma, \cht{x}{A}}}
      \qquad
      \infer[\rlabel{\rname{$\oplus_2$}}{rule:pill-oplus_2}]
        {\judge{}{\inr{x}{P}}{\Gamma, \cht{x}{A \oplus B}}}
        {\judge{}{P}{\Gamma, \cht{x}{B}}}
      \qquad
      \infer[\rlabel{\rname{$\with$}}{rule:pill-with}]
        {\judge{}{\choice{x}{P}{Q}}{\Gamma, \cht{x}{A \with B}}}
        {\judge{}{P}{\Gamma, \cht{x}{A}} 
        &\judge{}{Q}{\Gamma, \cht{x}{B}}}
      \\
      \infer[\rlabel{\rname{$\bang$}}{rule:pill-!}]
        {\judge{}{\server{x}{y}{P}}{\query\Gamma, \cht{x}{\bang A}}}
        {\judge{}{P}{\query\Gamma, \cht{y}{A}}}
      \qquad
      \infer[\rlabel{\rname{$\query$}}{rule:pill-?}]
        {\judge{}{\clientuse{x}{y}{P}}{\Gamma, \cht{x}{\query A}}}
        {\judge{}{P}{\Gamma, \cht{y}{A}}}
      \qquad
      \infer[\rlabel{\rname{W}}{rule:pill-weaken}]
        {\judge{}{\clientdisp{x}{P}}{\Gamma, \cht{x}{\query A}}}
        {\judge{}{P}{\Gamma}}
      \qquad
      \infer[\rlabel{\rname{C}}{rule:pill-contract}]
        {\judge{}{\clientdup{x}{x'}{x''}{P}}{\Gamma, \cht{x}{\query A}}}
        {\judge{}{P}{\Gamma, \cht{x'}{\query A}, \cht{x''}{\query A}}}
      \\
      \infer[\rlabel{\rname{$\exists$}}{rule:pill-exists}]
        {\judge{}{\sendtype{x}{A}{P}}{\Gamma,\cht{x}{\exists X.B}}}
        {\judge{}{P}{\Gamma,\cht{x}{B\{A/X\}}}}
      \qquad
      \infer[\rlabel{\rname{$\forall$}}{rule:pill-forall}]
       {\judge{}{\recvtype{x}{X}{P}}{\Gamma,\cht{x}{\forall X.B}}}
       {\judge{}{P}{\Gamma,\cht{x}{B}}
       &\thl{X \notin \ftv(\Gamma)}}
    \end{gather*}%
    \footer%
  \end{spreadlines}%
  \end{highlight}%
  \caption{\pill, typing rules.}
  \label{fig:pill-typing-rules}
\end{figure}

The inference rules for deriving typing judgements in \pill are displayed in \cref{fig:pill-typing-rules}.

\Cref{rule:pill-cut,rule:pill-mix,rule:pill-mix-0,rule:pill-tensor,rule:pill-one,rule:pill-bot} correspond to \pimll and are covered in detail in \cref{sec:mpill}. We discuss the new rules, which cover the additive, exponential, and first-order quantification fragments of \CLL.

\Cref{rule:pill-axiom} types a link (or ``forwarder'') between $x$ and $y$ by requiring that the types of $x$ and $y$ are the dual of each other. 
This ensures that any message on $x$ can be safely forwarded to $y$, and \viceversa.
\Cref{rule:pill-exists,rule:pill-forall} type polymorphic session behaviour by using quantifiers.

\Cref{rule:pill-oplus_1,rule:pill-oplus_2} type, respectively, the choice of the left or the right branch of an alternative behaviour offered on the other end of the session. Dually, \cref{rule:pill-with} types the offering of a choice between two behaviours.

All rules for typing channels enforce linear usage aside for client requests (typed with the exponential connective $\query$), for which contraction and weakening are allowed.
Specifically, contraction (\cref{rule:pill-contract}) allows for having multiple client requests for the same server endpoint, and weakening (\cref{rule:pill-weaken}) allows for having a client that does not use a server.
\Cref{rule:pill-!} types a server, where $\query\Gamma$ denotes that all session types in $\Gamma$ must be client requests, \ie, $\query \Gamma \Coloneqq \cht{x_1}{\query A_1}, \ldots, \cht{x_n}{\query A_n}$.
A server must be executable any number of times, since it does not know how many client requests it will have to support. 
This is guaranteed by requiring that all resources used by the server are acquired by communicating with the client (according to protocol $A$) and with other servers ($\query\Gamma$).

\begin{example}
The syntactic sugar for free output shown in \cref{ex:free-output} can be typed with a derivable rule, as shown below.
\[
\array{c}
\infer
  {\judge{}{\sendfree xyP}{\Gamma, \cht{y}{\dual A}, \cht{x}{A \tensor B}}}
  {\judge{}{P}{\Gamma, \cht{x}{B}}}
\endarray 
\quad \defeq \quad
\array{c}
\infer[\ref{rule:pill-tensor}]
  {\judge{}{\send{x}{z}{( \forward{y}{z} \pp P )}}{\Gamma, \cht{y}{\dual A}, \cht{x}{A \tensor B}}}
  {\infer[\ref{rule:pill-mix}]
    {\judge{}{\forward{y}{z} \pp P}{\Gamma, \cht{x}{B} \pp  \cht{y}{\dual A}, \cht{z}{A}}}
    {\infer[\ref{rule:pill-axiom}]
      {\judge{}{\forward{y}{z}}{\cht{y}{\dual A}, \cht{x}{A}}}
      {}
    &\judge{}{P}{\Gamma, \cht{x}{B}}}}
\endarray
\]
\end{example}

\begin{example}
Define the types for sending and receiving a bit, respectively.
\[
Bit \defeq \one \oplus \one \quad \emph{send a bit} \qquad\qquad \dual{Bit} \defeq \bot \with \bot \quad \emph{receive a bit}
\]
Then, we can type the server and client terms from \cref{ex:and-server} with dual types, as follows.
\[
\judge{}{Server_y}{\cht{y}{\bang(\dual{Bit} \parr \dual{Bit} \parr Bit \tensor \one)}} \qquad
\judge{}{Client^{b_1b_2}_{xz}}{
\cht{x}{\query ( Bit \tensor Bit \tensor \dual{Bit} \parr \bot), \cht{z}{Bit}}
}
\]
\label{ex:and-typing}
Thus, by \cref{rule:pill-cut,rule:pill-mix} we can type their composition for all distinct names $x$, $y$ and $z$, and any bits $b_1$ and $b_2$, \eg, to compute the logical AND of $0$ and $1$:
\[\judge{}{\res{xy}{\left( Client^{01}_{xz} \pp Server_y\right)}}{
\cht{z}{Bit}}\text{.}\]
\end{example}

\begin{remark}
\label{rem:pill-axiom-typing}
\Cref{rule:pill-axiom} is the standard axiom of \CLL.
It is well known that the axiom is admissible in the presence of a restricted form of it, as shown below, which accepts only atomic propositions (type variables).
\[
  \infer
    {\judge{}{\forward{x}{y}}{\cht{x}{\dual X}, \cht{y}{X}}}
    {}
\]
When translated to theories like \pill, the proof of admissibility of the axiom for general propositions corresponds to an $\eta$-expansion procedure: for every $A$, $x$, and $y$, we can construct a process $Fwd_{x,y}^{A}$ that implements the forwarding of protocol $A$ from $x$ to $y$. Below we show some illustrative cases (others are similar, see \citep{CLMSW16}).
\begin{gather*}
\array{c}
\infer
  {\judge{}{Fwd_{x,y}^{1}}{\cht{x}{\bot},\cht{y}{\one}}}
  {}
\endarray
\defeq
\array{c}
\infer[\ref{rule:pill-bot}]
  {\judge{}{\wait{x}{\close{y}{\nil}}}{\cht{x}{\bot},\cht{y}{\one}}}
  {\infer[\ref{rule:pill-one}]
    {\judge{}{\close{y}{\nil}}{\cht{y}{\one}}}
    {\infer[\ref{rule:pill-mix-0}]
      {\judge{}{\nil}{\emptyhyp}}
      {}}}
\endarray
=
\array{c}
\infer
  {\judge{}{Fwd_{y,x}^{\bot}}{\cht{x}{\bot},\cht{y}{\one}}}
  {}
\endarray
\\
\array{c}
\infer
  {\judge{}{Fwd_{x,y}^{\exists X.X}}{\cht{x}{\forall X.\dual{X}},\cht{y}{\exists X.X}}}
  {}
\endarray
\defeq
\array{c}
\infer[\ref{rule:pill-forall}]
  {\judge{}
    {\recvtype{x}{X}{\sendtype{y}{X}{\forward{x}{y}}}}
    {\cht{x}{\forall X.\dual{X}},\cht{y}{\exists X.X}}}
  {\infer[\ref{rule:pill-exists}]
    {\judge{}
        {\sendtype{y}{X}{\forward{x}{y}}}
        {\cht{x}{\dual{X}},\cht{y}{\exists X.X}}}
    {\infer[\ref{rule:pill-axiom}]
      {\judge{}
          {\forward{x}{y}}
          {\cht{x}{\dual{X}},\cht{y}{X}}}
      {}}}
\endarray
\end{gather*}
\end{remark}

\subsection{Operational Semantics of Derivations}
\label{sec:pill-sos}

\rlabelfwd{rule:mpill-mix-1}{rule:pill-mix-1}
\rlabelfwd{rule:mpill-mix-2}{rule:pill-mix-2}
\rlabelfwd{rule:mpill-mix-sync}{rule:pill-mix-sync}
\rlabelfwd{rule:mpill-cut-one}{rule:pill-cut-one}
\rlabelfwd{rule:mpill-cut-tensor}{rule:pill-cut-tensor}
\rlabelfwd{rule:mpill-cut-res}{rule:pill-cut-res}
\rlabelfwd{rule:mpill-alpha}{rule:pill-alpha}

We present the SOS specification for the new ingredients of \pill compared to \pimll.

\paragraph{Additives}
The derivation rules for selection (\ref{rule:pill-oplus_1}, \ref{rule:pill-oplus_2}) and choice (\ref{rule:pill-with}) are given below.
There are left and right rules for actions and communications. They are all symmetric.
\IfLabelExistsTF{sec:omitted-rules}
  {We omit the right cases, which are in \cref{sec:omitted-rules}.}
\begin{drules}
\plto
  {\infer[\ref{rule:pill-oplus_1}]
    {\judge{}{\inl{x}{P}}{\Gamma, \cht{x}{A \oplus B}}}
    {\deduce
      {\judge{}{P}{\Gamma, \cht{x}{A}}}
      {\der{D}}}}
  {\linl{x}}
  {\deduce
    {\judge{}{P}{\Gamma, \cht{x}{A}}}
    {\der{D}}}
\qquad
\plto
  {\infer[\ref{rule:pill-oplus_2}]
    {\judge{}{\inr{x}{P}}{\Gamma, \cht{x}{A \oplus B}}}
    {\deduce
      {\judge{}{P}{\Gamma, \cht{x}{B}}}
      {\der{D}}}}
  {\linr{x}}
  {\deduce
    {\judge{}{P}{\Gamma, \cht{x}{B}}}
    {\der{D}}}
\\
\plto
  {\infer[\ref{rule:pill-with}]
    {\judge{}{\choice{x}{P}{Q}}{\Gamma, \cht{x}{A \with B}}}
    {\deduce
      {\judge{}{P}{\Gamma, \cht{x}{A}}}
      {\der{D}}
    &\deduce
      {\judge{}{Q}{\Gamma, \cht{x}{B}}}
      {\der{E}}}}
  {\lcoinl{x}}
  {\deduce
    {\judge{}{P}{\Gamma, \cht{x}{A}}}
    {\der{D}}}
\\
\plto
  {\infer[\ref{rule:pill-with}]
    {\judge{}{\choice{x}{P}{Q}}{\Gamma, \cht{x}{A \with B}}}
    {\deduce
      {\judge{}{P}{\Gamma, \cht{x}{A}}}
      {\der{D}}
    &\deduce
      {\judge{}{Q}{\Gamma, \cht{x}{B}}}
      {\der{E}}}}
  {\lcoinr{x}}
  {\deduce
    {\judge{}{Q}{\Gamma, \cht{x}{B}}}
    {\der{E}}}
\\
\infer[\rlabel{\ref*{rule:pill-oplus_1}\ref*{rule:pill-with}}{rule:pill-cut-oplus_1}]
  {\plto
    {\infer[\ref{rule:pill-cut}]
     {\judge{}{\res{xy}{P}}{\hyp{G} \pp \Gamma, \Delta}}
     {\deduce
      {\judge{}{P}{\hyp{G}\pp\Gamma,\cht{x}{A \oplus B} \pp \Delta, \cht{y}{\dual{A} \parr \dual{B}}}}
      {\der{D}}}}
    {\tau}
    {\infer[\ref{rule:pill-cut}]
     {\judge{}{\res{xy}{P'}}{\hyp{G} \pp \Gamma, \Delta}}
     {\deduce
       {\judge{}{P'}{\hyp{G}\pp\Gamma,\cht{x}{A} \pp \Delta, \cht{y}{\dual{A}}}}
       {\der{D'}}}}}
  {\plto
    {\deduce
      {\judge{}{P}{\hyp{G}\pp\Gamma,\cht{x}{A \oplus B} \pp \Delta, \cht{y}{\dual{A} \parr \dual{B}}}}
      {\der{D}}}
    {\lsync
      {\linl{x}}
      {\lcoinl{y}}}
    {\deduce
     {\judge{}{P'}{\hyp{G}\pp\Gamma,\cht{x}{A} \pp \Delta, \cht{y}{\dual{A}}}}
     {\der{D'}}}}
\\
\infer[\rlabel{\ref*{rule:pill-oplus_2}\ref*{rule:pill-with}}{rule:pill-cut-oplus_2}]
  {\plto
    {\infer[\ref{rule:pill-cut}]
     {\judge{}{\res{xy}{P}}{\hyp{G} \pp \Gamma, \Delta}}
     {\deduce
      {\judge{}{P}{\hyp{G}\pp\Gamma,\cht{x}{A \oplus B} \pp \Delta, \cht{y}{\dual{A} \parr \dual{B}}}}
      {\der{D}}}}
    {\tau}
    {\infer[\ref{rule:pill-cut}]
     {\judge{}{\res{xy}{P'}}{\hyp{G} \pp \Gamma, \Delta}}
     {\deduce
       {\judge{}{P'}{\hyp{G}\pp\Gamma,\cht{x}{B} \pp \Delta, \cht{y}{\dual{B}}}}
       {\der{D'}}}}}
  {\plto
    {\deduce
      {\judge{}{P}{\hyp{G}\pp\Gamma,\cht{x}{A \oplus B} \pp \Delta, \cht{y}{\dual{A} \parr \dual{B}}}}
      {\der{D}}}
    {\lsync
      {\linr{x}}
      {\lcoinr{y}}}
    {\deduce
     {\judge{}{P'}{\hyp{G}\pp\Gamma,\cht{x}{B} \pp \Delta, \cht{y}{\dual{B}}}}
     {\der{D'}}}}
\end{drules}

\paragraph{Quantifiers}

Polymorphism is achieved by communicating types, according to the following transition axioms and rule.
\begin{drules}
\plto
  {\infer[\ref{rule:pill-exists}]
     {\judge{}{\sendtype{x}{A}{P}}{\Gamma,\cht{x}{\exists X.B}}}
     {\deduce
      {\judge{}{P}{\Gamma,\cht{x}{B\{A/X\}}}}
      {\der{D}}}}
  {\lsendtype{x}{A}}
  {\deduce
    {\judge{}{P}{\Gamma,\cht{x}{B\{A/X\}}}}
    {\der{D}}}
\\
\plto
  {\infer[\ref{rule:pill-forall}]
     {\judge{}{\recvtype{x}{X}{P}}{\Gamma,\cht{x}{\forall X.B}}}
     {\deduce
      {\judge{}{P}{\Gamma,\cht{x}{B}}}
      {\der{D}}}}
  {\lrecvtype{x}{A}}
  {\deduce
    {\judge{}{P\{A/X\}}{\Gamma,\cht{x}{B\{A/X\}}}}
    {\der{D}\{A/X\}}}
\\
\infer[\rlabel{\ref*{rule:pill-exists}\ref*{rule:pill-forall}}{rule:pill-cut-exists}]
  {\plto
    {\infer[\ref{rule:pill-cut}]
      {\judge{}{\res{xy}{P}}{\hyp{G} \pp \Gamma,\Delta}}
      {\deduce
        {\judge{}{P}{\hyp{G}\pp \Gamma,\cht{x}{\exists X.B} \pp \Delta, \cht{y}{\forall X.\dual{B}}}}
        {\der{D}}}}
    {\tau}
    {\infer[\ref{rule:pill-cut}]
      {\judge{}{\res{xy}{P}}{\hyp{G} \pp \Gamma,\Delta}}
      {\deduce
        {\judge{}{P}{\hyp{G}\pp \Gamma,\cht{x}{B\{A/X\}} \pp \Delta, \cht{y}{\forall X.\dual{B}\{A/X\}}}}
        {\der{D'}}}}}
  {\plto
    {\deduce
      {\judge{}{P}{\hyp{G}\pp \Gamma,\cht{x}{\exists X.B} \pp \Delta, \cht{y}{\forall X.\dual{B}}}}
      {\der{D}}}
    {\lsync
      {\lsendtype{x}{A}}
      {\lrecvtype{y}{A}}}
    {\deduce
      {\judge{}{P}{\hyp{G}\pp \Gamma,\cht{x}{B\{A/X\}} \pp \Delta, \cht{y}{\forall X.\dual{B}\{A/X\}}}}
      {\der{D'}}}}
\end{drules}

\paragraph{Axiom}
For derivations that consist of an application of \cref{rule:pill-axiom}, we have two (symmetric) transition axioms.
\begin{drules}
\plto
    {\infer[\ref{rule:pill-axiom}]
      {\judge{}{\forward{x}{y}}{\cht{x}{\dual{A}}, \cht{y}{A}}}
      {\phantom{\pp}}}
    {\lforward{x}{y}}
     {\infer[\ref{rule:pill-mix-0}]
      {\judge{}{\nil}{\emptyhyp}}
      {\vphantom{\pp}}}
\qquad
  \plto
     {\infer[\ref{rule:pill-axiom}]
       {\judge{}{\forward{x}{y}}{\cht{x}{\dual{A}}, \cht{y}{A}}}
       {\phantom{\pp}}}
     {\lforward{y}{x}}
     {\infer[\ref{rule:pill-mix-0}]
      {\judge{}{\nil}{\emptyhyp}}
      {\vphantom{\pp}}}
\intertext{
When the transition of an axiom interacts with a related cut, the cut disappears and a substitution is performed. This rule mimics the typical simplification of cuts applied to axioms found in linear logic \citep{W14}.
}
\infer[\rlabel{\rname{\ref*{rule:pill-axiom}\rname{Cut}}}{rule:pill-cut-axiom}]
  {\plto
    {\infer[\ref{rule:pill-cut}]
     {\judge{}{\res{yz}{P}}{\hyp{G} \pp \Gamma, \cht{x}{\dual{A}}}}
     {\deduce
       {\judge{}{P}{\hyp{G}\pp \cht{x}{\dual{A}},\cht{y}{A} \pp \Gamma, \cht{z}{\dual{A}}}}
       {\der{D}}}}
    {\tau}
    {\deduce
      {\judge{}{P'}{\hyp{G} \pp \Gamma,\cht{z}{\dual{A}}}}
      {\der{D'}}}
  }{\plto
    {\deduce
      {\judge{}{P}{\hyp{G}\pp \cht{x}{\dual{A}},\cht{y}{A} \pp \Gamma, \cht{z}{\dual{A}}}}
      {\der{D}}}
    {\lforward{x}{y}}
    {\deduce
      {\judge{}{P'}{\hyp{G} \pp \Gamma,\cht{z}{\dual{A}}}}
      {\der{D'}}}}
\end{drules}

\paragraph{Exponentials}
The semantics of exponentials models interactions between clients and servers.
A clients can use a server in three ways: request a single instance of the process provided by the server, duplicate the server, or dispose of the server.

Instance requests are modelled by the following rules, for the prefixes and their interaction.
\begin{drules}
\plto
  {\infer[\ref{rule:pill-?}]
    {\judge{}{\clientuse{x}{x'}{P}}{\Gamma, \cht{x}{\query A}}}
    {\deduce
      {\judge{}{P}{\Gamma, \cht{x'}{A}}}
      {\der{D}}}}
  {\luse{x}{x'}}
  {\deduce
    {\judge{}{P}{\Gamma, \cht{x'}{A}}}
    {\der{D}}}
\quad
\plto
  {\infer[\ref{rule:pill-!}]
    {\judge{}{\server{x}{x'}{P}}{\query\Gamma, \cht{x}{\bang A}}}
    {\deduce
      {\judge{}{P}{\query\Gamma, \cht{x'}{A}}}
      {\der{D}}}}
  {\lcouse{x}{x'}}
  {\deduce
    {\judge{}{P}{\query\Gamma, \cht{x'}{A}}}
    {\der{D}}}
\\
\infer[\rlabel{\ref*{rule:pill-!}\ref*{rule:pill-?}}{rule:pill-cut-?}]
{\plto
  {\infer[\ref{rule:pill-cut}]
    {\judge{}{\res{xy}{P}}{\hyp{G} \pp \Gamma, \Delta}}
    {\deduce
      {\judge{}{P}{\hyp{G} \pp \Gamma, \cht{x}{\query A} \pp \Delta,\cht{y}{\bang \dual A}}}
      {\der{D}}}}
  {\tau}
  {\infer[\ref{rule:pill-cut}]
    {\judge{}{\res{x'y'}{P'}}{\hyp{G} \pp \Gamma, \Delta}}
    {\deduce
      {\judge{}{P'}{\hyp{G} \pp \Gamma, \cht{x'}{A} \pp \Delta,\cht{y'}{\dual A}}}
      {\der{D'}}}}}
{\plto
  {\deduce
    {\judge{}{P}{\hyp{G} \pp \Gamma, \cht{x}{\query A} \pp \Delta,\cht{y}{\bang \dual A}}}
    {\der{D}}}
  {\lsync
    {\luse{x}{x'}}
    {\lcouse{y}{y'}}}
  {\deduce
    {\judge{}{P'}{\hyp{G} \pp \Gamma, \cht{x'}{A} \pp \Delta,\cht{y'}{\dual A}}}
    {\der{D'}}}}
\end{drules}

The following rules model server duplication.
A technicality: since a server can depend on other servers (the $\query\Gamma$ in \cref{rule:pill-!}), duplicating a server requires in turn to duplicate its dependencies.
In the following, we use $\sigma$ to denote a name substitution. We write $P\sigma$, $\Gamma\sigma$, $\der{D}\sigma$, and $x\sigma$ for the application of a substitution $\sigma$ to a process $P$, an environment $\Gamma$, a derivation $\der{D}$, and a name $x\sigma$, respectively.
\begin{drules}
{\plto
  {\infer[\ref{rule:pill-contract}]
    {\judge{}{\clientdup{x}{x_1}{x_2}{P}}{\Gamma, \cht{x}{\query A}}}
    {\deduce
      {\judge{}{P}{\Gamma, \cht{x_1}{\query A},\cht{x_2}{\query A}}}
      {\der{D}}}}
  {\ldup{x}{x_1}{x_2}}
  {\infer[\ref{rule:pill-parr}]
    {\judge{}{\recv{x_1}{x_2}{P}}{\Gamma, \cht{x_1}{\query A \parr \query A}}}
    {\deduce
      {\judge{}{P}{\Gamma, \cht{x_1}{\query A},\cht{x_2}{\query A}}}
      {\der{D}}}}}
\\
\infer
  {\plto*
    {\infer[\ref{rule:pill-!}]
      {\judge{}{\server{x}{x'}{P}}{\query\Gamma, \cht{x}{\bang A}}}
      {\deduce
        {\judge{}{P}{\bang\Gamma, \cht{x'}{A}}}
        {\der{D}}}}
    {\lcodup{x}{x_1}{x_2}}
    {\infer=[\ref{rule:pill-contract}]
      {\judge{}
        {\clientdup{z_1}{z_1\sigma_1}{z_1\sigma_2}{\dots\clientdup{z_n}{z_n\sigma_1}{z_n\sigma_2}{\send{x_1}{x_2}{(\server{x_1}{x'\sigma_1}{P_1} \pp \server{x_2}{x'\sigma_2}{P_2})}}}}
        {\query\Gamma, \cht{x_1}{\bang A \tensor \bang A}}}
      {\infer[\ref{rule:pill-tensor}]
        {\judge{}{\send{x_1}{x_2}{(\server{x_1}{x'\sigma_1}{P} \pp \server{x_2}{x'\sigma_2}{P})}}{\query\Gamma_1,\query\Gamma_2, \cht{x_1}{\bang A \tensor \bang A}}}
        {\infer[\ref{rule:pill-mix}]
          {\judge{}
            {\server{x_1}{x'\sigma_1}{P_1} \pp \server{x_2}{x'\sigma_2}{P_2}}
            {\query\Gamma_1,\cht{x_1}{\bang A} \pp \query\Gamma_2,\cht{x_2}{\bang A}}}
          {\infer[\ref{rule:pill-!}]
            {\judge{}{\server{x_1}{x'\sigma_1}{P_1}}{\query\Gamma_1, \cht{x_1}{\bang A}}}
            {\deduce
              {\judge{}{P_1}{\query\Gamma_1, \cht{x'\sigma_1}{A}}}
              {\der{D}\sigma_1}}
          &\infer[\ref{rule:pill-!}]
            {\judge{}{\server{x_2}{x'\sigma_2}{P_2}}{\query\Gamma_2, \cht{x_2}{\bang A}}}
            {\deduce
              {\judge{}{P_2}{\query\Gamma_2, \cht{x'\sigma_2}{A}}}
              {\der{D}\sigma_2}}}}}}}
  {\phl{P_i = P\sigma_i}
   &\thl{\Gamma_i = \Gamma\sigma_i}
   &\phl{\text{for } i \in \{1,2\}}
   &\phl{\fn(P_1) \cap \fn(P_2) = \emptyset}
   &\phl{\fn(P)\setminus\{x'\} = \{z_1,\dots,z_n\}}}
\\
\infer[\rlabel{\ref*{rule:pill-!}\ref*{rule:pill-contract}}{rule:pill-cut-contract}]
{\plto
  {\infer[\ref{rule:pill-cut}]
    {\judge{}{\res{xy}{P}}{\hyp{G} \pp \Gamma, \query \Delta}}
    {\deduce
      {\judge{}{P}{\hyp{G} \pp \Gamma, \cht{x}{\query A} \pp \query \Delta,\cht{y}{\bang \dual A}}}
      {\der{D}}}}
  {\tau}
  {\infer[\ref{rule:pill-cut}]
    {\judge{}{\res{x_1y_1}{P'}}{\hyp{G} \pp \Gamma, \query \Delta}}
    {\deduce
      {\judge{}{P'}{\hyp{G} \pp \Gamma, \cht{x_1}{\query A\parr\query A} \pp \query \Delta,\cht{y_1}{\bang \dual A \tensor\bang \dual A}}}
      {\der{D'}}}}}
{\plto
  {\deduce
    {\judge{}{P}{\hyp{G} \pp \Gamma, \cht{x}{\query A} \pp \query \Delta,\cht{y}{\bang \dual A}}}
    {\der{D}}}
  {\lsync
    {\ldup{x}{x_1}{x_2}}
    {\lcodup{y}{y_1}{y_2}}}
  {\deduce
    {\judge{}{P'}{\hyp{G} \pp \Gamma, \cht{x_1}{\query A\parr\query A} \pp \query \Delta,\cht{y_1}{\bang \dual A \tensor\bang \dual A}}}
    {\der{D'}}}}
\end{drules}

The disposal of a server is modelled by the following rules, again for the prefixes and their interaction.
Disposing of a server triggers the disposal of all its dependencies.
\begin{drules}
\plto
  {\infer[\ref{rule:pill-weaken}]
    {\judge{}{\clientdisp{x}{P}}{\Gamma, \cht{x}{\query A}}}
    {\deduce
      {\judge{}{P}{\Gamma}}
      {\der{D}}}}
  {\plbl{\ldisp{x}}{\lweaken{A}}}
  {\infer[\ref{rule:pill-bot}]
    {\judge{}{\wait{x}{P}}{\Gamma, \cht{x}{\bot}}}
    {\deduce
      {\judge{}{P}{\Gamma}}
      {\der{D}}}}
\\
\infer
  {\plto
    {\infer[\ref{rule:pill-!}]
      {\judge{}{\server{x}{x'}{P}}{\query \Gamma, \cht{x}{\bang A}}}
      {\deduce
        {\judge{}{P}{\query \Gamma, \cht{x'}{A}}}
        {\der{D}}}}
    {\plbl{\lcodisp{x}}{\lcoweaken{A}}}
    {\infer=[\ref{rule:pill-weaken}]
      {\judge{}
        {\clientdisp{z_1}{\dots\clientdisp{z_n}{\close{x}{\nil}}}}
        {\query \Gamma, \cht{x}{\one}}}
      {\infer[\ref{rule:pill-one}]
        {\judge{}{\close{x}{\nil}}{\cht{x}{\one}}}
        {}}}}
  {\phl{\fn(P)\setminus\{x'\} = \{z_1,\dots,z_n\}}}
\\
\infer[\rlabel{\ref*{rule:pill-!}\ref*{rule:pill-weaken}}{rule:pill-cut-weaken}]
{\plto
  {\infer[\ref{rule:pill-cut}]
    {\judge{}{\res{xy}{P}}{\hyp{G} \pp \Gamma, \query \Delta}}
    {\deduce
      {\judge{}{P}{\hyp{G} \pp \Gamma, \cht{x}{\query A} \pp \query \Delta,\cht{y}{\bang \dual A}}}
      {\der{D}}}}
  {\tau}
  {\infer[\ref{rule:pill-cut}]
    {\judge{}{\res{xy}{P'}}{\hyp{G} \pp \Gamma, \query \Delta}}
    {\deduce
      {\judge{}{P'}{\hyp{G} \pp \Gamma, \cht{x}{\bot} \pp \query \Delta,\cht{y}{\one}}}
      {\der{D'}}}}}
{\plto
  {\deduce
    {\judge{}{P}{\hyp{G} \pp \Gamma, \cht{x}{\query A} \pp \query \Delta,\cht{y}{\bang \dual A}}}
    {\der{D}}}
  {\lsync
    {\plbl{\ldisp{x}}{\lweaken{A}}}
    {\plbl{\lcodisp{y}}{\lcoweaken{\dual{A}}}}}
  {\deduce
    {\judge{}{P'}{\hyp{G} \pp \Gamma, \cht{x}{\bot} \pp \query \Delta,\cht{y}{\one}}}
    {\der{D'}}}}
\end{drules}%

\paragraph{LTS of Derivations}
We extend the sets $\ActSet$ of action labels and $\LblSet$ of all labels defined in \cref{sec:mpill-sos} to include the labels introduced in this section (for the axiom, additives, and exponentials).
\begin{align*}
\ActSet \defeq {} & \left\{\array{l} 
  \lclose{x}, \lwait{x}, \lsend{x}{y}, \lrecv{x}{y}, \linl{x}, \lcoinl{x}, \linr{x}, \lcoinr{x},\\ \luse{x}{y}, \lcouse{x}{y}, \ldup{x}{y}{z}, \lcouse{x}{y}{z}, \ldisp{x}, \lcodisp{x}, \lsendtype{x}{A}, \lrecvtype{x}{A}
  \endarray\middle\vert\, 
  x,y,z \text{ names}, A \text{ type} \right\}\\
\LblSet \defeq {} & \left\{\tau, \lforward{x}{y}, l, \lsync{l}{l'} \mid l,l' \in \ActSet, x, y \text{ names} \right\}
\end{align*}
\begin{definition}%
\label{def:pill-lts-derivations}
The \lts of derivations for \pill, denoted $lts_d$, is the triple $(\DerSet,\LblSet,\lto{})$, where:
\begin{itemize}
\item The set $\DerSet$ is the set of typing derivations for \pill.
\item The set $\LblSet$ is the set of transition labels.
\item The relation ${\lto{}} \subseteq \DerSet \times \LblSet \times \DerSet $ is the least relation closed under the SOS rules forming the speficication of \pimll (see \cref{sec:mpill-sos}) and the ones stated in this subsection.
\end{itemize}
\end{definition}

\subsection{Operational Semantics of Processes and Environments}
\label{sec:pill-proc-hyp}

To define the semantics for processes and typing environments of \pill, we follow the recipe introduced in \cref{sec:mpill-proc-hyp}.

The first step is to specify how derivations are projected to process terms and typing environments. Let $\ProcSet$ and $\EnvSet$ denote the sets of all process terms and typing environments of \pill. The projections 
$\proc\colon \DerSet \to \ProcSet$ and $\env\colon \DerSet \to \EnvSet$ are defined as in \cref{sec:mpill-proc-hyp}:
\begin{highlight}
\[
  \proc\begin{pmatrix}\deduce
              {\judge{}{P}{\hyp{G}}}
              {\der D}\end{pmatrix}
  = \phl{P}
  \qquad
  \env\begin{pmatrix}\deduce
            {\judge{}{P}{\hyp{G}}}
            {\der D}\end{pmatrix}
  = \thl{\hyp{G}}
  \text{.}
\]
\end{highlight}

The second step of the recipe is to formalise the notions of erasure and session fidelity.
Thanks to our characterisation in terms of homomorphisms, we only need to instantiate \cref{def:erasure,def:safety} with the $\LblSet$, $lts_d$, $\proc$, and $\env$ defined above.

\begin{figure}[t]
  \begin{highlight}\small
  \begin{spreadlines}{\termltsskipamount}
    \headertext{Actions}
    \begin{gather*}
      {\phl{\query x[].P} \lto{\query x[]} \phl{\wait xP}}
      \qquad
      {\phl{\ldup{x}{x_1}{x_2}.P} \lto{\ldup{x}{x_1}{x_2}} \phl{x_1(x_2).P}}
      \qquad
      {\phl{\recvtype{x}{X}{P}} \lto{\lrecvtype{x}{A}} \phl{P\{A/X\}}}
      \\
      \deduce
        {\phl{\choice xPQ} \lto{\lcoinr x} \phl{Q}}
        {\phl{\choice xPQ} \lto{\lcoinl x} \phl{P}}
      \qquad
      \infer
        {\phl{\pi.P} \lto{\vphantom{!(}\pi} \phl{P}}
        {\phl{\pi \neq \lrecvtype{x}{A},\ldisp{x},\ldup{x}{x_1}{x_2}}}
      \qquad
      \infer
        {\phl{\server{x}{x'}P} \lto{\lcodisp{x}} \phl{\clientdisp{z_1}{\cdots\clientdisp{z_n}{\close{x}{\nil}}}}}
        {\phl{\fn(P)\setminus\{x'\} = \{z_1, \ldots, z_n\}}}
      \\
      \infer
        {\phl{\server{x}{x'}P} \lto{\lcodup{x}{x_1}{x_2}} \phl{\clientdup{z_1}{z_1\sigma_1}{z_1\sigma_2}{\dots\clientdup{z_n}{z_n\sigma_1}{z_n\sigma_2}{\send{x_1}{x_2}{(\server{x_1}{x'\sigma_1}{P_1} \pp \server{x_2}{x'\sigma_2}{P_2})}}}}}
        {\phl{P_1 = P\sigma_1}
        & \phl{P_2 = P\sigma_2}
        & \phl{\fn(P_1) \cap \fn(P_2) = \emptyset}
        & \phl{\fn(P)\setminus\{x'\} = \{z_1,\ldots,z_n\}}}
     \end{gather*}%
    \headertext{Structural}
    \begin{gather*}
      {\phl{\forward xy} \lto{\lforward yx} \phl{\nil}}
      \qquad
      {\phl{\forward xy} \lto{\lforward xy} \phl{\nil}}
      \\
      \infer[\rlabel{\ref*{rule:pill-cut-axiom}}{rule:pill-proc-cut-forward}]
        {\phl{\res{xy}{P}} \lto{\phl\tau} \phl{P'\{x/z\}}}
        {\phl{P} \lto{\lforward{y}{z}} \phl{P'}}
      \quad
      \infer[\rrelabel{rule:mpill-proc-mix-1}{rule:pill-proc-mix-1}]
        {\phl{P \pp Q} \lto{l} \phl{P' \pp Q}}
        {\phl{P} \lto{l} \phl{P'}
        &\phl{\bn(l) \cap \fn(Q) = \emptyset}}
      \quad
      \infer[\rrelabel{rule:mpill-proc-mix-2}{rule:pill-proc-mix-2}]
        {\phl{P \pp Q} \lto{l} \phl{P \pp Q'}}
        {\phl{Q} \lto{l} \phl{Q'}
        &\phl{\bn(l) \cap \fn(P) = \emptyset}}
      \\
      \infer[\rrelabel{rule:mpill-proc-mix-sync}{rule:pill-proc-mix-sync}]
          {\phl{P \pp Q} \lto{\lsync{l}{l'}} \phl{P' \pp Q'}}
          {\phl{P} \lto{l} \phl{P'}
          &\phl{Q} \lto{l'} \phl{Q'}
          &\phl{\bn(l) \cap \bn(l') = \emptyset}}
      \qquad
        \infer[\rrelabel{rule:mpill-proc-cut-res}{rule:pill-proc-cut-res}]
          {\phl{\res{xy}{P}} \lto{l} \phl{\res{xy}{P'}}}
          {\phl{P} \lto{l} \phl{P'}
          &\phl{x,y \notin \cn(l)}}
      \qquad
       \infer[\rrelabel{rule:mpill-proc-alpha}{rule:pill-proc-alpha}]
         {\phl{P} \lto{l} \phl{R}}
         {\phl{P \aleq Q}
         &\phl{Q} \lto{l} \phl{R}}
    \end{gather*}%
    \headertext{Communications}
    \begin{gather*}
      \infer[\rrelabel{rule:mpill-proc-cut-tensor}{rule:pill-proc-cut-tensor}]
        {\phl{\res{xy}{P}} \lto{\tau} \phl{\res{xy}{\res{x'y'}{P'}}}}
        {\phl{P} \lto{\lsync{\lsend{x}{x'}}{\lrecv{y}{y'}}} \phl{P'}}
      \qquad
      \infer[\rrelabel{rule:mpill-proc-cut-one}{rule:pill-proc-cut-one}]
        {\phl{\res{xy}{P}} \lto{\tau} \phl{P'}}
        {\phl{P} \lto{\lsync{\lclose{x}}{\lwait{y}}} \phl{P'}}
      \qquad
      \infer[\rlabel{\ref*{rule:pill-cut-oplus_1}}{rule:pill-proc-cut-oplus_1}]
        {\phl{\res{xy}{P}} \lto{\tau} \phl{\res{xy}{P'}}}
        {\phl{P} \lto{\lsync{\linl{x}}{\lcoinl{y}}} \phl{P'}}
      \\
      \infer[\rlabel{\ref*{rule:pill-cut-oplus_2}}{rule:pill-proc-cut-oplus_2}]
         {\phl{\res{xy}{P}} \lto{\tau} \phl{\res{xy}{P'}}}
         {\phl{P} \lto{\lsync{\linr{x}}{\lcoinr{y}}} \phl{P'}}
      \qquad
        \infer[\rlabel{\ref*{rule:pill-cut-?}}{rule:pill-proc-cut-?}]
          {\phl{\res{xy}{P}} \lto{\tau} \phl{\res{x'y'}{P'}}}
          {\phl{P} \lto{\lsync{\luse{x}{x'}}{\lcouse{y}{y'}}} \phl{P'}}
      \qquad
        \infer[\rlabel{\ref*{rule:pill-cut-weaken}}{rule:pill-proc-cut-weaken}]
          {\phl{\res{xy}{P}} \lto{\phl{\tau}}
           \phl{\res{xy}{P'}}}
          {\phl{P} \lto{\lsync{\ldisp{x}}{\lcodisp{y}}} \phl{P'}}
      \\
        \infer[\rlabel{\ref*{rule:pill-cut-contract}}{rule:pill-proc-cut-contract}]
          {\phl{\res{xy}{P}} \lto{\tau} \phl{\phl{\res{x_1y_1}{P'}}}}
          {\phl{P} \lto{\lsync{\ldup{x}{x_1}{x_2}}{\lcodup{y}{y_1}{y_2}}} \phl{P'}}
      \qquad
         \infer[\rlabel{\ref*{rule:pill-cut-exists}}{rule:pill-proc-cut-exists}]
           {\phl{\res{xy}{P}} \lto{\tau} \phl{\res{xy}{P'}}}
           {\phl{P} \lto{\lsync{\lsendtype{x}{A}}{\lrecvtype{y}{A}}} \phl{P'}}
    \end{gather*}
     \footer%
    \end{spreadlines}%
\end{highlight}
\caption{\pill, process transitions.}
\label{fig:pill-sos-processes}
\end{figure}

\begin{figure}
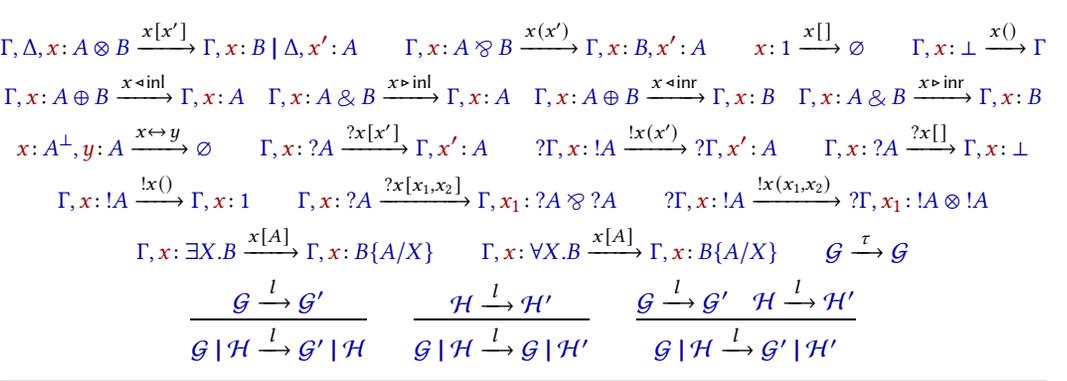

\begin{highlight}\small
\begin{spreadlines}{\envltsskipamount}
\header
\begin{gather*}
{\thl{\Gamma, \Delta, \cht{x}{A \tensor B}}
\lto{\lsend{x}{x'}}
\thl{\Gamma, \cht{x}{B} \pp \Delta, \cht{x'}{A}}}
\qquad
{\thl{\Gamma, \cht{x}{A \parr B}}
\lto{\lrecv{x}{x'}}
\thl{\Gamma, \cht{x}{B}, \cht{x'}{A}}}
\qquad
{\thl{\cht{x}{\one}}
\lto{\lclose{x}}
\thl{\emptyhyp}}
\qquad
{\thl{\Gamma, \cht{x}{\bot}}
\lto{\lwait{x}}
\thl{\Gamma}}
\\
{\thl{\Gamma, \cht{x}{A \oplus B}}
\lto{\linl{x}}
\thl{\Gamma, \cht{x}{A}}}
\quad
{\thl{\Gamma, \cht{x}{A \with B}}
\lto{\lcoinl{x}}
\thl{\Gamma, \cht{x}{A}}}
\quad
{\thl{\Gamma, \cht{x}{A \oplus B}}
\lto{\linr{x}}
\thl{\Gamma, \cht{x}{B}}}
\quad
{\thl{\Gamma, \cht{x}{A \with B}}
\lto{\lcoinr{x}}
\thl{\Gamma, \cht{x}{B}}}
\\
{\thl{\cht{x}{\dual{A}}, \cht{y}{A}}
\lto{\lforward{x}{y}}
\thl{\emptyhyp}}
\qquad
{\thl{\Gamma,\cht{x}{\query A}}
\lto{\luse{x}{x'}}
\thl{\Gamma,\cht{x'}{A}}}
\qquad
{\thl{\query\Gamma,\cht{x}{\bang A}}
\lto{\lcouse{x}{x'}}
\thl{\query\Gamma,\cht{x'}{A}}}
\qquad
{\thl{\Gamma,\cht{x}{\query A}}
\lto{\ldisp{x}}
\thl{\Gamma,\cht{x}{\bot}}}
\\
{\thl{\Gamma,\cht{x}{\bang A}}
\lto{\lcodisp{x}}
\thl{\Gamma,\cht{x}{1}}}
\qquad
{\thl{\Gamma,\cht{x}{\query A}}
\lto{\ldup{x}{x_1}{x_2}}
\thl{\Gamma,\cht{x_1}{\query A \parr \query A}}}
\qquad
{\thl{\query\Gamma,\cht{x}{\bang A}}
\lto{\lcodup{x}{x_1}{x_2}}
\thl{\query\Gamma,\cht{x_1}{\bang A \tensor \bang A}}}
\\
{\thl{\Gamma,\cht{x}{\exists X.B}}
\lto{\lsendtype{x}{A}}
\thl{\Gamma,\cht{x}{B\{A/X\}}}}
\qquad
{\thl{\Gamma,\cht{x}{\forall X.B}}
\lto{\lsendtype{x}{A}}
\thl{\Gamma,\cht{x}{B\{A/X\}}}}
\qquad
{\thl{\hyp{G}}
\lto{\tau}
\thl{\hyp{G}}}
\\
\infer
  {\thl{\hyp{G} \pp \hyp{H}} \lto{l} \thl{\hyp{G'} \pp \hyp{H}}}
  {\thl{\hyp{G}} \lto{l} \thl{\hyp{G'}}}
\qquad
\infer
  {\thl{\hyp{G} \pp \hyp{H}} \lto{l} \thl{\hyp{G} \pp \hyp{H'}}}
  {\thl{\hyp{H}} \lto{l} \thl{\hyp{H'}}}
\qquad
\infer
  {\thl{\hyp{G} \pp \hyp{H}} \lto{l} \thl{\hyp{G'} \pp \hyp{H'}}}
  {\thl{\hyp{G}} \lto{l} \thl{\hyp{G'}}
  &\thl{\hyp{H}} \lto{l} \thl{\hyp{H'}}}
\end{gather*}
\footer
\end{spreadlines}
\end{highlight}
  \caption{\pill, typing environment transitions.}
  \label{fig:pill-sos-environments}
\end{figure}

The third step is to apply the principles of ``erasing types'' and ``erasing processes'' to the SOS for \pill derivations. The resulting specifications are in \cref{fig:pill-sos-processes,fig:pill-sos-environments}, respectively.

\begin{definition}%
\label{def:pill-lts-processes}
The \lts of processes for \pill, denoted $lts_p$, is the triple $(\ProcSet,\LblSet,\lto{})$ where
\begin{itemize}
\item $\ProcSet$ is the set of process terms for \pill,
\item $\LblSet$ is the set of labels for \pill,
\item ${\lto{}} \subseteq \ProcSet \times \LblSet \times \ProcSet $ is the least relation closed under the SOS rules in \cref{fig:pill-sos-processes}.
\end{itemize}
\end{definition}

\begin{definition}
\label{def:pill-lts-environments}
The \lts of environments for \pill, denoted $lts_e$, is the triple $(\EnvSet,\LblSet,\lto{})$ where
\begin{itemize}
\item $\EnvSet$ is the set of typing hyperenvironments for \pill,
\item $\LblSet$ is the set of labels for \pill,
\item ${\lto{}} \subseteq \EnvSet \times \LblSet \times \EnvSet $ is the least relation closed under the SOS rules in \cref{fig:pill-sos-environments}.
\end{itemize}
\end{definition}

The fourth and final step of the recipe is to verify that semantics of processes and of typing environments enjoy erasure and session fidelity.

\begin{theorem}[Erasure]
\label{thm:pill-erasure}
$lts_p$ enjoys erasure wrt $lts_d$.
\end{theorem}

\begin{theorem}[session fidelity]
\label{thm:pill-safety}
$lts_p$ enjoys session fidelity wrt $lts_e$.
\end{theorem}

As for \pimll, erasure entails typability preservation.
\begin{corollary}[Typability Preservation]
Let $P$ be well-typed. Then, $P \lto{l} P'$ implies that $P'$ is well-typed.
\end{corollary}

\subsection{Metatheory}
All definitions and results presented in \cref{sec:mpill-metatheory} apply to \pill. Here we present explicitly the ones that are most relevant or that will be needed later.

\subsubsection{Behavioural Theory}
The standard definitions of behavioural equivalences and preorders (\cref{def:strong-bisimilarity,def:bisimilarity}), their laws (\cref{fact:bisimilarity-laws}), and definition of congruence (\cref{def:congruence}) apply to \pill without modification.
When we unfold \cref{def:congruence} using the grammar of \pill we obtain that a relation on processes $\asymp$ is a (pre)congruence if $P \asymp Q$ implies that:
\begin{enumerate}[nosep, noitemsep]
  \item ${P \pp R} \asymp {Q \pp R}$ for any $R$ (and the symmetric);
  \item $\prefixed{\pi}{P} \asymp \prefixed{\pi}{Q}$ for any prefix $\pi$;
  \item $\res{xy}{P} \asymp \res{xy}{Q}$ for any $x$, $y$;
  \item\label{def:pill-congruence-choice} $\choice{x}{R}{P} \asymp \choice{x}{R}{Q}$ for any $x$ and $R$ (and the symmetric).
\end{enumerate}
The only difference with respect to \pimll is the new case for external choice (\cref{def:pill-congruence-choice}) which is the only new term contrusctor introduced by \pill that is not a prefix.
\begin{theorem}[Congruence]
  \label{thm:pill-congruence}
  On the \lts of processes, $\ssimby$ and $\simby$ are precongruences and $\sbis$ and $\bis$ are congruences.
\end{theorem}

The type checking and similarity checking for \pill coincide.
\begin{theorem}
  \label{thm:pill-typing-iff-sim}
  For $P$ well-typed and $\hyp{G}$ such that $\fn(P) = \cn(\hyp{G})$, $P \simby \hyp{G}$ iff $\judge{}{P}{\hyp{G}}$.
\end{theorem}

It follows that bisimilarity is sound with respect to type equivalence.
\begin{corollary}
For $P$ and $Q$ well-typed, if $P \bis Q$ and $\judge{}{Q}{\hyp{G}}$, then $\judge{}{P}{\hyp{G}}$.
\end{corollary}

\begin{remark}[Soundness of $\eta$-expansion]
For the first time, we can use bisimilarity to observe that $\eta$-expansion is behaviourally sound. Specifically, under the ``linking'' semantics of \pill for forwarders, $\forward{x}{y}$
is indistinguishible from its expansion (recall \cref{rem:pill-axiom-typing}) for any well-typed process, in the following sense.
\[
\res{xy}{(P \pp \forward{y}{z})} \bis \res{xy}{(P \pp Fwd_{y,z}^{A})}
\qquad\text{for any $A$ and $\judge{}{P}{\hyp{G} \pp \Gamma, \cht{x}{A}}$.}
\]
\end{remark}

\subsubsection{Metatheory of Parallelism}

\pill enjoys the diamond property, serialisation, non-interference, and readiness.

\begin{theorem}[Diamond property]
  \label{thm:pill-diamond}
  The \lts of derivations enjoys the diamond property.
\end{theorem}

\begin{theorem}[Serialisation]
  \label{thm:pill-serialisation}
  The \lts of derivations enjoys serialisation.
\end{theorem}

\begin{theorem}[Non-Interference]
  \label{thm:pill-non-interference}
  The \lts of derivations enjoys the non-interference.
\end{theorem}

The definition of disentanglement (\cref{thm:mpill-disentanglement}) applies to \pill without modification. As for \pimll, we can compute disentanglement directly using a procedure given by recursion on the structure of processes. The definition of this procedure follows the one given in \cref{thm:pill-disentanglement} for \pimll. We only need to add two cases, for
$\forward{x}{y}$ and $\choice{x}{P}{Q}$, which are immediate.
\begin{lemma}
\label{thm:pill-disentanglement}
The disentanglement $\disen{P}$ of a process $P$ is the process recursively defined as follows:
\begin{gather*}
\disen{\prefixed{\pi}{P}} = \prefixed{\pi}{\disen{P}}
\qquad
\disen{\choice{x}{P}{Q}} = \choice{x}{\disen{P}}{\disen{Q}}
\\
\disen{\forward{x}{y}} = \forward{x}{y}
\qquad
\disen{\nil} = \nil 
\qquad
\qquad
\disen{P \pp Q} = \begin{cases}
  \disen{P} & \text{if } Q = \nil\\
  \disen{P} \pp \disen{Q}
  & \text{otherwise}
\end{cases}
\\
\disen{\res{xy}{P}} =  \res{xy}{(P_i \pp P_j)} \pp (\disen{P}\setminus P_i,P_j)
\text{ where } \disen{P} = P_1 \pp \dots \pp P_{n}\text{, } 
x \in \fn(P_i) \text{, } 
y \in \fn(P_j)
\end{gather*}
\end{lemma}

\begin{theorem}[Readiness]
\label{thm:pill-readiness}
Let $\judge{}{P}{\Gamma_1 \pp \cdots \pp \Gamma_n}$. For every $i \in [1,n]$, there exist $x\in\cn(\Gamma_i)$ and $l$ such that $x \in \fn(l)$ and $P \slto{l}$.
\end{theorem}

\subsubsection{Relation with Linear Logic and Classical Processes}
\label{sec:pill-cp}

All the ingredients we used in \cref{sec:mpill-mcp} to relate \pimll to the multiplicative fragment of \CLL and \CP extend without modifications to relate \pill to \CLL and \CP: the internalisation of ``$,$'' and ``$\pp$'' and the definition of packing rely only on the multiplicative fragment of \pill and both are unaffected by the introduction of the axiom, additives, exponentials, and quantifiers.

\begin{theorem}
\label{thm:cp-pill}
$\judge{}{P}{\hyp G}$ in \pill iff $\cpjudge{\pack{x}{\judge{}{P}{\hyp G}}}{\cht{x}{\bigtensor \hyp G}}$ in \CP.
\end{theorem}

\begin{corollary}\label{cor:cll-search}
If $\der{D}$ is a cut-free derivation of $\bigtensor \hyp{G}$ in \CLL, then 
$\psi(\der{D})$ is a cut-free derivation of $\hyp{G}$ in \pill.
\end{corollary}

%% file: hopill.tex
\section{Higher-Order \pill}
\label{sec:hopill}

In this section, we extend our approach to process mobility, by enhancing \pill with higher-order communication primitives.
In particular, we investigate rules that
\begin{enumerate*}[label=\em(\roman*)]
\item type new process terms that capture the feature of process mobility, and \item enable proof transformations that yield their expected semantics.
\end{enumerate*}
We call this language \expandacronym{hopill} (abbreviated as \hopill).

\subsection{Processes and Typing}
\paragraph{Processes}
To obtain process terms that enable process mobility, we can get direct inspiration from the higher-order $\pi$-calculus (\HOpi) \cite{SW01}. We need three basic features:
\begin{itemize}
\item A term for sending process code over a channel.
\item A term for receiving process code over a channel and storing it in a process variable.
\item A term for running the process code stored in a variable.
\end{itemize}
We thus extend the syntax of \pill as follows, where \hl{$\phl{p}$, $\phl{q}$, $\phl{r}$, \dots} range over process variables.
\def\bnfpimll{\small(\pimll)}
\def\bnfpill{\small(\pill)}
\begin{bnftable2}
	{\phl P, \phl Q \Coloneqq {}}%
  {\choice{x}{P}{Q}}
  {(\bnfpimll)}
    \phl P, \phl Q \Coloneqq {}
            & \cdots & all terms in \pill\\
    \mid {} & \sendho{x}{\abstr \rho P}{Q}  & output abstraction $\abstr{\rho}{P}$ on $x$ and continue as $Q$&\\
    \mid {} & \recvho{x}{p}{P}     & input an abstraction as $p$ on $x$ and continue as $P$&\\
    \mid {} & \invoke[\hyp{G}]{p}{\rho}     & run abstraction $p$ instantiated with $\rho$&\\
    \mid {} & \chop{q}{\abstr \rho Q}{P}          & explicit process variable substitution, ``chop''&
\end{bnftable2}%
We describe the new terms.

The higher-order output term $\sendho{x}{\abstr{\rho}{P}}{Q}$ sends the \emph{abstraction} $\abstr{\rho}{P}$ over channel $x$ and then continues as $Q$. 
An abstraction is a parameterised process term, enabling the reuse of process code in different contexts. 
The concept of abstraction is standard in the literature of \HOpi \citep{SW01}; here, we apply a slight twist and use \emph{named} formal parameters $\rho$ rather than just a parameter list, to make the invocation of abstractions not dependent on the order in which actual parameters are passed. 
This plays well with the typing contexts of linear logic, used for typing process terms in \pill, since they are order-independent as well (exchange is allowed).
We use $f$ to range over formal parameter names. 
These are constants, and are thus not affected by $\alpha$-renaming. 
Formally, $\rho$ is a record that maps parameter names to channels used inside $P$, \ie, $\rho = \{ f_i = x_i \}_{i\in I}$ (where $I$ is finite). 
We omit curly brackets for records in the remainder when they are clear from the context, \eg, as in $\invoke[\hyp{G}]{p}{f_1 = x, f_2 = y}$.
Given a record $\rho = \{ f_i = x_i \}_{i\in I}$, we call the set $\{f_i\}_{i\in I}$ the preimage of $\rho$ and the set $\{x_i\}_{i\in I}$ the image of $\rho$. 
An abstraction $\abstr{\rho}{P}$ binds all names in the image of $\rho$ in $P$.
We require abstractions to be closed with respect to channels, in the sense that the image of $\rho$ needs to be exactly the set $\fn(P)$ of all free channel names in $P$.

Dual to higher-order output, the higher-order input term $\recvho{x}{p}{Q}$ receives an abstraction over channel $x$ and stores it in the process variable $p$, which can be used in the continuation $Q$.

Abstractions stored in a process variable $p$ can be invoked (we also say run) using term $\invoke[\hyp{G}]{p}{\rho}$, where $\rho$ are the actual named parameters to be used by the process.

\begin{remark}
The choice of having (named) parameters and abstractions is justified by the desire to use received processes to implement behaviours. 
For example, we may want to receive a channel and then use a process to implement the necessary behaviour on that channel: $\recv{x}{y}{\recvho{x}{p}{\invoke[\hyp{G}]{p}{f=y}}}$.

A different way of achieving the same result would be to support the communication of processes with free names (not abstractions), and then allow invoke terms to dynamically rename free names of received processes. For example, if we knew from typing that any process received for $p$ above has $z$ as a free name, we could write $\recv{x}{y}{\recvho{x}{p}{\invoke[\hyp{G}]{p}{y/z}}}$.

Such a \emph{dynamic binding} mechanism would make our syntax simpler, since invocations would simply be $\invoke[\hyp{G}]{p}{\sigma}$ ($\sigma$ is a name substitution) and the higher-order output term would be $\sendho{x}{P}{Q}$. 
However, dynamic binding is undesirable in a programming model, since the scope of free names can change due to higher-order communications. 
The reader unfamiliar with dynamic binding may consult \citep[p.~376]{SW01} for a discussion of this issue in \HOpi.
\end{remark}

\paragraph{Types}
Types in \hopill extend those of \pill to include types for higher-order outputs and inputs. These types are inspired to \cite{M18}, the only difference the explicit treatment of parallelism with the use of hyperenvironments instead of environments. %
\begin{bnftable2}[\thl]%
  {\thl{A},\thl{B} \Coloneqq {}}%
  {A \tensor B}%
  {(\bnfpimll)}
  \thl{A},\thl{B} \Coloneqq {} & \dots & all types from \pill &\\
    \mid {} & {\provide{\hyp{G}}} & higher-order output&\\
    \mid {} & {\assume{\hyp{G}}} & higher-order input&
\end{bnftable2}%
From the perspective of typing derivations, the new type constructor ${\provide{\hyp{G}}}$ can be interpreted as ``assumes a derivation of $\hyp{G}$'' and, dually, ${\assume{\hyp{G}}}$ as ``provides a derivation of $\hyp{G}$''. 

Duality is as in \pill and extended the new higher-order types.
\begin{highlight}
\begin{gather*}
\thl{\dual{\provide{\hyp{G}}}} = \thl{\assume{\hyp{G}}}
\qquad
\thl{\dual{\assume{\hyp{G}}}} = \thl{\provide{\hyp{G}}}
\end{gather*}
\end{highlight}

\paragraph{Typing Environments}

Environments and hyperenvironments are defined as in \pimll (but contain types in the extended syntax). 

A process environment ($\Theta$, $\Pi$, \dots) associates process symbols to hyperenvironments.
We write process environments as lists; for example, $\Theta = \cht{p_1}{\hyp{G}_1}, \ldots, \cht{p_n}{\hyp{G}_n}$ associates each $p_i$ to its respective hyperenvironment $\hyp{G}_i$, for $1 \leq i \leq n$. 
Notice, that since each $p_i$ is going to be instantiated by an abstraction, we are not typing channels in each $\hyp{G}_i$, but the names of 
formal parameters ($f$).
We write $\pv(\Theta)$ for the set of process symbols associated by $\Theta$.
All process symbols in an environment are distinct. Process environments allow for exchange, that is, order in environments is ignored. 
Process environments can be combined when they do not share process symbols: assuming that $\Theta$ and $\Pi$ do not share names ($\pv(\Theta)\cap\pv(\Pi) = \emptyset$), $\Theta, \Pi$ is the process environment that consists exactly of all the associations in $\Theta$ and those in $\Pi$.
Combining processes is a commutative, associative partial operations with $\emptyho$, the empty process environment, acting as unit.

\begin{figure}[t]
  \begin{highlight}\small 
  \begin{spreadlines}{\typerulesskipamount}%
    \headertext{Structural rules}%
    \def\sep{\qquad}
    \begin{gather*}
      \def\sep{\qquad}
      \infer[\rrelabel{rule:pill-mix}{rule:hopill-mix}]
        {\judge{\Theta,\Pi}{P \pp Q}{\hyp{G} \pp \hyp{H}}}
        {\judge{\Theta}{P}{\hyp{G}}
        &\judge{\Pi}{Q}{\hyp{H}}}
      \sep
      \infer[\rrelabel{rule:pill-cut}{rule:hopill-cut}]
        {\judge{\Theta}{\res{xy}{P}}{\hyp{G} \pp \Gamma, \Delta}}
        {\judge{\Theta}{P}{\hyp{G}\pp\Gamma, \cht{x}{A} \pp \Delta, \cht{y}{\dual{A}}}}
      \sep
      \infer[\rlabel{\rname{Chop}}{rule:hopill-chop}]
        {\judge{\Theta,\Pi}{\chop{p}{\abstr \rho P}{Q}}{\hyp{G}}}
        {\judge{\Pi}{P}{\hyp{H}\rho}
        &\judge{\Theta,\pvt{p}{\hyp{H}}}{Q}{\hyp{G}}}
      \\
      \def\sep{\qquad}
      \infer[\rrelabel{rule:pill-mix-0}{rule:hopill-mix-0}]
        {\judge{\emptyho}{\nil}{\emptyhyp}}
        {}
      \sep
      \infer[\rrelabel{rule:pill-axiom}{rule:hopill-axiom}]
        {\judge{\emptyho}{\forward{x}{y}}{\cht{x}{A^\bot}, \cht{y}{A}}}
        {}
      \sep
      \infer[\rlabel{\rname{Id}}{rule:hopill-id}]
        {\judge{\pvt{p}{\hyp{G}}}{\invoke{p}{\rho}}{\hyp{G}\rho}}
        {}
    \end{gather*}%
    \headertext{Logical rules}%
    \begin{gather*}
      \def\sep{\quad}
      \infer[\rrelabel{rule:pill-tensor}{rule:hopill-tensor}]
        {\judge{\Theta}{\send{x}{y}{P}}{\Gamma,\Delta,\cht{x}{A \tensor B}}}
        {\judge{\Theta}{P}{\Gamma,\cht{y}{A} \pp \Delta,\cht{x}{B}}}
      \sep
      \infer[\rrelabel{rule:pill-one}{rule:hopill-one}]
        {\judge{\Theta}{\close{x}{P}}{\cht{x}{\one}}}
        {\judge{\Theta}{P}{\emptyhyp}}
      \sep
      \infer[\rrelabel{rule:pill-parr}{rule:hopill-parr}]
        {\judge{\Theta}{\recv{x}{y}{P}}{\Gamma, \cht{x}{A \parr B}}}
        {\judge{\Theta}{P}{\Gamma, \cht{y}{A}, \cht{x}{B}}}
      \sep
      \infer[\rrelabel{rule:pill-bot}{rule:hopill-bot}]
        {\judge{\Theta}{\wait{x}{P}}{\Gamma, \cht{x}{\bot}}}
        {\judge{\Theta}{P}{\Gamma}}
      \\
      \def\sep{\qquad}
      \infer[\rrelabel{rule:pill-oplus_1}{rule:hopill-oplus_1}]
        {\judge{\Theta}{\inl{x}{P}}{\Gamma, \cht{x}{A \oplus B}}}
        {\judge{\Theta}{P}{\Gamma, \cht{x}{A}}}
      \sep
      \infer[\rrelabel{rule:pill-oplus_2}{rule:hopill-oplus_2}]
        {\judge{\Theta}{\inr{x}{P}}{\Gamma, \cht{x}{A \oplus B}}}
        {\judge{\Theta}{P}{\Gamma, \cht{x}{B}}}
      \sep
      \infer[\rrelabel{rule:pill-with}{rule:hopill-with}]
        {\judge{\Theta}{\choice{x}{P}{Q}}{\Gamma, \cht{x}{A \with B}}}
        {\judge{\Theta}{P}{\Gamma, \cht{x}{A}} 
        &\judge{\Theta}{Q}{\Gamma, \cht{x}{B}}}
      \\
      \def\sep{\qquad}
      \infer[\rrelabel{rule:pill-!}{rule:hopill-!}]
        {\judge{\emptyho}{\server{x}{y}{P}}{\query\Gamma, \cht{x}{\bang A}}}
        {\judge{\emptyho}{P}{\query\Gamma, \cht{y}{A}}}
      \sep
      \infer[\rrelabel{rule:pill-?}{rule:hopill-?}]
        {\judge{\Theta}{\clientuse{x}{y}{P}}{\Gamma, \cht{x}{\query A}}}
        {\judge{\Theta}{P}{\Gamma, \cht{y}{A}}}
      \sep
      \infer[\rrelabel{rule:pill-weaken}{rule:hopill-weaken}]
        {\judge{\Theta}{\clientdisp{x}{P}}{\Gamma, \cht{x}{\query A}}}
        {\judge{\Theta}{P}{\Gamma}}
      \\
      \def\sep{\qquad}
      \infer[\rrelabel{rule:pill-contract}{rule:hopill-contract}]
        {\judge{\Theta}{\clientdup{x}{x'}{x''}{P}}{\Gamma, \cht{x}{\query A}}}
        {\judge{\Theta}{P}{\Gamma, \cht{x'}{\query A}, \cht{x''}{\query A}}}
      \sep
      \infer[\rrelabel{rule:pill-exists}{rule:hopill-exists}]
        {\judge{\Theta}{\sendtype{x}{A}{P}}{\Gamma,\cht{x}{\exists X.B}}}
        {\judge{\Theta}{P}{\Gamma,\cht{x}{B\{A/X\}}}}
      \sep
      \infer[\rrelabel{rule:pill-forall}{rule:hopill-forall}]
         {\judge{\Theta}{\recvtype{x}{X}{P}}{\Gamma,\cht{x}{\forall X.B}}}
         {\judge{\Theta}{P}{\Gamma,\cht{x}{B}}
         &\thl{X \notin \ftv(\Gamma)}}
      \\
      \def\sep{\qquad}
      \infer[\rlabel{\rname{$\provide{}$}}{rule:hopill-provide}]
        {\judge{\Theta}{\sendho{x}{\abstr{\rho}{P}}{Q}}{\cht{x}{\provide{\hyp{G}}}}}
        {\judge{\Theta}{P}{\hyp{G}\rho}
        &\judge{\emptyho}{Q}{\emptyhyp}}
      \sep
      \infer[\rlabel{\rname{$\assume{}$}}{rule:hopill-assume}]
        {\judge{\Theta}{\recvho{x}{p}{P}}{\Gamma,\cht{x}{\assume{\hyp{G}}}}}
        {\judge{\Theta,\pvt{p}{\hyp{G}}}{P}{\Gamma}}
    \end{gather*}%
    \footer%
  \end{spreadlines}%
  \end{highlight}%
  \caption{\hopill, typing rules.}
  \label{fig:hopill-typing-rules}
\end{figure}

\paragraph{Typing Judgements}

Typing judgements in \hopill have the form $\judge{\Theta}{P}{\hyp G}$ and state that process $P$ uses its free names according to $\hyp G$ and its free process variables represent processes that use their names according to the corresponding type in $\Theta$. 
Judgements can be derived by using the inference rules displayed in \cref{fig:hopill-typing-rules}.

Each typing rule, has a distinct process environment for each of its premises and these are combined into the in the rule conclusion. The only exeception in this process is \cref{rule:hopill-!} which requires an empty process environment to ensure the linear usage of process symbols.
We obtained the remaining rules for typing terms forming \pill (\ie, all rules \cref{fig:hopill-typing-rules} except \cref{rule:hopill-!,rule:hopill-id,rule:hopill-chop,rule:hopill-assume,rule:hopill-provide}) by applying this discipline to the typing rules of \pill. We refer to this process as ``lifting''. 

\Cref{rule:hopill-id} says that if we run $p$ by providing all the name parameters (collected in $\rho$) required by the abstraction that it represents, then we implement exactly the session types that type the code that may be associated with in $p$ (modulo the renaming of the formal parameters performed with $\rho$, hence the $\hyp{G}\rho$). 
Note that we use \emph{all} our available ``resources''---the endpoint names in the image of $\rho$---to run $p$, passing them as parameters. This ensures linearity.
\Cref{rule:hopill-assume} says that if we receive a process of type $\hyp{G}$ over endpoint $x$ and associate it with $p$, we can use $p$ later in the continuation $Q$ assuming that it implements $\hyp{G}$ hence the $x$ is $\assume{\hyp{G}}$.
\Cref{rule:hopill-provide} says that if we send a process of type $P$ over channel $x$ and $P$ implements $\hyp{G}$, then $x$ has type $\provide{\hyp{G}}$. Note that we require $P$ to use all the process symbols available in the context ($\Theta$).
The association of process symbols and abstractions is maintained by the term for explicit substitution and this is typed by \cref{rule:hopill-chop}. 
This rule says that we can replace $p$ with $\abstr{\rho}{P}$ in $Q$, provided that $P$ and $Q$ have compatible typing (up to the formal named parameters). 
The idea is that a variable $p$ stands for a ``hole'' in a proof, which has to be filled as expected by the type for $p$.

The typing rules for sending and receiving processes is close to the ones for multiplicative units in their use of endpoints (say, instead of the primitives for sending and receiving channels): the endpoint is discarded in the continuation. This is design is aimed at keeping the language minimal in the sense that these variations can be recovered as syntax sugar.

\begin{remark}[Higher-order I/O with continuations]\label{ex:ho-continuations}

\newcommand{\sendhowc}[3]{\prefixed{{#1}\bm{[}#2\bm{]}}{#3}}
\newcommand{\recvhowc}[3]{\prefixed{{#1}\bm{(}#2\bm{)}}{#3}}

We can derive constructs for sending and receiving processes over channels and then allow us to continue using that channel. We distinguish these sugared counterparts from our output and input primitives by using bold brackets.
\begin{bnftable}
	{\phl P, \phl Q \Coloneqq {}}
	{\sendhowc{x}{\abstr \rho P}{Q}}
		\phl P, \phl Q \Coloneqq {} & \nohl{\dots} & \\
		\mid {} & \sendhowc{x}{\abstr \rho P}{Q} & output a process and continue \\
		\mid {} & \recvhowc{x}{p}{P} & input a process and continue.
\end{bnftable}
These constructs are desugared as follows (we show directly the proofs).
\begin{spreadlines}{\typerulesskipamount}%
\begin{alignat*}{3}
	\array{c}
		\infer
			{\judge{\Theta,\Pi}{\sendhowc{x}{\abstr \rho P}{Q}}{\Gamma,\cht{x}{\provide{\hyp{G}}\tensor A}}}
			{\judge{\Theta}{P}{\hyp{G}\rho}
			&\judge{\Pi}{Q}{\Gamma, \cht{x}{A}}}
	\endarray
	& \quad \defeq \quad &&
	\array{c}
	\infer[\ref{rule:hopill-tensor}]
		{\judge{\Theta,\Pi}{\send{x}{y}{(\sendho{y}{\abstr \rho P}{\nil} \pp Q)}}{\Gamma, \cht{x}{\provide{\hyp{G}}\tensor A}}}
		{\infer[\ref{rule:hopill-mix}]
			{\judge{\Theta,\Pi}{\sendho{y}{\abstr \rho P}{\nil} \pp Q}{\cht{y}{\provide{\hyp{G}}} \pp \Gamma, \cht{x}{A}}}
			{\infer[\ref{rule:hopill-provide}]
				{\judge{\Theta}{\sendho{y}{\abstr \rho P}{\nil}}{\cht{y}{\provide{\hyp{G}}}}}
				{\judge{\Theta}{P}{\hyp{G}\rho}
        &\infer[\ref{rule:hopill-mix-0}]
          {\judge{\emptyho}{\nil}{\emptyhyp}}
          {}}
			&\judge{\Pi}{Q}{\Gamma, \cht{x}{A}}}}
	\endarray
\\
	\array{c}
		\infer
			{\judge{\Theta}{\recvhowc{x}{p}{P}}{\Gamma, \cht{x}{\assume{\hyp{G}} \parr A}}}
			{\judge{\Theta, \pvt{p}{\hyp{G}}}{P}{\Gamma, \cht{x}{A}}}
	\endarray
	& \quad \defeq \quad &&
	\array{c}
	\infer[\ref{rule:hopill-parr}]
		{\judge{\Theta}{\recv{x}{y}{\recv{y}{p}{P}}}{\Gamma, \cht{x}{\assume{\hyp{G}} \parr A}}}
		{\infer[\ref{rule:hopill-assume}]
			{\judge{\Theta}{\recv{y}{p}{P}}{\Gamma, \cht{y}{\assume{\hyp{G}}}, \cht{x}{A}}}
			{\judge{\Theta, \pvt{p}{\hyp{G}}}{P}{\Gamma, \cht{x}{A}}}}
	\endarray
\end{alignat*}
\end{spreadlines}
\end{remark}

\newcommand{\defproc}[3]{{\textsf{def}\; #1 \coloneqq #2 \;\textsf{in}\;} #3}
\newcommand{\callproc}[2]{{#1 \langle #2 \rangle}}
\begin{remark}[Procedures]
\label{ex:procedures}
If the abstraction that we send over a channel does not refer to any free 
process variable, then we can always replicate it as many times as we wish. Here is the proof.
\[
\infer[\ref{rule:hopill-!}]
	{\judge{\emptyho}{\server{x}{y}{\sendho{y}{\abstr \rho P}{\nil}}}{\cht{x}{\bang \provide{\hyp{G}}}}}
	{\infer[\ref{rule:hopill-provide}]
    {\judge{\emptyho}{\sendho{y}{\abstr \rho P}{\nil}}{\cht{y}{\provide{\hyp{G}}}}}
    {\judge{\emptyho}{P}{\hyp{G}\rho}
    &\infer[\ref{rule:hopill-mix-0}]
      {\judge{\emptyho}{\nil}{\emptyhyp}}
      {}}}
\]
We use this property to build a notion of procedures that can be used at will. 
We denote procedure names with $K$, for readability. We will later use it as a channel name in our desugaring.
\begin{bnftable}
	{\phl P, \phl Q \Coloneqq {}}
	{\defproc{K}{\abstr{\rho}{P}}{Q}}
		\phl P, \phl Q \Coloneqq {} & \nohl{\dots} & \\
		\mid {} & \defproc{K}{\abstr{\rho}{P}}{Q} & procedure definition \\
		\mid {} & \callproc{K}{\rho} & procedure invocation
\end{bnftable}
A term $\defproc{K}{\abstr{\rho}{P}}{Q}$ defines procedure $K$ as $\abstr{\rho}{P}$ in the scope of $Q$, and a term $\callproc{K}{\rho}$ invokes procedure $K$ by passing the parameters $\rho$.

Below is the desugaring of both constructs. For simplicity of presentation, we assume that $K$ in term $\defproc{K}{\abstr{\rho}{P}}{Q}$ is used at least once in $Q$. The generalisation to the case where $Q$ does not use $K$ at all is straightforward (thanks to rule \cref{rule:hopill-weaken}).
\begin{spreadlines}{\typerulesskipamount}%
\begin{alignat*}{3}
\mspace{-10mu}\array{c}
		\infer
			{\judge{\Theta}{\defproc{K}{\abstr{\rho}{P}}{Q}}{\hyp{G} \pp \Gamma}}
			{\judge{\emptyho}{P}{\Delta\rho}
			&\judge{\Theta}{Q}{\hyp{G} \pp \Gamma, \cht{K}{\query\assume{\Delta}}}}
	\endarray\mspace{-10mu}
	&\defeq &&
	\mspace{-3mu}\array{c}
	\infer[\ref{rule:hopill-cut}]
		{\judge{\Theta}{\res{xK}{\left(\server{x}{y}{\sendho{y}{\abstr \rho P}{\nil}} \pp Q \right)}}{\hyp{G} \pp \Gamma}}
		{\infer[\ref{rule:hopill-mix}]
			{\judge{\Theta}{\server{x}{y}{\sendho{y}{\abstr \rho P}{\nil}} \pp Q}{\cht{x}{\bang \provide{\Delta}} \pp \hyp{G} \pp \Gamma,\cht{K}{\query\assume{\Delta}}}}
			{\infer[\ref{rule:hopill-!}]
				{\judge{\emptyho}{\server{x}{y}{\sendho{y}{\abstr \rho P}{\nil}}}{\cht{x}{\bang \provide{\Delta}}}}
				{\infer[\ref{rule:hopill-provide}]
					{\judge{\emptyho}{\sendho y{\abstr \rho P}{\nil}}{\cht{y}{\provide{\Delta}}}}
					{\judge{\emptyho}{P}{\Delta\rho}
          &\infer[\ref{rule:hopill-mix-0}]
            {\judge{\emptyho}{\nil}{\emptyhyp}}
            {}}}
			  &\judge{\Theta}{Q}{\hyp{G} \pp \Gamma, \cht{K}{\query\assume{\Delta}}}}}
	\endarray\mspace{-10mu}
\\
	\array{c}
		\infer
			{\judge{\emptyho}{\callproc{K}{\rho}}{\Gamma\rho,\cht{K}{\query\assume{\Gamma}}}}
			{}
	\endarray
	& \defeq &&
	\array{c}
	\infer[\ref{rule:hopill-?}]
		{\judge{\emptyho}{\clientuse{K}{y}{\recvho{y}{p}{\invoke{p}\rho}}}{\Gamma\rho,\cht{K}{\query\assume{\Gamma}}}}
		{\infer[\ref{rule:hopill-assume}]
			{\judge{\emptyho}{\recvho{y}{p}{\invoke{p}{\rho}}}{\Gamma\rho,\cht{y}{\assume{\Gamma}}}}
			{\infer[\ref{rule:hopill-id}]
				{\judge{\pvt{p}{\Gamma}}{\invoke{p}{\rho}}{\Gamma\rho}}
				{}}}
	\endarray
\end{alignat*}
\end{spreadlines}
It is possible to derive a more general rule for procedure definitions that allows the type of the body $P$ to have any number of sequents (instead of exactly one). The same generalisation is not possible for procedure calls since \cref{rule:hopill-?} expects exactly one sequent.

Observe that even if procedures can be used at will, typing ensures that each usage respects linearity (\ie, every usage ``consumes'' the necessary linear resources available in the context). Note also that self-invocations are not supported, as typing forbids them (types must be finite).
\end{remark}

\begin{remark}[Higher-order parameters]
We chose not to make abstractions parametric on process variables for economy of the calculus.
This feature can be reconstructed with the following syntactic sugar.
\begin{bnftable}
	{\phl P, \phl Q \Coloneqq {}}
	{\defproc{K}{\abstr{\rho}{P}}{Q}}
	\phl P, \phl Q \Coloneqq {} & \nohl{\dots} & \\
	\mid {} & \prefixed{x\lambda q}{P} & named higher-order parameter \\
	\mid {} & \invoke{P}{x={\abstr{\rho} Q}} & application
\end{bnftable}
The desugaring is simple, interpreting a named higher-order parameter as a channel on which we perform a single higher-order input.
\begin{spreadlines}{\typerulesskipamount}%
\begin{alignat*}{3}
	\begin{array}{c}
		\infer
			{\judge{\Theta}{\prefixed{x\lambda p}{P}}{\Gamma, \cht{x}{\assume{\Delta}}}}
			{\judge{\Theta,\pvt{p}{\Delta}}{P}{\Gamma}}
	\end{array}
	&\defeq &&
	\begin{array}{c}
	\infer[\ref{rule:hopill-assume}]{
		\judge{\Theta}{\recvho{x}{p}{P}}{\Gamma, \cht{x}{\assume{\Delta}}}
	}{
		\judge{\Theta,\pvt{p}{\Delta}}{P}{\Gamma}
	}
	\end{array}
\\
	\begin{array}{c}
		\infer
			{\judge{\Theta,\Pi}{\invoke P{x={\abstr{\rho} Q}}}{\hyp{G} \pp \Gamma}}
			{\judge{\Theta}{Q}{\Delta\rho}
			&\judge{\Pi}{P}{\Gamma,\cht{x}{\assume\Delta}}}
	\end{array}
	&\defeq &&
	\begin{array}{c}
	\infer[\ref{rule:hopill-cut}]
		{\judge{\Theta,\Pi}{\res{yx}{(\sendho{y}{\abstr{\rho}{Q}}{\nil} \pp P)}}{\hyp{G} \pp \Gamma}}
		{\infer[\ref{rule:hopill-mix}]
			{\judge{\Theta,\Pi}{\sendho{y}{\abstr{\rho}{Q}}{\nil} \pp P}{\hyp{G} \pp \Gamma,\cht{x}{\assume\Delta} \pp \cht{y}{\provide{\Delta}}}}
			{\infer[\ref{rule:hopill-provide}] 
				{\judge{\Theta}{\sendho{y}{\abstr \rho Q}{\nil}}{\cht{y}{\provide{\Delta}}}}
				{\judge{\Theta}{Q}{\Delta\rho}
        &\infer[\ref{rule:hopill-mix-0}]
          {\judge{\emptyho}{\nil}{\emptyhyp}}
          {}}
			&\judge{\Theta'}{P}{\hyp{G} \pp \Gamma,\cht{x}{\assume\Delta}}}}
	\end{array}
\end{alignat*}
\end{spreadlines}
The desugaring of $\invoke{P}{x=\abstr \rho Q}$ yields a process that allows for reductions to happen in $P$ before the application takes place, since the corresponding higher-order named parameter term may be nested inside of $P$.
Also, this desugaring cannot be implemented by merely using an application of \cref{rule:hopill-chop}, since $p$ is \emph{bound} to $P$ in term $\prefixed{x\lambda p}{P}$ (as can be observed by its desugaring), and \cref{rule:hopill-chop} acts on free process variables.
\end{remark}

\subsection{Example: A cloud server}
\label{sec:cloud_server}
We illustrate the expressiveness of \hopill by implementing a cloud server for running applications that require a database (this example is adapted from the technical report \citep{M18}).
The idea is that clients are able to choose between two options: run both the application and the database it needs in the server, or run just the application in the server and connect it to an externally-provided database (which we could imagine is run somewhere else in the cloud).

\paragraph{A cloud server}
The process for the cloud server follows. We assume that $A$ is the protocol (left unspecified) that applications have to use in order to communicate with databases, and $f$ is the named parameter used by both abstractions to represent this shared connection (the parameter names may be different, we do this only for a simpler presentation).
We also let applications and databases access some external services, \eg, loggers, through the parameters $\widetilde{f}$ and $\widetilde{g}$, respectively. 
We write $\widetilde{f = x}$ for $f_1 = x_1,\ldots,f_n = x_n$.
\[
	\server{cs}{x}{\choice*{x}
		{\recv{x}{x'}
			{\recvho{x}{app}
				{\recvho{x'}{db}
					{\res{zw}{\left(\invoke{app}{f=z,\widetilde{f = u}} \pp \invoke{db}{f=w,\widetilde{g = v}}\right)}}}}}
		{\recv{x}{extdb}
			{\recvho{x}{app}
				{\res{zw}{\left(\invoke{app}{f=w,\widetilde{f = u}} \pp \forward{extdb}{w}\right)}}}}}
\]

The cloud server waits for client requests on channel $cs$. 
Then, it communicates with the client on the established channel $x$. 
It offers the options that we mentioned through a choice, respectively with the
left and right branches. 
In the left branch, we first receive an auxiliary channel $x'$. 
We then receive the application $app$ on $x$ and the database $db$ on $x'$.
Then, we compose and run $app$ and $db$ in parallel, connecting them through the endpoints $z$ and $w$.
In the right branch, we first receive a channel $extdb$ for communicating with the external database, and then receive the application $app$. 
Then, we compose $app$ with a link term $\forward{extdb}{w}$, which connects the application to the external database.

\paragraph{Typing the cloud server}
We now illustrate how to use our typing rules to prove that the cloud server is well-typed.
Let $\query \Gamma = (\cht{f_i}{\query A_i})_i$, $\query\Delta = (\cht{g_i}{B_i})_i$, $\rho = \{f=z,\widetilde{f = u}\}$, and $\rho' = \{f=w,\widetilde{g = v}\}$.
For readability, we first type the left and right branches in the choice offered through $x$. 
Here is the proof for the left branch.
\[\small
\infer[\ref{rule:hopill-parr}]
	{\judge{\emptyho}
		{\recv{x}{x'}{\recvho{x}{app}{\recvho{x'}{db}{\res{zw}{\left(\invoke{app}{\rho} \pp \invoke{db}{\rho'}\right)}}}}}
		{\query\Gamma\rho,\query\Delta\rho',\cht{x}{\assume{\query\Delta, \cht{f}{\dual A}}	\parr	\assume{\query\Gamma, \cht{f}{A}}}}}
	{\infer[\ref{rule:hopill-assume}]
		{\judge{\emptyho}
			{\recvho{x}{app}{\recvho{x'}{db}{\res{zw}{\left(\invoke{app}{\rho} \pp \invoke{db}{\rho'}\right)}}}}
			{\query\Gamma\rho,\query\Delta\rho',\cht{x'}{\assume{\query\Delta, \cht{f}{\dual{A}}}}, \cht{x}{\assume{\query\Gamma,\cht{f}{A}}}}}
		{\infer[\ref{rule:hopill-assume}]
			{\judge{\pvt{app}{(\query\Gamma,\cht{l}{A})}}
				{\recvho{x'}{db}{\res{zw}{\left(\invoke{app}{\rho} \pp \invoke{db}{\rho'}\right)}}}
				{\query\Gamma\rho,\query\Delta\rho',\cht{x'}{\assume{\query\Delta, \cht{f}{\dual{A}}}}}}
			{\infer[\ref{rule:hopill-cut}]
				{\judge{\pvt{app}{(\query\Gamma,\cht{f}{A})},\pvt{db}{(\query\Delta, \cht{f}{\dual{A}})}}
					{\res{zw}{\left(\invoke{app}{\rho} \pp \invoke{db}{\rho'}\right)}}
					{\query\Gamma\rho,\query\Delta\rho'}}
				{\infer[\ref{rule:hopill-mix}]
					{\judge{\pvt{app}{(\query\Gamma,\cht{f}{A})},\pvt{db}{(\query\Delta, \cht{f}{\dual{A}})}}
						{\invoke{app}{\rho} \pp \invoke{db}{\rho'}}
						{\query\Gamma\rho,\cht{z}{A} \pp \query\Delta\rho',\cht{w}{\dual{A}}}}
					{\infer[\ref{rule:hopill-id}]
						{\judge{\pvt{app}{(\query\Gamma,\cht{f}{A})}}
							{\invoke{app}{\rho}}
							{\query\Gamma\rho,\cht{z}{A}}}
						{}
					&\infer[\ref{rule:hopill-id}]
						{\judge{\pvt{db}{(\query\Delta, \cht{f}{\dual{A}})}}
							{\invoke{db}{\rho'}}
							{\query\Delta\rho',\cht{w}{\dual{A}}}}
						{}}}}}}
\]
And this is the proof for the right branch.
\[\small
\infer[\ref{rule:hopill-parr}]
	{\judge{\emptyho}
		{\recv{x}{extdb}{\recvho{x}{app}{\res{zw}{\left(\invoke{app}{\rho} \pp \forward{extdb}{w}\right)}}}}
		{\query\Gamma\rho,\cht{x}{A \parr \assume{\query\Gamma, \cht{f}{A}}}}}
	{\infer[\ref{rule:hopill-assume}]
		{\judge{\emptyho}
			{\recvho{x}{app}{\res{zw}{\left(\invoke{app}{\rho} \pp \forward{extdb}{w}\right)}}}
			{\query\Gamma\rho,\cht{extdb}{A},\cht{x}{\assume{\query\Gamma, \cht{f}{A}}}}}
		{\infer[\ref{rule:hopill-cut}]
			{\judge{\pvt{app}{(\query\Gamma, \cht{f}{A})}}
				{\res{zw}{\left(\invoke{app}{\rho} \pp \forward{extdb}{w}\right)}}
				{\query\Gamma\rho,\cht{extdb}{A}}}
			{\infer[\ref{rule:hopill-mix}]
				{\judge{\pvt{app}{(\query\Gamma, \cht{f}{A})}}
					{\invoke{app}{\rho} \pp \forward{extdb}{w}}
					{\query\Gamma\rho, \cht{z}{A} \pp \cht{extdb}{A},\cht{w}{\dual{A}}}}
				{\infer[\ref{rule:hopill-id}]
					{\judge{\pvt{app}{(\query\Gamma, \cht{f}{A})}}
						{\invoke{app}{\rho}}
						{\query\Gamma\rho, \cht{z}{A}}}
					{}
				&\infer[\ref{rule:hopill-axiom}]
					{\judge{\emptyho}
						{\forward{extdb}{w}}
						{\cht{extdb}{A},\cht{w}{\dual{A}}}}
					{}}}}}
\]
Now that we have proofs for the two branches, we can use them to type the entire cloud server.
Let $\mathcal{L}$ and $\mathcal{R}$ be the two proofs above, respectively, and $P_L$ and $P_R$ the processes that they type (the left and right branches in the cloud server).
Then, we type the cloud server as follows.
\[\small
\infer[\ref{rule:hopill-!}]
	{\judge{\emptyho}
		{\server{cs}{x}{\choice{x}{P_L}{P_R}}}
		{\query\Gamma\rho, \query\Delta\rho', \cht{cs}{\bang\left(\left(\assume{\query\Delta.\cht{f}{\dual{A}}}\parr\assume{\query\Gamma,\cht{f}{A}}\right)\with\left(A\parr\assume{\query\Gamma,\cht{f}{A}}\right)\right)}}}
	{\infer[\ref{rule:hopill-with}]
		{\judge{\emptyho}
			{\choice{x}{P_L}{P_R}}
			{\query\Gamma\rho, \query\Delta\rho', \cht{x}{\left(\assume{\query\Delta.\cht{f}{\dual{A}}}\parr\assume{\query\Gamma,\cht{f}{A}}\right)\with\left(A\parr\assume{\query\Gamma,\cht{f}{A}}\right)}}}
		{\mathcal{L}
		&\mathcal{R}}}
\]
The types for our cloud server follows the intuition that we initially
discussed for this example. 
The typing derivation also illustrates the interplay between our new features (process mobility and the usage of process variables) with the other features of the calculus---in this example: channel mobility, exponentials (replicated services), choices, and links.

\paragraph{A generic improvement}
\def\hlchanged#1{{\fboxsep=0pt\colorbox{yellow!30}{\vphantom{{\ensuremath{\overline{#1}}}}{\ensuremath{#1}}}}}
The code of our cloud server implementation would not depend on how clients and database communicate, were it not for the hardcoded protocol $A$.
We can get rid of the hardcoded $A$ and reach a generic implementation using polymorphism.
Here is the improved implementation, where we $\hlchanged{\text{underline}}$ the improvements.
\[
	\server{cs}{x}
		{\prefixed{\hlchanged{\arecvtype{x}{X}}}
			{\choice*{x}
				{\recv{x}{x'}
					{\recvho{x}{app}
						{\recvho{x'}{db}
							{\res{zw}{\left(\invoke{app}{f=z,\widetilde{f = u}} \pp \invoke{db}{f=w,\widetilde{g = v}}\right)}}}}}
				{\recv{x}{extdb}
					{\recvho{x}{app}
						{\res{zw}{\left(\invoke{app}{f=w,\widetilde{f = u}} \pp \forward{extdb}{w}\right)}}}}}}
\]
In the improved cloud server, the client must also send us the protocol $X$ that the application will use to communicate with the database. The cloud server is thus now generic, and the type of channel $cs$ is the following.
\[
\bang
\left(
	\hlchanged{\forall X.}\left(
		\left( 
			\assume{\query\Delta, \cht{f}{\hlchanged{\dual{X}}}}
			\parr
			\assume{\query\Gamma, \cht{f}{\hlchanged{X}}} 
		\right)
		\with
		\left( 
			\hlchanged{X} 
			\parr 
			\assume{\query\Gamma, \cht{f}{\hlchanged{X}}}
		\right)
	\right)
\right)
\]
Here is the proof.
\[\small
\infer[\ref{rule:hopill-!}]
	{\judge{\emptyho}
		{\server{cs}{x}{\recvtype{x}{X}{\choice{x}{P_L}{P_R}}}}
		{\query\Gamma\rho, \query\Delta\rho', \cht{cs}{\bang\left(\forall X.\left(\left(\assume{\query\Delta.\cht{f}{\dual{X}}}\parr\assume{\query\Gamma,\cht{f}{X}}\right)\with\left(X\parr\assume{\query\Gamma,\cht{f}{X}}\right)}\right)\right)}}
	{\infer[\ref{rule:hopill-forall}]
		{\judge{\emptyho}
			{\recvtype{x}{X}{\choice{x}{P_L}{P_R}}}
			{\query\Gamma\rho, \query\Delta\rho', \cht{x}{\forall X.\left(\left(\assume{\query\Delta.\cht{f}{\dual{X}}}\parr\assume{\query\Gamma,\cht{f}{X}}\right)\with\left(X\parr\assume{\query\Gamma,\cht{f}{X}}\right)\right)}}}
		{\infer[\ref{rule:hopill-with}]
			{\judge{\emptyho}
				{\choice{x}{P_L}{P_R}}
				{\query\Gamma\rho, \query\Delta\rho', \cht{x}{\left(\assume{\query\Delta.\cht{f}{\dual{X}}}\parr\assume{\query\Gamma,\cht{f}{X}}\right)\with\left(X\parr\assume{\query\Gamma,\cht{f}{X}}\right)}}}
			{\mathcal{L}\{X/A\}
			&\mathcal{R}\{X/A\}}}}
\]
The proofs $\mathcal{L}\{X/A\}$ and $\mathcal{R}\{X/A\}$ used above are as $\mathcal{L}$ and $\mathcal{R}$ respectively, but wherever we had $A$ we now have $X$.
The interplay between type variables and process variables merits illustration, so we show $\mathcal{R}\{X/A\}$ in full.
\[\small
\infer[\ref{rule:hopill-parr}]
	{\judge{\emptyho}
		{\recv{x}{extdb}{\recvho{x}{app}{\res{zw}{\left(\invoke{app}{\rho} \pp \forward{extdb}{w}\right)}}}}
		{\query\Gamma\rho,\cht{x}{X \parr \assume{\query\Gamma, \cht{f}{X}}}}}
	{\infer[\ref{rule:hopill-assume}]
		{\judge{\emptyho}
			{\recvho{x}{app}{\res{zw}{\left(\invoke{app}{\rho} \pp \forward{extdb}{w}\right)}}}
			{\query\Gamma\rho,\cht{extdb}{X},\cht{x}{\assume{\query\Gamma, \cht{f}{X}}}}}
		{\infer[\ref{rule:hopill-cut}]
			{\judge{\pvt{app}{(\query\Gamma, \cht{f}{X})}}
				{\res{zw}{\left(\invoke{app}{\rho} \pp \forward{extdb}{w}\right)}}
				{\query\Gamma\rho,\cht{extdb}{X}}}
			{\infer[\ref{rule:hopill-mix}]
				{\judge{\pvt{app}{(\query\Gamma, \cht{f}{X})}}
					{\invoke{app}{\rho} \pp \forward{extdb}{w}}
					{\query\Gamma\rho, \cht{z}{X} \pp \cht{extdb}{X},\cht{w}{\dual{X}}}}
				{\infer[\ref{rule:hopill-id}]
					{\judge{\pvt{app}{(\query\Gamma, \cht{f}{X})}}
						{\invoke{app}{\rho}}
						{\query\Gamma\rho, \cht{z}{X}}}
					{}
				&\infer[\ref{rule:hopill-axiom}]
					{\judge{\emptyho}
						{\forward{extdb}{w}}
						{\cht{extdb}{X},\cht{w}{\dual{X}}}}
					{}}}}}
\]
Observe the application of \cref{rule:hopill-assume} in the proof. 
If we read it bottom-up, we are moving the type variable $X$ from the typing of a channel in the conclusion---$\cht{x}{X \parr \assume{\query\Gamma, \cht{f}{X}}}$---to the typing of a process in the premise---$\pvt{app}{( \query\Gamma, \cht{f}{X})}$. 
This is then carried over to the application of \cref{rule:hopill-id}, which is thus able to type the usage of a process variable that is generic on the behaviour that will be enacted.

\subsection{Operational Semantics of Derivations}
\label{sec:hopill-sos}

We present the SOS specification for the new ingredients of \hopill compared to \pill. The full SOS specification can be found in \cref{sec:hopill-full}.

The semantics of the fragment corresponding to \pill does not interact with the process environment ($\Theta$) in any significant way and can thus be defined by lifting the SOS specification of \pill (\cref{sec:pill-sos}). Intuitively, this ammounts to adding process environments following the same recepie we used to lift the typing rules of \pill. For instance, the semantics of \cref{rule:hopill-bot} is given by the following axiom.
\begin{drules}
\plto
  {\infer[\ref{rule:hopill-bot}]
      {\judge{\Theta}{\wait{x}{P}}{\Gamma, \cht{x}{\bot}}}
      {\deduce
        {\judge{\Theta}{P}{\Gamma}}
        {\der{D}}}}
  {\lwait{x}}
  {\deduce
    {\judge{\Theta}{P}{\Gamma}}
    {\der{D}}}
\end{drules}

\paragraph{Process mobility}
The derivation rules for higher-order output (\ref{rule:hopill-provide}) and input (\ref{rule:hopill-provide}) are similar to the rules of other action prefixes and are given below.
\begin{drules}
\plto
	{\infer[\ref{rule:hopill-provide}]
        {\judge{\Theta}{\sendho{x}{\abstr{\rho}{P}}{Q}}{\cht{x}{\provide{\hyp{G}}}}}
        {\deduce
          {\judge{\Theta}{P}{\hyp{G}\rho}}
          {\der{D}}
        &\deduce
          {\judge{\emptyho}{Q}{\emptyhyp}}
          {\der{E}}}}
	{\lsendho{x}{\abstr{\rho}{\der{D}}}}
  {\deduce
    {\judge{\emptyho}{Q}{\emptyhyp}}
    {\der{E}}}
\quad
\plto
	{\infer[\ref{rule:hopill-assume}]
    {\judge{\Theta}{\recvho{x}{p}{P}}{\Gamma,\cht{x}{\assume{\hyp{G}}}}}
    {\deduce
      {\judge{\Theta,\pvt{p}{\hyp{G}}}{P}{\Gamma}}
      {\der{D}}}}
	{\lrecvho{x}{p}}
  {\deduce
    {\judge{\Theta,\pvt{p}{\hyp{G}}}{P}{\Gamma}}
    {\der{D}}}
\end{drules}
The derivation rule for higher-order communication synchronises a send and receive over connected endpoints and records the association between the process abstraction from the sender and the process symbol at the receiver with an explicit substitution (\ref{rule:hopill-chop}). This is similar to how in endpoint communication, the association between the output and input endpoints is recorded in a new restriction term (\ref{rule:hopill-cut}) by \cref{rule:hopill-cut-tensor}.
\begin{drules}
\infer[\rlabel{\ref*{rule:hopill-provide}\ref*{rule:hopill-assume}}{rule:hopill-cut-provide}]
  {\plto
    {\infer[\ref{rule:hopill-cut}]
      {\judge{\Theta,\Pi}{\res{xy}{Q}}{\hyp{G} \pp \Gamma}}
      {\deduce
        {\judge{\Theta,\Pi}{Q}{\hyp{G}\pp \cht{x}{\provide{\hyp{H}}} \pp \Gamma, \cht{y}{\assume{\hyp{H}}}}}
        {\der{D}}}}
    {\tau}
    {\infer[\ref{rule:hopill-chop}]
      {\judge{\Theta,\Pi}{\chop{p}{\abstr \rho P}{Q'}}{\hyp{G} \pp \Gamma}}
      {\deduce
        {\judge{\Pi}{P}{\hyp{H}\rho}}
        {\der{E}}
      &\deduce
        {\judge{\Theta,\pvt{p}{\hyp{H}}}{Q'}{\hyp{G} \pp \Gamma}}
        {\der{D'}}}}}
  {\plto
    {\deduce
      {\judge{\Theta,\Pi}{Q}{\hyp{G}\pp \cht{x}{\provide{\hyp{H}}} \pp \Gamma, \cht{y}{\assume{\hyp{H}}}}}
      {\der{D}}}
    {\lsync
      {\lsendho{x}{\abstr{\rho}{\der{E}}}}
      {\lrecvho{y}{p}}}
    {\deduce
      {\judge{\Theta,\pvt{p}{\hyp{H}}}{Q'}{\hyp{G} \pp \Gamma}}
      {\der{D'}}}}
\end{drules}

\paragraph{Explicit substitution}
To invoke the process bound to a symbol variable we need to obtain this process from the context of the invocation, \ie, the closest substitution term binding the process symbol (which can then be discarded thank to the linear use of process symbols). This mechanism is captured by the derivation rules below.
\begin{drules}
\plto
  {\infer[\ref{rule:hopill-id}]
    {\judge{\pvt{p}{\hyp{G}}}{\invoke[\hyp{G}]{p}{\sigma}}{\hyp{G}\sigma}}
    {\vphantom{\hyp{G}}}}
  {\lsubst[\hyp{G}]{p}{\abstr{\rho}{\der{D}}}}
  {\deduce
    {\judge{\Pi}{P\{\sigma\circ\rho^{-1}\}}{\hyp{G}\sigma}}
    {\der{D}}}
\\
\infer[\rlabel{\rname{Run}}{rule:hopill-chop-id}]
  {\plto
    {\infer[\ref{rule:hopill-chop}]
      {\judge{\Theta,\Pi}{\chop{p}{\abstr \rho P}{Q}}{\hyp{G}}}
      {\deduce
        {\judge{\Pi}{P}{\hyp{H}\rho}}
        {\der{E}}
      &\deduce
        {\judge{\Theta,\pvt{p}{\hyp{H}}}{Q}{\hyp{G}}}
        {\der{D}}}}
    {\tau}
    {\deduce
      {\judge{\Theta,\Pi}{Q'}{\hyp{G}}}
      {\der{D}'}}}
  {\plto
    {\deduce
      {\judge{\Theta,\pvt{p}{\hyp{H}}}{Q}{\hyp{G}}}
      {\der{D}}}
    {\lsubst[\hyp{H}]{p}{\abstr{\rho}{\der{E}}}}
    {\deduce
      {\judge{\Theta,\Pi}{Q'}{\hyp{G}}}
      {\der{D}'}}}
\end{drules}
The following derivation rule captures scope extrusion for process symbols that occur in the payload of a higher-order output (\cref{rule:hopill-provide} allows for free process symbols in abstractions).
\begin{drules}
\infer[\rlabel{\rname{Ext}}{rule:hopill-chop-provide}]
  {\plto
    {\infer[\ref{rule:hopill-chop}]
      {\judge{\Theta,\Pi}{\chop{p}{\abstr \rho P}{Q}}{\hyp{G}}}
      {\deduce
        {\judge{\Pi}{P}{\hyp{H}\rho}}
        {\der{E}}
      &\deduce
        {\judge{\Theta,\pvt{p}{\hyp{H}}}{Q}{\hyp{G}}}
        {\der{D}}}}
    {\lsendho{x}{\abstr{\sigma}{\der{F'}}}}
    {\deduce
      {\judge{\Theta'}{Q'}{\hyp{G'}}}
      {\der{D'}}}}
  {\plto
    {\deduce
      {\judge{\Theta,\pvt{p}{\hyp{H}}}{Q}{\hyp{G}}}
      {\der{D}}}
    {\lsendho{x}{\abstr{\sigma}{\der{F}}}}
    {\deduce
      {\judge{\Theta'}{Q'}{\hyp{G'}}}
      {\der{D'}}}
  &\der{F'} = \pbox{
  {\infer[\ref{rule:hopill-chop}]
    {\judge{\Xi,\Pi}{\chop{p}{\abstr \rho P}{R}}{\hyp{I}}}
    {\deduce
      {\judge{\Pi}{P}{\hyp{H}\rho}}
      {\der{E}}
    &\deduce
      {\judge{\Xi,\pvt{p}{\hyp{H}}}{R}{\hyp{I}}}
      {\der{F}}}}}
  &\phl{p \in \fpv(R)}}
\end{drules}
Similarly to \cref{rule:hopill-cut-res}, the following rule performs the propagation (or lifting) of actions that do not involve the process symbol bound by the explicit substitution term.
\begin{drules}
\infer[\rlabel{\rname{Lift}}{rule:hopill-chop-res}]
  {\plto
    {\infer[\ref{rule:hopill-chop}]
      {\judge{\Theta,\Pi}{\chop{p}{\abstr \rho P}{Q}}{\hyp{G}}}
      {\deduce
        {\judge{\Pi}{P}{\hyp{H}\rho}}
        {\der{E}}
      &\deduce
        {\judge{\Theta,\pvt{p}{\hyp{H}}}{Q}{\hyp{G}}}
        {\der{D}}}}
    {l}
    {\infer[\ref{rule:hopill-chop}]
      {\judge{\Theta',\Pi}{\chop{p}{\abstr \rho P}{Q'}}{\hyp{G'}}}
      {\deduce
        {\judge{\Pi}{P}{\hyp{H}\rho}}
        {\der{E}}
      &\deduce
        {\judge{\Theta',\pvt{p}{\hyp{H}}}{Q'}{\hyp{G'}}}
        {\der{D'}}}}}
  {\plto
    {\deduce
      {\judge{\Theta,\pvt{p}{\hyp{H}}}{Q}{\hyp{G}}}
      {\der{D}}}
    {l}
    {\deduce
      {\judge{\Theta',\pvt{p}{\hyp{H}}}{Q'}{\hyp{G'}}}
      {\der{D'}}}
    &\phl{p \notin \fpv(l)}}
\end{drules}%

\paragraph{LTS of Derivations}
We extend the sets $\ActSet$ of action labels and $\LblSet$ of all labels defined in \cref{sec:pill-sos} to include the labels introduced in this section.
To avoid ``leaking'' derivations to the labels used by the transition systems of processes and environments we parametrise the definition of these sets in a $S$ of higher-order payloads--for derivations $S = \DerSet$, for processes $S = \ProcSet$, \etc.
\begin{align*}
\ActSet(S) \defeq {} & \left\{\array{l} 
  \lclose{x}, \lwait{x}, \lsend{x}{y}, \lrecv{x}{y}, \linl{x}, \lcoinl{x}, \linr{x}, \lcoinr{x},\\ \luse{x}{y}, \lcouse{x}{y}, \ldup{x}{y}{z}, \lcouse{x}{y}{z}, \ldisp{x}, \lcodisp{x},\\ \lsendtype{x}{A}, \lrecvtype{x}{A}, \lsendho{x}{\abstr{\sigma}{s}}, \lrecvho{x}{p}
  \endarray\middle\vert
  \array{l}
    x,y,z \text{ names}, A \text{ type,}\\
    p \text{ process variable, }\\
    \abstr{\sigma}{s} \text{ abstraction for } s \in S
  \endarray\right\}\\
\LblSet(S) \defeq {} & \left\{\tau, \lforward{x}{y}, l, \lsync{l}{l'},\lsubst[\hyp{G}]{p}{\abstr{\sigma}{s}} 
  \,\middle\vert\array{l}
   l,l' \in \ActSet(S)\text{, } x, y \text{ names},
      p \text{ process variable,}\\
      \hyp{G} \in \EnvSet\text{, } 
      \abstr{\sigma}{s} \text{ abstraction for } s \in S
    \endarray\right\}
\end{align*}
Every function $f\colon S \to S'$ induces a function $\LblSet(f)\colon \LblSet(S) \to \LblSet({S'})$ that replaces payloads from $S$ with payloads from $S'$. This function acts as the identity on every label except for higher-order output and explicit substitution where it acts as follows:
\[
\LblSet(f)(\lsubst[\hyp{G}]{p}{\abstr{\sigma}{s}}) \defeq 
\lsubst[\hyp{G}]{p}{\abstr{\sigma}{f(s)}}
\qquad
\LblSet(f)(\lsendho{x}{\abstr{\sigma}{s}}) \defeq 
\lsendho{x}{\abstr{\sigma}{f(s)}}
\text{.}
\]
\begin{definition}
\label{def:hopill-lts-derivations}
The \lts of derivations for \hopill, denoted $lts_d$, is the triple $(\DerSet,\LblSet(\DerSet),\lto{})$:
\begin{itemize}
\item The set $\DerSet$ is the set of typing derivations for \pill.
\item The set $\LblSet(\DerSet)$ is the set of transition labels.
\item The relation ${\lto{}} \subseteq \DerSet \times \LblSet(\DerSet) \times \DerSet $ is the least relation closed under the SOS rules forming the speficication of \hopill (see \cref{sec:hopill-sos-full}).
\end{itemize}
\end{definition}

\subsection{Operational Semantics of Processes and Environments}

To define the semantics for processes and typing environments of \hopill, we follow the recipe introduced in \cref{sec:mpill-proc-hyp}.

The first step is to specify how derivations are projected to process terms and typing environments. Let $\ProcSet$ and $\EnvSet$ denote the sets of all process terms and typing environments of \hopill. The projections 
$\proc\colon \DerSet \to \ProcSet$ and $\env\colon \DerSet \to \EnvSet$ are defined as in \cref{sec:mpill-proc-hyp} (typing environments now include process environemnts):
\begin{highlight}
\[
  \proc\begin{pmatrix}\deduce
              {\judge{\Theta}{P}{\hyp{G}}}
              {\der D}\end{pmatrix}
  = \phl{P}
  \qquad
  \env\begin{pmatrix}\deduce
            {\judge{\Theta}{P}{\hyp{G}}}
            {\der D}\end{pmatrix}
  = \hoenv{\Theta}{\hyp{G}}
  \text{.}
\]
\end{highlight}

The second step of the recipe is to formalise the notions of erasure and session fidelity.
Because the set labels is parametrised in the set of payloads for higher-order operatioions, we need slightly update the definition of $\lb(-)$ to account for this parameter (which is exactly the state space of the \lts):
\[
\lb(X) = X \times \LblSet(X) \qquad
\lb(f)(x,l) = (f(x),\LblSet(f)(l))
\text{.}
\]
Then, we can instantiate \cref{def:erasure,def:safety} with the $lts_d$, $\proc$, $\env$, and $\lb(-)$ defined above.
The resulting notion of erasure for \hopill states that for any derivation $\der{D}$ and $l \in \LblSet(\DerSet)$:
\begin{itemize}
\item
if $\der{D} \lto{l} \der{D'}$, then $\proc(\der{D}) \lto{\LblSet(\proc)(l)} \proc(\der{D'})$, and
\item
if $\proc(\der{D}) \lto{\LblSet(\proc)(l)} P'$, then $\der{D} \lto{l} \der{D'}$ for some $\proc(\der{D'}) = P'$.
\end{itemize}
Observe that the second clause ignores higher-order operations with ill-typed payloads (if $l$ has a payload, then it is a derivation hence $\proc$ yields a well-typed process). This kind of restriction is standard for higher-order calculi and many typed process calculi \eg, it can be found in the subject reduction theorems of the Simply Typed $\pi$-calculus and \HOpi \cite[p.~245,p.~377]{SW01}.
Session fidelity for \hopill states that given $\judge{\Theta}{P}{\hyp{G}}$ and $l \in \LblSet(\DerSet)$ if $P \lto{\LblSet(\proc)(l)} P'$ for some $P'$ then $\hoenv{\Theta}{\hyp{G}} \lto{\LblSet(\env)(l)} \hoenv{\Theta'}{\hyp{G'}}$ for some $\Theta'$ and $\hyp{G'}$ such that $\judge{\Theta'}{P'}{\hyp{G'}}$.

\begin{figure}[t]
  \begin{highlight}\small
  \begin{minipage}{.68\textwidth}
    \begin{spreadlines}{\termltsskipamount}
      \headertext{Structural}
      \begin{gather*}
        \infer[\rlabel{\ref*{rule:hopill-chop-id}}{rule:hopill-proc-chop-id}]
          {\phl{\chop{q}{\abstr{\rho}{Q}}{P}} \lto{\tau} \phl{P'}}
          {\phl{P} \lto{\lsubst[\hyp{G}]{q}{\abstr{\rho}{Q}}} \phl{P'}}
        \quad
        \infer[\rlabel{\ref*{rule:hopill-chop-res}}{rule:hopill-proc-chop-res}]
          {\phl{\chop{q}{\abstr{\rho}{Q}}{P}} \lto{l} \phl{\chop{q}{\abstr{\rho}{Q}}{P'}}}
          {\phl{P} \lto{l} \phl{P'}
          &\phl{q \notin \fpv(l)}}
        \\
        \infer[\rlabel{\ref*{rule:hopill-chop-provide}}{rule:hopill-proc-chop-provide}]
          {\phl{\chop{q}{\abstr{\rho}{Q}}{P}} \lto{\lsendho{x}{\abstr{\sigma}{\chop{q}{\abstr{\rho}{Q}}{R}}}} \phl{P'}}
          {\phl{P} \lto{\lsendho{x}{\abstr{\sigma}{R}}} \phl{P'}} 
        \quad
        {\phl{\invoke[\hyp{G}]{p}{\sigma}} \lto{\lsubst[\hyp{G}]{p}{\abstr{\rho}{P}}} \phl{P\{\sigma \circ \rho^{-1}\}}}
      \end{gather*}%
      \footer%
    \end{spreadlines}
  \end{minipage}\hfill%
  \begin{minipage}{.29\textwidth}
    \begin{spreadlines}{\termltsskipamount}
      \headertext{Communications}
      \begin{gather*}
        \infer[\rlabel{\ref*{rule:hopill-cut-provide}}{rule:hopill-proc-cut-provide}]
          {\phl{\res{xy}{P}} \lto{\tau} \phl{\chop{q}{\abstr{\rho}{Q}}{P'}}}
          {\phl{P} \lto{\lsync{\lsendho{x}{\abstr{\rho}{Q}}}{\lrecvho{y}{q}}} \phl{P'}}
       \\\vphantom{\infer{P\lto{q^G}}{P\lto{q^G}}}
      \end{gather*}
      \footer%
    \end{spreadlines}%
  \end{minipage}
\end{highlight}
\caption{\hopill, process transitions, new rules.}
\label{fig:hopill-sos-processes}
\end{figure}

\begin{figure}
\begin{highlight}\small
\begin{spreadlines}{\envltsskipamount}
\header
\begin{gather*}
{\hoenv{\Theta}{\Gamma, \Delta, \cht{x}{A \tensor B}}
\lto{\lsend{x}{x'}}
\hoenv{\Theta}{\Gamma, \cht{x}{B} \pp \Delta, \cht{x'}{A}}}
\qquad
{\hoenv{\Theta}{\Gamma, \cht{x}{A \parr B}}
\lto{\lrecv{x}{x'}}
\hoenv{\Theta}{\Gamma, \cht{x}{B}, \cht{x'}{A}}}
\\
{\hoenv{\Theta}{\cht{x}{\one}}
\lto{\lclose{x}}
\hoenv{\Theta}{\emptyhyp}}
\qquad
{\hoenv{\Theta}{\Gamma, \cht{x}{\bot}}
\lto{\lwait{x}}
\hoenv{\Theta}{\Gamma}}
\qquad
{\hoenv{\Theta}{\Gamma, \cht{x}{A \oplus B}}
\lto{\linl{x}}
\hoenv{\Theta}{\Gamma, \cht{x}{A}}}
\\
{\hoenv{\Theta}{\Gamma, \cht{x}{A \with B}}
\lto{\lcoinl{x}}
\hoenv{\Theta}{\Gamma, \cht{x}{A}}}
\quad
{\hoenv{\Theta}{\Gamma, \cht{x}{A \oplus B}}
\lto{\linr{x}}
\hoenv{\Theta}{\Gamma, \cht{x}{B}}}
\quad
{\hoenv{\Theta}{\Gamma, \cht{x}{A \with B}}
\lto{\lcoinr{x}}
\hoenv{\Theta}{\Gamma, \cht{x}{B}}}
\\
{\hoenv{\emptyho}{\cht{x}{\dual{A}}, \cht{y}{A}}
\lto{\lforward{x}{y}}
\hoenv{\emptyho}{\emptyhyp}}
\qquad
{\hoenv{\Theta}{\Gamma,\cht{x}{\query A}}
\lto{\luse{x}{x'}}
\hoenv{\Theta}{\Gamma,\cht{x'}{A}}}
\qquad
{\hoenv{\emptyho}{\query\Gamma,\cht{x}{\bang A}}
\lto{\lcouse{x}{x'}}
\hoenv{\emptyho}{\query\Gamma,\cht{x'}{A}}}
\\
{\hoenv{\Theta}{\Gamma,\cht{x}{\query A}}
\lto{\ldisp{x}}
\hoenv{\Theta}{\Gamma,\cht{x}{\bot}}}
\quad
{\hoenv{\emptyho}{\Gamma,\cht{x}{\bang A}}
\lto{\lcodisp{x}}
\hoenv{\emptyho}{\Gamma,\cht{x}{1}}}
\quad
{\hoenv{\Theta}{\Gamma,\cht{x}{\query A}}
\lto{\ldup{x}{x_1}{x_2}}
\hoenv{\Theta}{\Gamma,\cht{x_1}{\query A \parr \query A}}}
\\
{\hoenv{\emptyho}{\query\Gamma,\cht{x}{\bang A}}
\lto{\lcodup{x}{x_1}{x_2}}
\hoenv{\emptyho}{\query\Gamma,\cht{x_1}{\bang A \tensor \bang A}}}
\qquad
{\hoenv{\Theta}{\Gamma,\cht{x}{\exists X.B}}
\lto{\lsendtype{x}{A}}
\hoenv{\Theta}{\Gamma,\cht{x}{B\{A/X\}}}}
\\
{\hoenv{\Theta}{\Gamma,\cht{x}{\forall X.B}}
\lto{\lsendtype{x}{A}}
\hoenv{\Theta}{\Gamma,\cht{x}{B\{A/X\}}}}
\qquad
{\hoenv{\Theta,\Pi}{\cht{x}{\provide{\hyp{G}}}}}
\lto{\lsendho{x}{\abstr{\sigma}{\hoenv{\Pi}{\hyp{G}\sigma}}}}
{\hoenv{\Theta}{\emptyhyp}}
\\
{\hoenv{\Theta}{\Gamma,\cht{x}{\assume{\hyp{G}}}}}
\lto{\lrecv{x}{p}}
{\hoenv{\Theta,\pvt{p}{\hyp{G}}}{\Gamma}}
\qquad
{\hoenv{\pvt{p}{\hyp{G}}}{\hyp{G}\rho}}
\lto{\lsubst[\hyp{G}]{p}{\abstr{\sigma}{\hoenv{\Pi}{\hyp{G}\sigma}}}}
{\hoenv{\Pi}{\hyp{G}{\rho}}}
\qquad
{\hoenv{\Theta}{\hyp{G}}
\lto{\tau}
\hoenv{\Theta}{\hyp{G}}}
\\
\infer
  {\hoenv{\Theta,\Pi}{\hyp{G} \pp \hyp{H}} \lto{l} \hoenv{\Theta',\Pi}{\hyp{G'} \pp \hyp{H}}}
  {\hoenv{\Theta}{\hyp{G}} \lto{l} \hoenv{\Theta}{\hyp{G'}}}
\qquad
\infer
  {\hoenv{\Theta,\Pi}{\hyp{G} \pp \hyp{H}} \lto{l} \hoenv{\Theta,\Pi'}{\hyp{G} \pp \hyp{H'}}}
  {\hoenv{\Pi}{\hyp{H}} \lto{l} \hoenv{\Pi'}{\hyp{H'}}}
\\
\infer
  {\hoenv{\Theta,\Pi}{\hyp{G} \pp \hyp{H}} \lto{l} \hoenv{\Theta',\Pi'}{\hyp{G'} \pp \hyp{H'}}}
  {\hoenv{\Theta}{\hyp{G}} \lto{l} \hoenv{\Theta'}{\hyp{G'}}
  &\hoenv{\Pi}{\hyp{H}} \lto{l} \hoenv{\Pi'}{\hyp{H'}}}
\end{gather*}
\footer
\end{spreadlines}
\end{highlight}
  \caption{\hopill, typing environment transitions.}
  \label{fig:hopill-sos-environments}
\end{figure}

The third step is to apply the principles of ``erasing types'' and ``erasing processes'' to the SOS for \pill derivations. The resulting specifications are in \cref{fig:pill-sos-processes,fig:pill-sos-environments}, respectively---
for processes we include only the reles that are new \wrt \pill.

\begin{definition}
\label{def:hopill-lts-processes}
The \lts of processes for \hopill, denoted $lts_p$, is the triple $(\ProcSet,\LblSet(\ProcSet),{\lto{}})$ where
\begin{itemize}
\item $\ProcSet$ is the set of process terms for \hopill,
\item $\LblSet(\ProcSet) $ is the set of labels,
\item ${\lto{}} \subseteq \ProcSet \times \LblSet(\ProcSet) \times \ProcSet $ is the least relation closed under the SOS rules in \cref{fig:hopill-sos-processes,fig:pill-sos-processes}.
\end{itemize}
\end{definition}

\begin{definition}
\label{def:hopill-lts-environments}
The \lts of environments for \hopill, denoted $lts_e$, is the triple $(\EnvSet, \LblSet(\EnvSet), {\lto{}})$ where
\begin{itemize}
\item $\EnvSet$ is the set of typing erenvironments for \hopill,
\item $\LblSet(EnvSet)$ is the set of labels,
\item ${\lto{}} \subseteq \EnvSet \times \LblSet(EnvSet) \times \EnvSet $ is the least relation closed under the SOS rules in \cref{fig:hopill-sos-environments}.
\end{itemize}
\end{definition}

The fourth and final step of the recipe is to verify that semantics of processes and of typing environments enjoy erasure and session fidelity.

\begin{theorem}[Erasure]
\label{thm:pill-erasure}
$lts_p$ enjoys erasure wrt $lts_d$.
\end{theorem}

\begin{theorem}[Safety]
\label{thm:pill-safety}
$lts_p$ enjoys session fidelity wrt $lts_e$.
\end{theorem}

As for \pimll, erasure entails typability preservation.
\begin{corollary}[Typability Preservation]
Let $P$ be well-typed. Then, $P \lto{l} P'$ implies that $P'$ is well-typed.
\end{corollary}

\subsection{Metatheory}

\subsubsection{Behavioural Theory}

The standard definitions of behavioural equivalences and preorders (\cref{def:strong-bisimilarity,def:bisimilarity}), their laws (\cref{fact:bisimilarity-laws}), and definition of congruence (\cref{def:congruence}) apply to \hopill without modification save for minor changes to handle the parametric definitions of labels.

When we consider higher-order parameters in labels and instantiate \cref{def:strong-bisimilarity} we obtain the following definition.
Let $lts_1 = (S_1,\LblSet(S_1),\lto{}_1)$ and $lts_2 = (S_2,\LblSet(S_2),\lto{}_2)$ be two labelled transition systems and let $\crel{R}$ be a relation between $S_1$ and $S_2$.
\begin{itemize}
  \item $\crel{R}$ is a strong forward simulation from $lts_1$ to $lts_2$ if $s_1 \crel{R} s_2$ implies that 
  \begin{itemize}
  \item if $s_1 \lto{\lsendho{x}{\abstr{\sigma_1}{s_1''}}}_1 s_1'$ then  $s_2 \lto{\lsendho{x}{\abstr{\sigma_2}{s_2''}}}_2 s_2'$ for some $s_2'$, $s_2''$, $\sigma_2$ such that $s_1' \crel{R} s_2'$ and $s_1'' \crel{R} s_2''$,
  \item if $s_1 \lto{\lsubst[\hyp{G}]{p}{\abstr{\sigma_1}{s_1''}}}_1 s_1'$ then  $s_2 \lto{\lsubst[\hyp{G}]{p}{\abstr{\sigma_2}{s_2''}}}_2 s_2'$ for some $s_2'$, $s_2''$, $\sigma_2$ such that $s_1' \crel{R} s_2'$ and $s_1'' \crel{R} s_2''$,
  \item otherwise if $s_1 \lto{l}_1 s_1'$ then $s_2 \lto{l}_2 s_2'$ and $s_1' \crel{R} s_2'$.
  \end{itemize}
  \item $\crel{R}$ is a strong backward simulation from $lts_1$ to $lts_2$ if its symmetric $\crel{R}^{-1}$ is a strong forward simulation from $lts_2$ to $lts_1$.
  \item $\crel{R}$ is a strong bisimulation for $lts_1$ and $lts_2$ if it is both a strong forward and backward simulation from $lts_1$ to $lts_2$;
\end{itemize}

When we unfold \cref{def:congruence} using the grammar of \hopill we obtain that a relation on processes $\asymp$ is a (pre)congruence if $P \asymp Q$ implies that:
\begin{enumerate}[nosep, noitemsep]
  \item ${P \pp R} \asymp {Q \pp R}$ for any $R$ (and the symmetric);
  \item $\prefixed{\pi}{P} \asymp \prefixed{\pi}{Q}$ for any prefix $\pi$;
  \item $\res{xy}{P} \asymp \res{xy}{Q}$ for any $x$, $y$;
  \item $\choice{x}{R}{P} \asymp \choice{x}{R}{Q}$ for any $x$ and $R$ (and the symmetric);
  \item\label{def:hopill-cong-5} $\chop{p}{\abstr{\rho}{R}}{P} \asymp \chop{p}{\abstr{\rho}{R}}{Q}$ for any $p$ and $\rho$;
  \item\label{def:hopill-cong-6} $\chop{p}{\abstr{\rho}{P}}{R} \asymp \chop{p}{\abstr{\sigma}{Q}}{R}$ for any $p$, $\rho$, and $\sigma$;
  \item\label{def:hopill-cong-7} $\sendho{x}{\abstr{\rho}{P}}{R} \asymp \sendho{x}{\abstr{\sigma}{Q}}{R}$ for any $x$, $\rho$, and $\sigma$.
\end{enumerate}
The only difference with respect to \pill is the new case for explicit substitution and higher-order output (\cref{def:hopill-cong-5,def:hopill-cong-6,def:hopill-cong-7}) which are the only new term contrusctors with a process that is not their continuation.
\begin{theorem}[Congruence]
  \label{thm:hopill-congruence}
  On the \lts of processes, $\ssimby$ and $\simby$ are precongruences and $\sbis$ and $\bis$ are congruences.
\end{theorem}

Type checking and similarity checking for \hopill coincide.
\begin{theorem}
  \label{thm:pill-typing-iff-sim}
  For $P$ well-typed, $\Theta$, and $\hyp{G}$ such that $\fn(P) = \cn(\hyp{G})$, $\fpv(P) = \pv(\Theta)$, $P \simby \hoenv{\Theta}{\hyp{G}}$ iff $\judge{\Theta}{P}{\hyp{G}}$.
\end{theorem}

It follows that bisimilarity is sound with respect to type equivalence.
\begin{corollary}
For $P$ and $Q$ well-typed, if $P \bis Q$ and $\judge{\Theta}{Q}{\hyp{G}}$, then $\judge{\Theta}{P}{\hyp{G}}$.
\end{corollary}

\subsubsection{Metatheory of Parallelism}

Like \pill, \hopill enjoys the diamond property, serialisation, non-interference, and readiness.

\begin{theorem}[Diamond property]
  \label{thm:hopill-diamond}
  The \lts of derivations enjoys the diamond property.
\end{theorem}

\begin{theorem}[Serialisation]
  \label{thm:hopill-serialisation}
  The \lts of derivations enjoys serialisation.
\end{theorem}

\begin{theorem}[Non-Interference]
  \label{thm:hopill-non-interference}
  The \lts of derivations enjoys the non-interference.
\end{theorem}

If a \hopill process does not have free process variables, then it enjoys the same notion of readiness of \pill.

\begin{theorem}[First-Order Readiness]
\label{thm:hopill-readiness-fo}
Let $\judge{\emptyho}{P}{\Gamma_1 \pp \cdots \pp \Gamma_n}$. For every $i \in [1,n]$, there exist $x\in\cn(\Gamma_i)$ and $l$ such that $x \in \fn(l)$ and $P \slto{l}$.
\end{theorem}

If a \hopill process has free process variables, then it either enjoys the same notion of readiness of \pill or is ready to invoke a process variable.

\begin{theorem}[Higher-Order Readiness]
\label{thm:hopill-readiness-ho}
Let $\judge{\Theta}{P}{\Gamma_1 \pp \cdots \pp \Gamma_n}$. Either of the following statements hols.
\begin{itemize}
\item For every $i \in [1,n]$, there exist $x\in\cn(\Gamma_i)$ and $l$ such that $x \in \fn(l)$ and $P \slto{l}$.
\item There exist $\pvt{p}{\hyp{G}} \in \Theta$ and $l$ such that $l = \lsubst[\hyp{G}]{p}{\abstr{\sigma}{Q}}$ and $P \slto{l}$.
\end{itemize}
\end{theorem}

\subsubsection{Relation with \pill}

\pill can be regarded as the first-order fragment of \hopill (we simply have to require all process environments in the relevant typing rules to be empty). 

Is \hopill strictily more expressive than its first-order fragment? 
To answer this question we present a fully-abstract translation $\enc{-}$ of \hopill to \pill. 

The translation of session types acts as the identity in all cases except for provide and assume where we first neet to apply ``packing'' (\cf \cref{sec:mpill-metatheory}) as follows.
\[
\enc{\provide{\hyp{G}}} \defeq \enc{\bigtensor\hyp{G}}
\qquad
\enc{\assume{\hyp{G}}} \defeq \dual{\enc{\bigtensor\hyp{G}}}
\]
The translation of (hyper)environemnts is given by point-wise extension of the translation of session types.
\[
\enc{\Gamma_1 \pp \dots \pp \Gamma_n} \defeq \enc{\Gamma_1} \pp \dots \pp \enc{\Gamma_n}
\qquad
\enc{\cht{x_1}{A_1}, \dots, \cht{x_n}{A_n}} \defeq \cht{x_1}{\enc{A_1}}, \dots, \cht{x_n}{\enc{A_n}}
\]
The translation of processes requires packing which in turn depends on typing information. For simplicity, we define the translation of well-typed processes by recursion on their typing derivation. 
For conciseness, let $\hyp{G} = \Gamma_1 \pp \dots \pp \Gamma_n$  and assume$x^p_i$ fresh as needed.
\begin{alignat*}{1}
\enc*{\small\infer[\ref{rule:hopill-provide}]
  {\judge{\Theta}{\sendho{x}{\abstr{\rho}{P}}{Q}}{\cht{x}{\provide{\hyp{G}}}}}
  {\judge{\Theta}{P}{\hyp{G}\rho}
  &\judge{\emptyho}{Q}{\emptyhyp}}}
\defeq {} &
  \pack[\tensor]{x}{\hyp{G}\rho}[P_1,\dots,P_n] \pp \enc{\judge{\emptyho}{Q}{\emptyhyp}}
\\ & 
  \text{where }
  \disen{\enc{\judge{\Theta}{P}{\hyp{G}\rho}}} = P_1 \pp \dots \pp P_n
\\
\enc*{\small\infer[\ref{rule:hopill-assume}]
  {\judge{\Theta}{\recvho{x}{p}{P}}{\Gamma,\cht{x}{\assume{\hyp{G}}}}}
  {\judge{\Theta,\pvt{p}{\hyp{G}}}{P}{\Gamma}}}
\defeq {} &
  \recv{x}{x^p_1}{\cdots\recv{x}{x^p_n}{\wait{x}{\enc{\judge{\Theta,\pvt{p}{\hyp{G}}}{P}{\Gamma}}}}}
\\
\enc*{\small\infer[\ref{rule:hopill-id}]
  {\judge{\pvt{p}{\hyp{G}}}{\invoke{p}{\rho}}{\hyp{G}\rho}}
  {\vphantom{A}}} 
\defeq {} &
	\unpack{x^p_1}{\enc{\Gamma_1}}
   \pp \dots \pp
   \unpack{x^p_n}{\enc{\Gamma_n}}
\\&\text{where } \unpack{x}{\emptyseq} \defeq \close{x}{\nil}
\\&\text{and }
  \unpack{x}{\Gamma,\cht{y}{A}} \defeq \sendfree{x}{y}{\unpack{x}{\Gamma}}
\\
\enc*{\small\infer[\ref{rule:hopill-chop}]
  {\judge{\Theta,\Pi}{\chop{p}{\abstr{\rho}{P}}{Q}}{\hyp{H}}}
  {\judge{\Pi}{P}{\hyp{G}\rho}
  &\judge{\Theta,\pvt{p}{\hyp{G}}}{Q}{\hyp{H}}}} 
\defeq {} &
  \res{x^{p}_1y^{p}_1}{\dots\res{x^{p}_ny^{p}_n}{(
  Q_1 \pp \dots \pp Q_n \pp 
  \enc{\judge{\Theta,\pvt{p}{\hyp{G}}}{Q}{\hyp{H}}})}}
\\& \text{where }
Q_i = \pack[\parr]{x^p_i}{\bigparr\enc{\Gamma_i\rho}}[P_i]
\text{ and }
\\& 
\disen{\enc{\judge{\Pi}{P}{\hyp{G}\rho}}} = P_1 \pp \dots \pp P_n
\end{alignat*}
Remaning cases are translated homomorphically.

The ``unpacking'' processes $\unpack{x}{\Gamma}$ introduced above has type $\Gamma,\cht{x}{\dual{(\bigparr{\Gamma})}}$ and in can be regarded as symmetric to the packing of $\Gamma$ (henve the name) in the following sense.
\begin{lemma}
Let $\judge{\Theta}{P}{\Gamma}$.
\begin{itemize}
\item $\judge{\Theta}{\res{xy}{(\pack[\parr]{x}{\Gamma}[P] \pp \unpack{y}{\Gamma})}}{\Gamma}$ and 
\item $\res{xy}{(\pack[\parr]{x}{\Gamma}[P] \pp \unpack{y}{\Gamma}) \bis P}$.
\end{itemize}
\end{lemma}

If a processes has no free process symbols, say $\judge{\emptyho}{P}{\hyp{G}}$, then its translation $\enc{\judge{\emptyho}{P}{\hyp{G}}}$ has type $\enc{\hyp{G}}$ in \pill. A similar result holds for processes with free process variables, the only difference is the introduction of new channels $x^p_i$ for representing process variables. The distribution of these names in the type depends on the structure of the process. For instance, $\enc{{\judge{\pvt{p}{\hyp{G}}}{\invoke{p}{\rho}}{\hyp{G}\rho}}}$ has type $\enc{\Gamma_1},\cht{x_1^p}{\dual{(\bigparr\enc{\Gamma_1})}} \pp \dots \enc{\Gamma_n},\cht{x_n^p}{\dual{(\bigparr\enc{\Gamma_n})}}$ where $\hyp{G} = \Gamma_1 \pp \dots \pp \Gamma_n$.

The translation of \hopill to \pill is fully abstract in the sense it is invariant under bisimilarity.
\begin{theorem}[Full Abstraction]
For $\judge{\Theta}{P,Q}{\hyp{G}}$, $P \bis Q$ iff $\enc{\judge{\Theta}{P}{\hyp{G}}} \bis \enc{\judge{\Theta}{Q}{\hyp{G}}}$.
\end{theorem}

%% file: repill.tex
\section{Recursion}
\label{sec:repill}

We briefly discuss how recursion could be added to \pill and \hopill. For conciseness we use \hopill for our discussion.

\newcommand{\rec}[3]{{\textrm{rec}\, #1 \coloneqq #2 \,\textrm{in}\,} #3}

Finiteness of types in \hopill prevents well-typed processes from exhibiting infinite behaviours.
To extend \hopill with recursion and infinite (regular) behaviours we extend the type theory of \hopill with equi-recursive types (similarly for \pill).

We formalise recursive types as equi-recursive types instead of iso-recursive ones because the former allows us to introduce recursion in \pill changing neither the neither the syntax nor the semantics of the language (\eg iso-recursive types require explicit folding/unfolding of recursive types).

We write $\Sigma$ for a set of (recursive) type definitions of form $T \coloneqq E$ where $T$ is a type name and $E$ is a type expression obtained extending the grammar of \pill types with type names. We implicitly assume $\Sigma$ is well-formed in the sense that all type names in its expressions are defined and that every defined name has a unique definition. The dual $\dual T$ of a recursive type given by $T \coloneqq E$ is defined by the expression $\dual T \coloneqq \dual E$. We write $A =_\Sigma B$ to signify that types $A$ and $B$ are definitionally equal given the set of definitions $\Sigma$ ($\Sigma$-equivalent, for short). For instance, given $\Sigma = \{T \coloneqq 1 \otimes \dual{T}\}$ we have that $\dual{T} =_\Sigma \bot \parr T$ and $T =_\Sigma 1 \otimes (\bot \parr T)$.

Typing judgements in \pill with recursive types have form $\judge[\Sigma]{\Theta}{P}{\hyp{G}}$ and which reads ``process $P$ uses process variables according to $\Theta$ and channels according to $\hyp{G}$ given the recursive type definitions in $\Sigma$''. Typing rules are exactly those of \pill but in derivations $\Sigma$-equivalent types can be arbitrarily replaced.
For readability, we elicit the use of $\Sigma$-equivalence by means of \cref{rule:eqrec} below where an instance of type $B$ in the premise is replaced with one of type $A$ provided they are definitionally equivalent under $\Sigma$.
\begin{highlight}
\begin{spreadlines}{.5\bigskipamount}%
	\begin{gather*}
		\infer[\rlabel{\rname{$\Sigma$-Eq}}{rule:eqrec}]
	     {\judge[\Sigma]{\Theta}{P}{\hyp{G}[A]}}
	     {\judge[\Sigma]{\Theta}{P}{\hyp{G}[B]}
	     &\thl{A =_\Sigma B}}
	\end{gather*}
\end{spreadlines}
\end{highlight}%
For simplicity, we also adopt the following rule that allow us introduce local (\ie limited to the premises of the rule) recursive type definitions in $\Sigma$.
\begin{highlight}
\[
		\infer[\rlabel{\rname{$\Sigma$-S}}{rule:eqstrengthen}]
			{\judge[\Sigma]{\Theta}{P}{\hyp{G}}}
			{\judge[\Sigma,\Sigma']{\Theta}{P}{\hyp{G}}}
\]
\end{highlight}%

\paragraph{Recursive procedures}

\newcommand{\recproc}[3]{{\textsf{rec}\; #1 \coloneqq #2 \;\textsf{in}\;} #3}
\newcommand{\callrecproc}[2]{{#1 \langle #2 \rangle}}
 
Recursive types allows us to send over a channel abstractions that refer to a channel able to provide the same abstraction (or any abstraction of the same type). Here is a proof.
\begin{equation}
\label{eq:rec-output}
{\infer[\ref{rule:eqrec}]
	{\judge[T=\query\assume{\Gamma,\cht{K}{T}}]{\emptyho}{\server{x}{y}{\sendho{y}{\abstr \rho P}{\nil}}}{\cht{x}{\dual{T}}}}
	{\infer[\ref{rule:hopill-!}]
		{\judge[T=\query\assume{\Gamma,\cht{K}{T}}]{\emptyho}{\server{x}{y}{\sendho{y}{\abstr \rho P}{\nil}}}{\cht{x}{\bang\provide{\Gamma,\cht{K}{T}}}}}
			{\infer[\ref{rule:hopill-provide}]
				{\judge[T=\query\assume{\Gamma,\cht{K}{T}}]{\emptyho}{\sendho{y}{\abstr \rho P}{\nil}}{\cht{y}{\provide{\Gamma,\cht{K}{T}}}}}
				{\judge[T=\query\assume{\Gamma,\cht{K}{T}}]{\emptyho}{P}{\Gamma\rho,\cht{K}{T}}}
        &\infer[\ref{rule:hopill-mix-0}]
          {\judge[T=\query\assume{\Gamma,\cht{K}{T}}]{\emptyho}{\nil}{\emptyhyp}}
          {}}}}
\end{equation}
We use this property to extend the notion of procedures introduced in \cref{ex:procedures} to support recursion. 
\begin{bnftable}
	{\phl P, \phl Q \Coloneqq {}}
	{\recproc{K}{\abstr{\rho}{P}}{Q}}
		\phl P, \phl Q \Coloneqq {} & \nohl{\dots} & \\
		\mid {} & \rec{K}{\abstr{\rho}{P}}{Q} & recursive procedure definition \\
		\mid {} & \callrecproc{K}{\rho} & procedure invocation
\end{bnftable}
A term $\recproc{K}{\abstr{\rho}{P}}{Q}$ defines a recursive procedure $K$ as $\abstr{\rho}{P}$ in the scope of $Q$, and a term $\callrecproc{K}{\rho}$ invokes procedure $K$ by passing the parameters $\rho$.

Below are (derivable) rules for typing recursive procedure definitions and their invocations. 
As in the non-recursive case, in term $\proc{K}{\abstr{\rho}{P}}{Q}$ we assume that $K$ is used at least once in $Q$. 
The generalisation to the case where $Q$ does not use $K$ at all is straightforward (thanks to \cref{rule:hopill-weaken}).
\begin{spreadlines}{\typerulesskipamount}%
\begin{gather*}
	\infer
		{\judge[\Sigma]{\Theta}{\recproc{K}{\abstr{\rho}{P}}{Q}}{\hyp{G} \pp \Gamma}}
		{\judge[\Sigma,T=\query\assume{\Delta,\cht{K}{T}}]{\emptyho}{P}{\Delta\rho,\cht{K}{T}}
		&\judge[\Sigma,T=\query\assume{\Delta,\cht{K}{T}}]{\Theta}{Q}{\hyp{G} \pp \Gamma, \cht{K}{T}}}
\quad
	\infer
		{\judge[\Sigma,T=\query\assume{\Gamma,\cht{K}{T}}]{\emptyho}{\callrecproc{K}{\rho}}{\Gamma\rho,\cht{K}{T}}}
		{}
\end{gather*}
\end{spreadlines}
Below is the desugaring of $\recproc{K}{\abstr{\rho}{P}}{Q}$ as well as the proof that the corresponding typing rule is derivable ($\der{D}$ stands for the derivation in \eqref{eq:rec-output}).
\begin{gather*}
\infer[\ref{rule:eqstrengthen}]
	{\judge[\Sigma]{\Theta}{\res{xK}{\left(\server{x}{y}{\sendho{y}{\abstr{\rho}{P}}{\nil}} \pp Q \right)}}{\hyp{G} \pp \Gamma}}
	{\infer[\ref{rule:hopill-cut}]
		{\judge[\Sigma,T=\query\assume{\Delta,\cht{K}{T}}]{\Theta}{\res{xK}{\left(\server{x}{y}{\sendho{y}{\abstr{\rho}{P}}{\nil}} \pp Q \right)}}{\hyp{G} \pp \Gamma}}
		{\infer[\ref{rule:hopill-mix}]
			{\judge[\Sigma,T=\query\assume{\Delta,\cht{K}{T}}]{\Theta}{\server{x}{y}{\sendho{y}{\abstr \rho P}{\nil}} \pp Q}{\cht{x}{\dual{T}} \pp \hyp{G} \pp \Gamma,\cht{K}{T}}}
			{\deduce
				{\judge[\Sigma,T=\query\assume{\Delta,\cht{K}{T}}]{}{\server{x}{y}{\sendho{y}{\abstr \rho P}{\nil}}}{\cht{x}{\dual{T}}}}
				{\der{D}}
			&\judge[\Sigma,T=\query\assume{\Delta,\cht{K}{T}}]{\Theta}{Q}{\hyp{G} \pp \Gamma, \cht{K}{T}}}}}
\end{gather*}
Below is the desugaring of $\callrecproc{K}{\rho}$ as well as the proof that the corresponding typing rule is derivable.
\begin{gather*}
\infer[\ref{rule:eqrec}]
	{\judge[\Sigma,T=\query\assume{\Gamma,\cht{K}{T}}]{\emptyho}{\clientdup{K}{K}{K'}{ \clientuse{K'}{y}{\recvho{y}{p}{\invoke{p}\rho}}}}{\Gamma\rho,\cht{K}{T}}}
	{\infer[\ref{rule:hopill-contract}]
			{\judge[\Sigma,T=\query\assume{\Gamma,\cht{K}{T}}]{\emptyho}
				{\clientdup{K}{K}{K'}{ \clientuse{K'}{y}{\recvho{y}{p}{\invoke{p}\rho}}}}
				{\Gamma\rho,\cht{K}{\query\assume{\Gamma,\cht{K}{T}}}}}
			{\infer[\ref{rule:hopill-?}]
				{\judge[\Sigma,T=\query\assume{\Gamma,\cht{K}{T}}]{\emptyho}
								{\clientuse{K'}{y}{\recvho{y}{p}{\invoke{p}\rho}}}
								{\Gamma\rho,\cht{K}{\query\assume{\Gamma,\cht{K}{T}}},\cht{K'}{\query\assume{\Gamma,\cht{K}{T}}}}}
				{\infer[\ref{rule:hopill-assume}]
					{\judge[\Sigma,T=\query\assume{\Gamma,\cht{K}{T}}]{\emptyho}{\recvho{y}{p}{\invoke{p}{\rho}}}{\Gamma\rho,\cht{K}{\query\assume{\Gamma,\cht{K}{T}}},\cht{K}{\assume{\Gamma,\cht{K}{T}}}}}
					{\infer[\ref{rule:hopill-id}]
						{\judge[\Sigma,T=\query\assume{\Gamma,\cht{K}{T}}]{\pvt{p}{\Gamma,\cht{K}{T}}}{\invoke{p}{\rho}}{\Gamma\rho,\cht{K}{T}}}
						{}}}}}
\end{gather*}

\paragraph{Divergence}
Recursive types and \pill are enough to write divergent processes.
For instance, we can mimic the diverging $\lambda$-term $\Omega = (\lambda x.xx)(\lambda x.xx)$ using exponentials:
\begin{align*}
  \omega_{\query} \defeq
    \clientdup{x}{x_1}{x_2}{                   %
      \clientuse{x_1}{w_1}{                     %
      	\sendfree{w_1}{x_2}{\close{w_1}{\nil}}}}
  \qquad
  \omega_{\bang} \defeq
  	\server{y}{z}{
  			\recv{z}{x}{
  				\wait{z}{\omega_{\query}}
  			}
  		}
  \qquad
  \Omega \defeq {} & \res{xy}{(\omega_{\bang} \pp \omega_{\query})}
\end{align*}
Process $\Omega$ does not terminate since every sequence of transitions eventually recreates $\Omega$ as shown below (for each transition we indicate the relevant cut-elimination rule):
\begin{spreadlines}{-1pt}
\def\vlto{\left\downarrow\vbox to 1.6ex{\vfill\hbox to 2ex{\,$\tau$}\vfill}\right.}
\begin{alignat*}{2}
&\Omega & \\
&\vlto& \qquad \text{(\cref{rule:pill-cut-contract})} \\
&\res{x_1y_1}{(\send{y_1}{y_2}{(\omega_{\bang}\{y_1/y\}\pp\omega_{\bang}\{y_2/y\})} \pp \recv{x_1}{x_2}{\clientuse{x_1}{w_1}{\sendfree{w_1}{x_2}{\close{w_1}{\nil}}}})} & \\
&\vlto& \qquad \text{(\cref{rule:pill-cut-tensor})} \\
&\res{x_1y_1}{\res{x_2y_2}{(\omega_{\bang}\{y_1/y\}\pp\omega_{\bang}\{y_2/y\} \pp \clientuse{x_1}{w_1}{\sendfree{w_1}{x_2}{\close{w_1}{\nil}}})}} & \\
&\vlto& \qquad \text{(\cref{rule:pill-cut-?})} \\
&\res{z_1w_1}{\res{x_2y_2}{(\recv{z_1}{x}{\wait{z_1}{\omega_{\query}}}\pp\omega_{\bang}\{y_2/y\} \pp \sendfree{w_1}{x_2}{\close{w_1}{\nil}})}} & \\
&\vlto& \qquad \text{(\cref{rule:pill-cut-tensor})} \\
&\res{z_1w_1}{\res{w_2x}{\res{x_2y_2}{(\wait{z_1}{\omega_{\query}}\pp\omega_{\bang}\{y_2/y\} \pp \forward{w_2}{x_2} \pp \close{w_1}{\nil})}}} & \\
&\vlto& \qquad \text{(\cref{rule:pill-cut-one})} \\
&\res{w_2x}{\res{x_2y_2}{(\omega_{\query}\pp\omega_{\bang}\{y_2/y\} \pp \forward{w_2}{x_2})}} & \\
&\vlto& \qquad \text{(\cref{rule:pill-cut-axiom})} \\
&\res{x_2y_2}{(\omega_{\query}\{x_2/x\}\pp\omega_{\bang}\{y_2/y\})} \aleq \Omega &
\end{alignat*}
\end{spreadlines}
All other executions for $\Omega$ differ from the one above exclusively for the interleaving of transitions derived using \cref{rule:pill-cut-axiom,rule:pill-cut-one}.

Let $\Sigma = \{T = \bang(\bot \parr \dual{T})\}$ and $\mathcal{D}$ be the following derivation.
\[
  \infer[\ref{rule:eqrec}]
    {\judge[\Sigma]{}{\clientdup{x}{x_1}{x_2}{\clientuse{x_1}{w_1}{\send{w_1}{w_2}{(\forward{w_2}{x_2}\pp\close{w_1}{\nil})}}}}{\cht{x}{\dual{T}}}}
    {\infer[\ref{rule:pill-contract}]
      {\judge[\Sigma]{}{\clientdup{x}{x_1}{x_2}{\clientuse{x_1}{w_1}{\send{w_1}{w_2}{(\forward{w_2}{x_2}\pp\close{w_1}{\nil})}}}}{\cht{x}{\query(\one \tensor T)}}}
      {\infer[\ref{rule:eqrec}]
        {\judge[\Sigma]{}{\clientuse{x_1}{w_1}{\send{w_1}{w_2}{(\forward{w_2}{x_2}\pp\close{w_1}{\nil})}}}{\cht{x_1}{\query(\one \tensor T)},\cht{x_2}{\query(\one \tensor T)}}}
        {\infer[\ref{rule:pill-?}] {\judge[\Sigma]{}{\clientuse{x_1}{w_1}{\send{w_1}{w_2}{(\forward{w_2}{x_2}\pp\close{w_1}{\nil})}}}{\cht{x_1}{\query(\one \tensor T)},\cht{x_2}{\dual{T}}}}
          {\infer[\ref{rule:pill-tensor}] {\judge[\Sigma]{}{\send{w_1}{w_2}{(\forward{w_2}{x_2}\pp\close{w_1}{\nil})}}{\cht{w_1}{\one \tensor T},\cht{x_2}{\dual{T}}}}
            {\infer[\ref{rule:pill-mix}] {\judge[\Sigma]{}{\forward{w_2}{x_2}\pp\close{w_1}{\nil}}{\cht{w_1}{\one} \pp \cht{w_2}{T},\cht{x_2}{\dual{T}}}}
              {\infer[\ref{rule:pill-axiom}] {\judge[\Sigma]{}{\forward{w_2}{x_2}}{\cht{w_2}{T},\cht{x_2}{\dual{T}}}}
                {}
              &\infer[\ref{rule:pill-one}]
                {\judge[\Sigma]{}{\close{w_1}{\nil}}{\cht{w_1}{\one}}}
                {}}}}}}}
\]
Process $\Omega$ is well-typed.
\[
	{\infer[\ref{rule:pill-cut}]
    {\judge[\Sigma]{}{\res{xy}{(\server{y}{z}{\recv{z}{x}{\wait{z}{\omega_{\query}}}}\pp\omega_{\query})}}{\emptyseq}}
    {\infer[\ref{rule:pill-mix}]
      {\judge[\Sigma]{}{\server{y}{z}{\recv{z}{x}{\wait{z}{\omega_{\query}}}}\pp\omega_{\query}}{\cht{y}{T}\pp\cht{x}{\dual{T}}}}
      {\infer[\ref{rule:eqrec}]
        {\judge[\Sigma]{}{\server{y}{z}{\recv{z}{x}{\wait{z}{\omega_{\query}}}}}{\cht{y}{T}}}
        {\infer[\ref{rule:pill-!}]
          {\judge[\Sigma]{}{\server{y}{z}{\recv{z}{x}{\wait{z}{\omega_{\query}}}}}{\cht{y}{\bang(\bot \parr \dual{T})}}}
          {\infer[\ref{rule:pill-parr}]
            {\judge[\Sigma]{}{\recv{z}{x}{\wait{z}{\omega_{\query}}}}{\cht{z}{\bot \parr \dual{T}}}}
            {\infer[\ref{rule:pill-bot}]
              {\judge[\Sigma]{}{\wait{z}{\omega_{\query}}}{\cht{x}{\dual{T}},\cht{z}{\bot}}}
              {\infer
                {\judge[\Sigma]{}{\omega_{\query}}{\cht{x}{\dual{T}}}}
                {\mathcal{D}}}}}}
        &\infer
          {\judge[\Sigma]{}{\omega_{\query}}{\cht{x}{\dual{T}}}}
          {\mathcal{D}}}}}
\]
Observe that the empty sequent $\emptyseq$ is not inhabited by any process written in \pill without recursive types. Indeed, $\seq \emptyseq$ is not derivable in \CLL.

%% file: conclusions.tex
\section{Related work}
\label{sec:related}

Our work stands on previous efforts in the research line of Proofs as Processes. \citet{A94} and \citet{BS94} took the first steps in setting up this line, by connecting proofs in linear logic to processes in the $\pi$-calculus.
Even though the Proofs as Processes agenda has met many challenges along the way, the idea of adopting linearity in types for processes spawned interesting research lines, like the seminal theories of linear types for the $\pi$-calculus \cite{KPT99}
and session types \cite{HVK98}. Even though these theories do not use exactly linear logic, \citet{DGS17} showed that the underlying link is still strong enough that session types can be encoded into linear types.

\citet{CP10} established that propositions in Intuitionistic Linear Logic (ILL) can be interpreted as session types. Adopting ILL means having to distinguish between ``required'' and ``provided'' endpoints, depending on whether an endpoint appears on the left- or the right-hand side of a sequent. \citet{W14} removed this distinction by reformulating the correspondence with session types in Classical Linear Logic (\CLL), yielding a more standard presentation of session types.

The usage of hyperenvironments to capture the parallel operator of the $\pi$-calculus originates from \citet{CMS18}, who used the composition of hypersequents that share a name to model connected processes. Hypersequents were originally investigated by \citet{A91}.
Later, \citet{MP18} and \citet{KMP19} revisited the approach by \citet{CMS18} to formulate hyperenvironments and a labelled transition system.
The calculus by \citet{KMP19} is based on non-blocking I/O, which introduces equivalences that are not present in the canonical $\pi$-calculus, such as $\wait x{\wait yP} \sbis \wait y{\wait xP}$. Similar observations hold for previous calculi based on linear logic, including \citep{CP10} and \citep{W14}.
\pill dispenses with these additions for the sake of minimality, and in particular for showing that the expected observable behaviour of the $\pi$-calculus can be reconstructed in Proofs as Processes.
Furthermore, the non-blocking I/O in \citep{KMP19} breaks the property that separation of hyperenvironments guarantees independence (formally, for example, the diamond property does not hold).
If wanted, non-blocking I/O can be added to \pill by extending the logical typing rules to allow for additional environments in the hyperenvironment of interest, as in \citep{KMP19}.

Our transition rules for communications evoke cut reductions in \CLL: the way in which types are matched and deconstructed is similar.
The key difference is that we do not need to permute cuts in derivations (commuting conversions) until they reach the rule applications that formed the types being deconstructed.
This is because we can \emph{observe} the action corresponding to the deconstruction of a type within a subderivation from our transition labels, rather than having to inspect intensionally the structure of the derivation for the premises of our transition rules. Commuting conversions give rise to non-standard reductions in \citep{W14} as described above, so observing actions is an important component for achieving the standard process dynamics of the $\pi$-calculus.

To the best of our knowledge, \pill is the first calculus that comes with a logical reconstruction of session fidelity in terms of the expected transition systems, as in standard presentations of session types \citep{HYC16,KY14,MY13}. \citet{BSZ14} suggested a new behavioural relation for session types that recalls our behavioural characterisation of session fidelity, but in the case of \pill our characterisation uses standard similarity.

Our recipe for constructing \pill is inspired by dialgebraic interpretations of labelled transition systems (lts) \cite{CHM13}. Previous languages with transition systems rooted in linear logic do not feature polymorphism, nor higher-order communication \citep{MP18,KMP19}.

Other process calculi based on linear logic have been extended with primitives for moving processes by relying on a functional layer that can encapsulate process terms as values \citep{TCP13,TY18}. Instead, our \hopill offers the first logical reconstruction of (a linear variant of) \HOpi \cite{S93}, where mobile code is just processes, instead of functions (or values as intended in the $\lambda$-calculus). There are two key differences between \hopill and the calculi by \citet{TCP13} and \citet{TY18}. First, \hopill treats process variables linearly, following the intuition that it is a higher-order linear logic. This gives control on how process variables are used.
Second, thanks to our usage of hyperenvironments, the syntax and semantics of \hopill are nearer to those expected for the $\pi$-calculus: restriction and parallel are independent operators, rather than combined with other term constructors; and our extracted operational semantics is the expected one---the proof of subject reduction for the systems in \citet{TCP13,TY18} comes from \citet{CPT16}, which requires term rewritings in order to align the result of proof transformations with the expected result in the $\pi$-calculus.
Generally speaking, the ideas of process abstractions and functions are similar.
Our formulation based on process abstractions could be arguably seen as more direct, making the theory of \hopill simpler.
For example, we do not require the additional asymmetric connectives in session types used
in \citep{TCP13,TY18} for communicating processes ($\tau \supset A$ and $\tau \wedge A$).
The ``send a process and continue over channel $x$'' primitive found in \citep{TCP13,MY15,TY18} can be encoded in \hopill, as shown in \cref{ex:ho-continuations}.

The proof theory of the higher-order extension of \pill, \hopill, generalises linear logic by allowing to assume that some judgements can be proven, and to provide evidence for resolving these assumptions. Similar ideas have been used in the past in different contexts, for example for modal
logic \citep{NPP08} and logical frameworks \citep{BS15}.

Process mobility in the context of the $\pi$-calculus has been the subject of deep study, starting from the inception of the Higher-Order $\pi$-calculus (\HOpi) by \citet{S93}, who was also the first one to provide a fully-abstract encoding from \HOpi to the (first-order) $\pi$-calculus. \citet{LPSS08} showed that channel passing, restriction, and having continuations in the term for higher-order output are not necessary to achieve Turing-completeness in (untyped) \HOpi.
\citet{MY15} proposed a session-typed asynchronous \HOpi that is based on the original theory of session types \citep{HVK98}, rather than the Proofs as Processes correspondence. Their system does not support behavioural polymorphism like \hopill.
In this context, \citet{KPY16} discovered that name passing and recursion can be simulated by higher-order I/O in the session-typed \HOpi.
Their result is based on an encoding that requires \emph{open} abstractions (abstractions that include free names), whereas all abstractions in \hopill are closed. Hence the same technique does not apply to our calculus. Indeed, it would be surprising that adding closed abstractions as in \hopill would add behaviours that cannot be simulated with first-order features---the curious reader may consult other studies on the behavioural theory and expressivity of higher-order calculi, e.g., \citep{LPSS11,F13,F17}.
An interesting direction to generalise abstractions in \hopill might be to adopt a notion of sharing for linear logic, along the lines of the work by \citet{BP17}.

Our explicit substitutions are inspired by those originally developed to formalise
execution strategies for the $\lambda$-calculus that are more amenable to efficient
implementations \cite{ACCL91}.

Our treatment of recursion is different than in previous work.
\citet{LM16} extended \CP with (co)recursive types, following the formulation of fixed points in classical linear logic by \citet{B12}.
Our approach is more similar to the more recent works by \citet{BP17,BPT18,BTP19}, where typing judgements depend on a pre-populated signature $\Sigma$ of recursive procedure definitions.
In our case, $\Sigma$ needs to contain only (recursive) type definitions, since we can use higher-order primitives to capture procedures as derivable constructs.

\section{Conclusions}
\label{sec:conclusions}

Since its inception, linear logic has been described as the logic of concurrency \citep{G87}.
The Proofs as Processes agenda aims at grasping this potential for the study of concurrent processes.

In this article, we have presented a new calculus rooted in linear logic, \pill, along with a robust recipe for its construction and extension based on dialgebras. Thanks to our construction, for the first time, Proofs as Processes is in harmony with the expected metatheory of session types and the $\pi$-calculus, and we have gained substantial understanding on the implications of hyperenvironments about the behavioural theory of processes.
We have also extended the recent developments on hyperenvironments to deal with polymorphism, code mobility (higher-order communication), and recursion.
The higher-order variant of \pill, \hopill, extends linear logic to higher-order reasoning, viewing proofs as linear ``resources'' that can be used to assume premises in other proofs.

Our development provides guidelines for the future extension of the Proofs as Processes correspondence, recalling the Curry-Howard correspondence. We hope that \pill and its accompanying recipe can serve as an inspiration for the uniform integration of different results in the field of session types.

Interesting future directions include the integration of \hopill with multiparty session types---session types that consider more than two endpoints \citep{HYC16}---and choreographic programming---where choreographic descriptions of multiparty computations are translated to process implementations \citep{M13:phd}.
These topics have been studied in linear logic before \citep{CMSY17,CH15,CLMSW16,CP16,CMS18}, but we still lack (i) an understanding of higher-order communication rooted in linear logic for multiparty session types, and (ii) a choreographic programming model that supports code mobility at all.
\citet{MP17} augmented multiparty session types with higher-order I/O outside of the Proofs as Processes correspondence: a successful integration of \hopill with the results by \citet{CMSY17} might root this feature in linear logic for the first time.

Process mobility is the underlying concept behind the emerging interest on runtime adaptation, a mechanism by which processes can receive updates to their internal code at runtime.
Different attempts at formalising programming disciplines for runtime adaptations have been
made, e.g., by \citet{DP15,DGGLM16,BCDV17}, but none are rooted in a propositions as types correspondence and all offer different features and properties.
Leveraging \hopill to formulate runtime adaptation based on Proofs as Processes represent another interesting direction for future work.

The translation of \hopill into \pill shows how our higher-order processes that use mobile
code can be simulated by using reference passing instead.
This result is distinct from the original one by \citet{S93}, because we are operating in a typed 
setting.
Understanding what a calculus with behavioural types, such as session types, can express---i.e., 
what well-typed terms can model---is a nontrivial challenge in general \citep{P16}.
In practice, our translation has the usual implications: the translation gives us the possibility to write
programs that use code mobility and then choose later whether we should really use code mobility or 
translate it to an implementation basedon reference passing.
This choice depends on the application case. If we are modelling the transmission of an
application to be run somewhere else (as in cloud computing), then code mobility is necessary.
Otherwise, if we are in a situation where we can choose freely, then we should choose whichever 
implementation is more efficient. For example, code mobility is useful if two processes, say a 
client and a server, are operating on a slow connection; then, instead of performing many 
communications over the slow connection, the client may send an application to the server such that 
the server can communicate with
the application locally, and then send to the client only the final result.
Lastly, if we are using code mobility in an environment where communications are implemented in
local memory (as in many object-oriented language implementations or other emerging languages, like
Go), then the translation gives us a compilation technique towards a
simpler language without code mobility (\pill), which we can use to simplify runtime implementations.

%% file: hopill-full.tex
\section{\hopill, complete specification}
\label{sec:hopill-full}

\subsection{SOS of derivations}
\label{sec:hopill-full-sos}

Here we report the complete SOS specification for the \lts of derivations of \hopill.

\subsubsection{Actions}
\begin{drules}
\plto
  {\infer[\ref{rule:hopill-bot}]
      {\judge{\Theta}{\wait{x}{P}}{\Gamma, \cht{x}{\bot}}}
      {\deduce
        {\judge{\Theta}{P}{\Gamma}}
        {\der{D}}}}
  {\lwait{x}}
  {\deduce
    {\judge{\Theta}{P}{\Gamma}}
    {\der{D}}}
\\
\plto
  {\infer[\ref{rule:hopill-one}]
    {\judge{\Theta}{\close{x}{P}}{\cht{x}{\one}}}
    {\deduce
      {\judge{\Theta}{P}{\emptyhyp}}
      {\der{D}}}}
  {\lclose{x}}
  {\deduce
    {\judge{\Theta}{P}{\emptyhyp}}
    {\der{D}}}
\\
\plto
  {\infer[\ref{rule:hopill-tensor}]
    {\judge{\Theta}{\send{x}{x'}{P}}{\Gamma,\Delta,\cht{x}{A \tensor B}}}
    {\deduce
      {\judge{\Theta}{P}{\Gamma,\cht{x'}{A} \pp \Delta,\cht{x}{B}}}
      {\der{D}}}}
  {\lsend{x}{x'}}
  {\deduce
    {\judge{\Theta}{P}{\Gamma,\cht{x'}{A} \pp \Delta,\cht{x}{B}}}
    {\der{D}}}
\\
\plto
  {\infer[\ref{rule:hopill-parr}]
    {\judge{\Theta}{\recv{x}{x'}{P}}{\Gamma, \cht{x}{A \parr B}}}
    {\deduce
      {\judge{\Theta}{P}{\Gamma, \cht{x'}{A}, \cht{x}{B}}}
      {\der{D}}}}
  {\lrecv{x}{x'}}
  {\deduce
    {\judge{\Theta}{P}{\Gamma, \cht{x'}{A}, \cht{x}{B}}}
    {\der{D}}}
\\
\plto
  {\infer[\ref{rule:hopill-oplus_1}]
    {\judge{\Theta}{\inl{x}{P}}{\Gamma, \cht{x}{A \oplus B}}}
    {\deduce
      {\judge{\Theta}{P}{\Gamma, \cht{x}{A}}}
      {\der{D}}}}
  {\linl{x}}
  {\deduce
    {\judge{\Theta}{P}{\Gamma, \cht{x}{A}}}
    {\der{D}}}
\\
\plto
  {\infer[\ref{rule:hopill-oplus_2}]
    {\judge{\Theta}{\inr{x}{P}}{\Gamma, \cht{x}{A \oplus B}}}
    {\deduce
      {\judge{\Theta}{P}{\Gamma, \cht{x}{B}}}
      {\der{D}}}}
  {\linr{x}}
  {\deduce
    {\judge{\Theta}{P}{\Gamma, \cht{x}{B}}}
    {\der{D}}}
\\
\plto
  {\infer[\ref{rule:hopill-with}]
    {\judge{\Theta}{\choice{x}{P}{Q}}{\Gamma, \cht{x}{A \with B}}}
    {\deduce
      {\judge{\Theta}{P}{\Gamma, \cht{x}{A}}}
      {\der{D}}
    &\deduce
      {\judge{\Theta}{Q}{\Gamma, \cht{x}{B}}}
      {\der{E}}}}
  {\lcoinl{x}}
  {\deduce
    {\judge{\Theta}{P}{\Gamma, \cht{x}{A}}}
    {\der{D}}}
\\
\plto
  {\infer[\ref{rule:hopill-with}]
    {\judge{\Theta}{\choice{x}{P}{Q}}{\Gamma, \cht{x}{A \with B}}}
    {\deduce
      {\judge{\Theta}{P}{\Gamma, \cht{x}{A}}}
      {\der{D}}
    &\deduce
      {\judge{\Theta}{Q}{\Gamma, \cht{x}{B}}}
      {\der{E}}}}
  {\lcoinr{x}}
  {\deduce
    {\judge{\Theta}{Q}{\Gamma, \cht{x}{B}}}
    {\der{E}}}
\\
\plto
  {\infer[\ref{rule:hopill-exists}]
     {\judge{\Theta}{\sendtype{x}{A}{P}}{\Gamma,\cht{x}{\exists X.B}}}
     {\deduce
      {\judge{\Theta}{P}{\Gamma,\cht{x}{B\{A/X\}}}}
      {\der{D}}}}
  {\lsendtype{x}{A}}
  {\deduce
    {\judge{\Theta}{P}{\Gamma,\cht{x}{B\{A/X\}}}}
    {\der{D}}}
\\
\plto
  {\infer[\ref{rule:hopill-forall}]
     {\judge{\Theta}{\recvtype{x}{X}{P}}{\Gamma,\cht{x}{\forall X.B}}}
     {\deduce
      {\judge{\Theta}{P}{\Gamma,\cht{x}{B}}}
      {\der{D}}}}
  {\lrecvtype{x}{A}}
  {\deduce
    {\judge{\Theta}{P\{A/X\}}{\Gamma,\cht{x}{B\{A/X\}}}}
    {\der{D}\{A/X\}}}
\\
\plto
  {\infer[\ref{rule:hopill-?}]
    {\judge{\Theta}{\clientuse{x}{x'}{P}}{\Gamma, \cht{x}{\query A}}}
    {\deduce
      {\judge{\Theta}{P}{\Gamma, \cht{x'}{A}}}
      {\der{D}}}}
  {\luse{x}{x'}}
  {\deduce
    {\judge{\Theta}{P}{\Gamma, \cht{x'}{A}}}
    {\der{D}}}
\\
\plto
  {\infer[\ref{rule:hopill-!}]
    {\judge{\emptyho}{\server{x}{x'}{P}}{\query\Gamma, \cht{x}{\bang A}}}
    {\deduce
      {\judge{\emptyho}{P}{\query\Gamma, \cht{x'}{A}}}
      {\der{D}}}}
  {\lcouse{x}{x'}}
  {\deduce
    {\judge{\emptyho}{P}{\query\Gamma, \cht{x'}{A}}}
    {\der{D}}}
\\
{\plto
  {\infer[\ref{rule:hopill-contract}]
    {\judge{\Theta}{\clientdup{x}{x_1}{x_2}{P}}{\Gamma, \cht{x}{\query A}}}
    {\deduce
      {\judge{\Theta}{P}{\Gamma, \cht{x_1}{\query A},\cht{x_2}{\query A}}}
      {\der{D}}}}
  {\ldup{x}{x_1}{x_2}}
  {\infer[\ref{rule:hopill-parr}]
    {\judge{\Theta}{\recv{x_1}{x_2}{P}}{\Gamma, \cht{x_1}{\query A \parr \query A}}}
    {\deduce
      {\judge{\Theta}{P}{\Gamma, \cht{x_1}{\query A},\cht{x_2}{\query A}}}
      {\der{D}}}}}
\\
\infer
  {\plto*
    {\infer[\ref{rule:hopill-!}]
      {\judge{\emptyho}{\server{x}{x'}{P}}{\query\Gamma, \cht{x}{\bang A}}}
      {\deduce
        {\judge{\emptyho}{P}{\bang\Gamma, \cht{x'}{A}}}
        {\der{D}}}}
    {\lcodup{x}{x_1}{x_2}}
    {\infer=[\ref{rule:hopill-contract}]
      {\judge{\emptyho}
        {\clientdup{z_1}{z_1\sigma_1}{z_1\sigma_2}{\dots\clientdup{z_n}{z_n\sigma_1}{z_n\sigma_2}{\send{x_1}{x_2}{(\server{x_1}{x'\sigma_1}{P_1} \pp \server{x_2}{x'\sigma_2}{P_2})}}}}
        {\query\Gamma, \cht{x_1}{\bang A \tensor \bang A}}}
      {\infer[\ref{rule:hopill-tensor}]
        {\judge{\emptyho}{\send{x_1}{x_2}{(\server{x_1}{x'\sigma_1}{P} \pp \server{x_2}{x'\sigma_2}{P})}}{\query\Gamma_1,\query\Gamma_2, \cht{x_1}{\bang A \tensor \bang A}}}
        {\infer[\ref{rule:hopill-mix}]
          {\judge{\emptyho}
            {\server{x_1}{x'\sigma_1}{P_1} \pp \server{x_2}{x'\sigma_2}{P_2}}
            {\query\Gamma_1,\cht{x_1}{\bang A} \pp \query\Gamma_2,\cht{x_2}{\bang A}}}
          {\infer[\ref{rule:hopill-!}]
            {\judge{\emptyho}{\server{x_1}{x'\sigma_1}{P_1}}{\query\Gamma_1, \cht{x_1}{\bang A}}}
            {\deduce
              {\judge{\emptyho}{P_1}{\query\Gamma_1, \cht{x'\sigma_1}{A}}}
              {\der{D}\sigma_1}}
          &\infer[\ref{rule:hopill-!}]
            {\judge{\emptyho}{\server{x_2}{x'\sigma_2}{P_2}}{\query\Gamma_2, \cht{x_2}{\bang A}}}
            {\deduce
              {\judge{\emptyho}{P_2}{\query\Gamma_2, \cht{x'\sigma_2}{A}}}
              {\der{D}\sigma_2}}}}}}}
  {\phl{P_i = P\sigma_i}
   &\thl{\Gamma_i = \Gamma\sigma_i}
   &\phl{\text{for } i \in \{1,2\}}
   &\phl{\fn(P_1) \cap \fn(P_2) = \emptyset}
   &\phl{\fn(P)\setminus\{x'\} = \{z_1,\dots,z_n\}}}
\\
\plto
  {\infer[\ref{rule:hopill-weaken}]
    {\judge{\Theta}{\clientdisp{x}{P}}{\Gamma, \cht{x}{\query A}}}
    {\deduce
      {\judge{\Theta}{P}{\Gamma}}
      {\der{D}}}}
  {\plbl{\ldisp{x}}{\lweaken{A}}}
  {\infer[\ref{rule:hopill-bot}]
    {\judge{\Theta}{\wait{x}{P}}{\Gamma, \cht{x}{\bot}}}
    {\deduce
      {\judge{\Theta}{P}{\Gamma}}
      {\der{D}}}}
\\
\infer
  {\plto
    {\infer[\ref{rule:hopill-!}]
      {\judge{\emptyho}{\server{x}{x'}{P}}{\query \Gamma, \cht{x}{\bang A}}}
      {\deduce
        {\judge{\emptyho}{P}{\query \Gamma, \cht{x'}{A}}}
        {\der{D}}}}
    {\plbl{\lcodisp{x}}{\lcoweaken{A}}}
    {\infer=[\ref{rule:hopill-weaken}]
      {\judge{\emptyho}
        {\clientdisp{z_1}{\dots\clientdisp{z_n}{\close{x}{\nil}}}}
        {\query \Gamma, \cht{x}{\one}}}
      {\infer[\ref{rule:hopill-one}]
        {\judge{\emptyho}{\close{x}{\nil}}{\cht{x}{\one}}}
        {}}}}
  {\phl{\fn(P)\setminus\{x'\} = \{z_1,\dots,z_n\}}}
\\
\plto
	{\infer[\ref{rule:hopill-provide}]
        {\judge{\Theta,\Pi}{\sendho{x}{\abstr{\rho}{P}}{Q}}{\cht{x}{\provide{\hyp{G}}}}}
        {\deduce
          {\judge{\Theta}{P}{\hyp{G}\rho}}
          {\der{D}}
        &\deduce
          {\judge{\Pi}{Q}{\emptyhyp}}
          {\der{E}}}}
	{\lsendho{x}{\abstr{\rho}{\der{D}}}}
  {\deduce
    {\judge{\Pi}{Q}{\emptyhyp}}
    {\der{E}}}
\\
\plto
	{\infer[\ref{rule:hopill-assume}]
    {\judge{\Theta}{\recvho{x}{p}{P}}{\Gamma,\cht{x}{\assume{\hyp{G}}}}}
    {\deduce
      {\judge{\Theta,\pvt{p}{\hyp{G}}}{P}{\Gamma}}
      {\der{D}}}}
	{\lrecvho{x}{p}}
  {\deduce
    {\judge{\Theta,\pvt{p}{\hyp{G}}}{P}{\Gamma}}
    {\der{D}}}
\end{drules}

\subsubsection{Structural}
\begin{drules}
\plto
    {\infer[\ref{rule:hopill-axiom}]
      {\judge{\emptyho}{\forward{x}{y}}{\cht{x}{\dual{A}}, \cht{y}{A}}}
      {\phantom{\pp}}}
    {\lforward{x}{y}}
     {\infer[\ref{rule:hopill-mix-0}]
      {\judge{\emptyho}{\nil}{\emptyhyp}}
      {\vphantom{\pp}}}
\\
\plto
   {\infer[\ref{rule:hopill-axiom}]
     {\judge{\emptyho}{\forward{x}{y}}{\cht{x}{\dual{A}}, \cht{y}{A}}}
     {\phantom{\pp}}}
   {\lforward{y}{x}}
   {\infer[\ref{rule:hopill-mix-0}]
    {\judge{\emptyho}{\nil}{\emptyhyp}}
    {\vphantom{\pp}}}
\\
\def\regh{\vphantom{\hyp{G}'\pp}}
\infer[\rlabel{\ref*{rule:mpill-mix-1}}{rule:hopill-mix-1}]
  {\plto
    {\infer[\ref{rule:hopill-mix}]
      {\judge{\Theta,\Pi}{P \pp Q}{\hyp{G} \pp \hyp{H}}\regh}
      {\deduce
        {\judge{\Theta}{P}{\hyp{G}}\regh}
        {\der{D}}
      &\deduce
        {\judge{\Pi}{Q}{\hyp{H}}\regh}
        {\der{E}}}}
    {l}
    {\infer[\ref{rule:hopill-mix}]
      {\judge{\Theta',\Pi}{P' \pp Q}{\hyp{G}' \pp \hyp{H}}\regh}
      {\deduce
        {\judge{\Theta'}{P'}{\hyp{G}'}\regh}
        {\der{D}'}
      &\deduce
        {\judge{\Pi}{Q}{\hyp{H}}\regh}
        {\der{E}}}}}
  {\plto
    {\deduce
      {\judge{\Theta}{P}{\hyp{G}}\regh}
      {\der{D}}}
    {l}
    {\deduce
      {\judge{\Theta'}{P'}{\hyp{G}'}\regh}
      {\der{D}'}}
  &\phl{\bn(l) \cap \fn(Q) = \emptyset}}
\\
\def\regh{\vphantom{\pp\hyp{H}'}}
\infer[\rlabel{\ref*{rule:mpill-mix-2}}{rule:hopill-mix-2}]
  {\plto
    {\infer[\ref{rule:hopill-mix}]
      {\judge{\Theta,\Pi}{P \pp Q}{\hyp{G} \pp \hyp{H}}\regh}
      {\deduce
        {\judge{\Theta}{P}{\hyp{G}}\regh}
        {\der{D}}
      &\deduce
        {\judge{\Pi}{Q}{\hyp{H}}\regh}
        {\der{E}}}}
    {l}
    {\infer[\ref{rule:hopill-mix}]
      {\judge{\Theta,\Pi'}{P \pp Q'}{\hyp{G} \pp \hyp{H}'}\regh}
      {\deduce
        {\judge{\Theta}{P}{\hyp{G}}\regh}
        {\der{D}}
      &\deduce
        {\judge{\Pi'}{Q'}{\hyp{H}'}\regh}
        {\der{E}'}}}}
  {\plto
    {\deduce
      {\judge{\Pi}{Q}{\hyp{H}}\regh}
      {\der{E}}}
    {l}
    {\deduce
      {\judge{\Pi'}{Q'}{\hyp{H}'}\regh}
      {\der{E}'}}
  &\phl{\bn(l) \cap \fn(P) = \emptyset}}
\\
\def\regh{\vphantom{\hyp{G}'\pp\hyp{H}'}}
\infer[\rlabel{\ref*{rule:mpill-mix-sync}}{rule:hopill-mix-sync}]
  {\plto
    {\infer[\ref{rule:hopill-mix}]
      {\judge{\Theta,\Pi}{P \pp Q}{\hyp{G} \pp \hyp{H}}\regh}
      {\deduce
        {\judge{\Theta}{P}{\hyp{G}}\regh}
        {\der{D}}
      &\deduce
        {\judge{\Pi}{Q}{\hyp{H}}\regh}
        {\der{E}}}}
    {\lsync{l}{l'}}
    {\infer[\ref{rule:hopill-mix}]
      {\judge{\Theta',\Pi'}{P' \pp Q'}{\hyp{G}' \pp \hyp{H}'}\regh}
      {\deduce
        {\judge{\Theta'}{P'}{\hyp{G}'}\regh}
        {\der{D}'}
      &\deduce
        {\judge{\Pi'}{Q'}{\hyp{H}'}\regh}
        {\der{E}'}}}}
  {\plto
    {\deduce
      {\judge{\Theta}{P}{\hyp{G}}\regh}
      {\der{D}}}
    {l}
    {\deduce
      {\judge{\Theta'}{P'}{\hyp{G}'}\regh}
      {\der{D}'}}
  &\plto
    {\deduce
      {\judge{\Pi}{Q}{\hyp{H}}\regh}
      {\der{E}}}
    {l}
    {\deduce
      {\judge{\Pi'}{Q'}{\hyp{H}'}\regh}
      {\der{E}'}}
  & \phl{\bn(l) \cap \bn(l') = \emptyset}}
\\
\infer[\rlabel{\ref*{rule:mpill-cut-res}}{rule:hopill-cut-res}]
  {\plto
    {\infer[\ref{rule:hopill-cut}]
      {\judge{\Theta}{\res{xy}{P}}{\hyp{G} \pp \Gamma,\Delta}}
      {\deduce
        {\judge{\Theta}{P}{\hyp{G} \pp \Gamma,\cht{x}{A} \pp \Delta,\cht{y}{\dual{A}}}}
        {\der{D}}}}
    {l}
    {\infer[\ref{rule:hopill-cut}]
      {\judge{\Theta'}{\res{xy}{P'}}{\hyp{G}' \pp \Gamma',\Delta'}}
      {\deduce
        {\judge{\Theta'}{P'}{\hyp{G}' \pp \Gamma',\cht{x}{A} \pp \Delta',\cht{y}{\dual{A}}}}
        {\der{D'}}}}}
  {\plto
    {\deduce
      {\judge{\Theta}{P}{\hyp{G} \pp \Gamma,\cht{x}{A} \pp \Delta,\cht{y}{\dual{A}}}}
      {\der{D}}}
    {l}
    {\deduce
      {\judge{\Theta'}{P'}{\hyp{G}' \pp \Gamma',\cht{x}{A} \pp \Delta',\cht{y}{\dual{A}}}}
      {\der{D'}}}
  &\phl{x,y \notin \cn(l)}}
\\
\infer[\rlabel{\ref*{rule:mpill-alpha}}{rule:hopill-alpha}]
  {\plto
    {\deduce
      {\judge{\Theta}{P}{\hyp{G}}}
      {\der{D}}}
    {l}
    {\deduce
      {\judge{\Theta'}{Q'}{\hyp{G'}}}
      {\der{E'}}}}
  {\phl{P \aleq Q}
  &\plto
    {\deduce
      {\judge{\Theta}{Q}{\hyp{G}}}
      {\der{E}}}
    {l}
    {\deduce
      {\judge{\Theta'}{Q'}{\hyp{G'}}}
      {\der{E'}}}}
\\
\infer[\rlabel{\ref*{rule:pill-cut-axiom}}{rule:hopill-cut-axiom}]
  {\plto
    {\infer[\ref{rule:hopill-cut}]
     {\judge{\Theta}{\res{yz}{P}}{\hyp{G} \pp \Gamma, \cht{x}{\dual{A}}}}
     {\deduce
       {\judge{\Theta}{P}{\hyp{G}\pp \cht{x}{\dual{A}},\cht{y}{A} \pp \Gamma, \cht{z}{\dual{A}}}}
       {\der{D}}}}
    {\tau}
    {\deduce
      {\judge{\Theta}{P'}{\hyp{G} \pp \Gamma,\cht{z}{\dual{A}}}}
      {\der{D'}}}
  }{\plto
    {\deduce
      {\judge{\Theta}{P}{\hyp{G}\pp \cht{x}{\dual{A}},\cht{y}{A} \pp \Gamma, \cht{z}{\dual{A}}}}
      {\der{D}}}
    {\lforward{x}{y}}
    {\deduce
      {\judge{\Theta}{P'}{\hyp{G} \pp \Gamma,\cht{z}{\dual{A}}}}
      {\der{D'}}}}
\\
\plto
  {\infer[\ref{rule:hopill-id}]
    {\judge{\pvt{p}{\hyp{G}}}{\invoke[\hyp{G}]{p}{\sigma}}{\hyp{G}\sigma}}
    {\vphantom{\hyp{G}}}}
  {\lsubst[\hyp{G}]{p}{\abstr{\rho}{\der{D}}}}
  {\deduce
    {\judge{\Pi}{P\{\sigma\circ\rho^{-1}\}}{\hyp{G}\sigma}}
    {\der{D}}}
\\
\infer[\ref*{rule:hopill-chop-id}]
  {\plto
    {\infer[\ref{rule:hopill-chop}]
      {\judge{\Theta,\Pi}{\chop{p}{\abstr \rho P}{Q}}{\hyp{G}}}
      {\deduce
        {\judge{\Pi}{P}{\hyp{H}\rho}}
        {\der{E}}
      &\deduce
        {\judge{\Theta,\pvt{p}{\hyp{H}}}{Q}{\hyp{G}}}
        {\der{D}}}}
    {\tau}
    {\deduce
      {\judge{\Theta,\Pi}{Q'}{\hyp{G}}}
      {\der{D}'}}}
  {\plto
    {\deduce
      {\judge{\Theta,\pvt{p}{\hyp{H}}}{Q}{\hyp{G}}}
      {\der{D}}}
    {\lsubst[\hyp{H}]{p}{\abstr{\rho}{\der{E}}}}
    {\deduce
      {\judge{\Theta,\Pi}{Q'}{\hyp{G}}}
      {\der{D}'}}}
\\
\infer[\ref*{rule:hopill-chop-provide}]
  {\plto
    {\infer[\ref{rule:hopill-chop}]
      {\judge{\Theta,\Pi}{\chop{p}{\abstr \rho P}{Q}}{\hyp{G}}}
      {\deduce
        {\judge{\Pi}{P}{\hyp{H}\rho}}
        {\der{E}}
      &\deduce
        {\judge{\Theta,\pvt{p}{\hyp{H}}}{Q}{\hyp{G}}}
        {\der{D}}}}
    {\lsendho{x}{\abstr{\sigma}{\der{F'}}}}
    {\deduce
      {\judge{\Theta'}{Q'}{\hyp{G'}}}
      {\der{D'}}}}
  {\plto
    {\deduce
      {\judge{\Theta,\pvt{p}{\hyp{H}}}{Q}{\hyp{G}}}
      {\der{D}}}
    {\lsendho{x}{\abstr{\sigma}{\der{F}}}}
    {\deduce
      {\judge{\Theta'}{Q'}{\hyp{G'}}}
      {\der{D'}}}
  &\der{F'} = \pbox{
  {\infer[\ref{rule:hopill-chop}]
    {\judge{\Xi,\Pi}{\chop{p}{\abstr \rho P}{R}}{\hyp{I}}}
    {\deduce
      {\judge{\Pi}{P}{\hyp{H}\rho}}
      {\der{E}}
    &\deduce
      {\judge{\Xi,\pvt{p}{\hyp{H}}}{R}{\hyp{I}}}
      {\der{F}}}}}
  &\phl{p \in \fpv(R)}}
\\
\infer[\ref*{rule:hopill-chop-res}]
  {\plto
    {\infer[\ref{rule:hopill-chop}]
      {\judge{\Theta,\Pi}{\chop{p}{\abstr \rho P}{Q}}{\hyp{G}}}
      {\deduce
        {\judge{\Pi}{P}{\hyp{H}\rho}}
        {\der{E}}
      &\deduce
        {\judge{\Theta,\pvt{p}{\hyp{H}}}{Q}{\hyp{G}}}
        {\der{D}}}}
    {l}
    {\infer[\ref{rule:hopill-chop}]
      {\judge{\Theta',\Pi}{\chop{p}{\abstr \rho P}{Q'}}{\hyp{G'}}}
      {\deduce
        {\judge{\Pi}{P}{\hyp{H}\rho}}
        {\der{E}}
      &\deduce
        {\judge{\Theta',\pvt{p}{\hyp{H}}}{Q'}{\hyp{G'}}}
        {\der{D'}}}}}
  {\plto
    {\deduce
      {\judge{\Theta,\pvt{p}{\hyp{H}}}{Q}{\hyp{G}}}
      {\der{D}}}
    {l}
    {\deduce
      {\judge{\Theta',\pvt{p}{\hyp{H}}}{Q'}{\hyp{G'}}}
      {\der{D'}}}
    &\phl{p \notin \fpv(l)}}
\end{drules}

\subsubsection{Communication}
\begin{drules}
\infer[\rlabel{\ref*{rule:mpill-cut-one}}{rule:hopill-cut-one}]
  {\plto
    {\infer[\ref{rule:hopill-cut}]
      {\judge{\Theta}{\res{xy}{P}}{\hyp{G} \pp \Gamma}}
      {\deduce
        {\judge{\Theta}{P}{\hyp{G}\pp \cht{x}{\one} \pp \Gamma, \cht{y}{\bot}}}
        {\der{D}}}}
    {\tau}
    {\deduce
      {\judge{\Theta}{P'}{\hyp{G}\pp \Gamma}}
      {\der{D'}}}}
  {\plto
    {\deduce
      {\judge{\Theta}{P}{\hyp{G}\pp \cht{x}{\one} \pp \Gamma, \cht{y}{\bot}}}
      {\der{D}}}
    {\lsync
      {\lclose{x}}
      {\lwait{y}}}
    {\deduce
      {\judge{\Theta}{P'}{\hyp{G}\pp \Gamma}}
      {\der{D'}}}}
\\
\infer[\rlabel{\ref*{rule:mpill-cut-tensor}}{rule:hopill-cut-tensor}]
  {\plto
    {\infer[\ref{rule:hopill-cut}]
      {\judge{\Theta}{\res{xy}{P}}{\hyp{G} \pp \Gamma,\Delta, \Xi}}
      {\deduce
        {\judge{\Theta}{P}{\hyp{G}\pp\Gamma,\Delta, \cht{x}{A \tensor B} \pp \Xi, \cht{y}{\dual{A} \parr \dual{B}}}}
        {\der{D}}}}
    {\tau}
    {\infer[\ref{rule:hopill-cut}]
      {\judge{\Theta}{\res{xy}{\res{x'y'}{P'}}}{\hyp{G} \pp \Gamma, \Delta, \Xi}}
      {\infer[\ref{rule:hopill-cut}]
        {\judge{\Theta}{\res{x'y'}{P'}}{\hyp{G} \pp \Gamma,\cht{x}{B} \pp \Delta, \Xi,\cht{y}{\dual{B}}}}
        {\deduce
          {\judge{\Theta}{P'}{\hyp{G} \pp \Gamma, \cht{x}{B} \pp \Delta,\cht{x'}{A} \pp \Xi, \cht{y}{\dual{B}}, \cht{y'}{\dual{A}}}}
          {\der{D}'}}}}}
  {\plto
    {\deduce
      {\judge{\Theta}{P}{\hyp{G}\pp\Gamma,\Delta, \cht{x}{A \tensor B} \pp \Xi, \cht{y}{\dual{A} \parr \dual{B}}}}
      {\der{D}}}
    {\lsync
      {\lsend{x}{x'}}
      {\lrecv{y}{y'}}}
    {\deduce
      {\judge{\Theta}{P'}{\hyp{G} \pp \Gamma, \cht{x}{B} \pp \Delta,\cht{x'}{A} \pp \Xi, \cht{y}{\dual{B}}, \cht{y'}{\dual{A}}}}
      {\der{D}'}}}
\\
\infer[\rlabel{\ref*{rule:pill-cut-oplus_1}}{rule:hopill-cut-oplus_1}]
  {\plto
    {\infer[\ref{rule:hopill-cut}]
     {\judge{\Theta}{\res{xy}{P}}{\hyp{G} \pp \Gamma, \Delta}}
     {\deduce
      {\judge{\Theta}{P}{\hyp{G}\pp\Gamma,\cht{x}{A \oplus B} \pp \Delta, \cht{y}{\dual{A} \parr \dual{B}}}}
      {\der{D}}}}
    {\tau}
    {\infer[\ref{rule:hopill-cut}]
     {\judge{\Theta}{\res{xy}{P'}}{\hyp{G} \pp \Gamma, \Delta}}
     {\deduce
       {\judge{\Theta}{P'}{\hyp{G}\pp\Gamma,\cht{x}{A} \pp \Delta, \cht{y}{\dual{A}}}}
       {\der{D'}}}}}
  {\plto
    {\deduce
      {\judge{\Theta}{P}{\hyp{G}\pp\Gamma,\cht{x}{A \oplus B} \pp \Delta, \cht{y}{\dual{A} \parr \dual{B}}}}
      {\der{D}}}
    {\lsync
      {\linl{x}}
      {\lcoinl{y}}}
    {\deduce
     {\judge{\Theta}{P'}{\hyp{G}\pp\Gamma,\cht{x}{A} \pp \Delta, \cht{y}{\dual{A}}}}
     {\der{D'}}}}
\\
\infer[\rlabel{\ref*{rule:pill-cut-oplus_2}}{rule:hopill-cut-oplus_2}]
  {\plto
    {\infer[\ref{rule:hopill-cut}]
     {\judge{\Theta}{\res{xy}{P}}{\hyp{G} \pp \Gamma, \Delta}}
     {\deduce
      {\judge{\Theta}{P}{\hyp{G}\pp\Gamma,\cht{x}{A \oplus B} \pp \Delta, \cht{y}{\dual{A} \parr \dual{B}}}}
      {\der{D}}}}
    {\tau}
    {\infer[\ref{rule:hopill-cut}]
     {\judge{\Theta}{\res{xy}{P'}}{\hyp{G} \pp \Gamma, \Delta}}
     {\deduce
       {\judge{\Theta}{P'}{\hyp{G}\pp\Gamma,\cht{x}{B} \pp \Delta, \cht{y}{\dual{B}}}}
       {\der{D'}}}}}
  {\plto
    {\deduce
      {\judge{\Theta}{P}{\hyp{G}\pp\Gamma,\cht{x}{A \oplus B} \pp \Delta, \cht{y}{\dual{A} \parr \dual{B}}}}
      {\der{D}}}
    {\lsync
      {\linr{x}}
      {\lcoinr{y}}}
    {\deduce
     {\judge{\Theta}{P'}{\hyp{G}\pp\Gamma,\cht{x}{B} \pp \Delta, \cht{y}{\dual{B}}}}
     {\der{D'}}}}
\\
\infer[\rlabel{\ref*{rule:pill-cut-exists}}{rule:hopill-cut-exists}]
  {\plto
    {\infer[\ref{rule:hopill-cut}]
      {\judge{\Theta}{\res{xy}{P}}{\hyp{G} \pp \Gamma,\Delta}}
      {\deduce
        {\judge{\Theta}{P}{\hyp{G}\pp \Gamma,\cht{x}{\exists X.B} \pp \Delta, \cht{y}{\forall X.\dual{B}}}}
        {\der{D}}}}
    {\tau}
    {\infer[\ref{rule:hopill-cut}]
      {\judge{\Theta}{\res{xy}{P}}{\hyp{G} \pp \Gamma,\Delta}}
      {\deduce
        {\judge{\Theta}{P}{\hyp{G}\pp \Gamma,\cht{x}{B\{A/X\}} \pp \Delta, \cht{y}{\forall X.\dual{B}\{A/X\}}}}
        {\der{D'}}}}}
  {\plto
    {\deduce
      {\judge{\Theta}{P}{\hyp{G}\pp \Gamma,\cht{x}{\exists X.B} \pp \Delta, \cht{y}{\forall X.\dual{B}}}}
      {\der{D}}}
    {\lsync
      {\lsendtype{x}{A}}
      {\lrecvtype{y}{A}}}
    {\deduce
      {\judge{\Theta}{P}{\hyp{G}\pp \Gamma,\cht{x}{B\{A/X\}} \pp \Delta, \cht{y}{\forall X.\dual{B}\{A/X\}}}}
      {\der{D'}}}}
\\
\infer[\rlabel{\ref*{rule:pill-cut-?}}{rule:hopill-cut-?}]
{\plto
  {\infer[\ref{rule:hopill-cut}]
    {\judge{\Theta}{\res{xy}{P}}{\hyp{G} \pp \Gamma, \Delta}}
    {\deduce
      {\judge{\Theta}{P}{\hyp{G} \pp \Gamma, \cht{x}{\query A} \pp \Delta,\cht{y}{\bang \dual A}}}
      {\der{D}}}}
  {\tau}
  {\infer[\ref{rule:hopill-cut}]
    {\judge{\Theta}{\res{x'y'}{P'}}{\hyp{G} \pp \Gamma, \Delta}}
    {\deduce
      {\judge{\Theta}{P'}{\hyp{G} \pp \Gamma, \cht{x'}{A} \pp \Delta,\cht{y'}{\dual A}}}
      {\der{D'}}}}}
{\plto
  {\deduce
    {\judge{\Theta}{P}{\hyp{G} \pp \Gamma, \cht{x}{\query A} \pp \Delta,\cht{y}{\bang \dual A}}}
    {\der{D}}}
  {\lsync
    {\luse{x}{x'}}
    {\lcouse{y}{y'}}}
  {\deduce
    {\judge{\Theta}{P'}{\hyp{G} \pp \Gamma, \cht{x'}{A} \pp \Delta,\cht{y'}{\dual A}}}
    {\der{D'}}}}
\\
\infer[\rlabel{\ref*{rule:pill-cut-contract}}{rule:hopill-cut-contract}]
{\plto
  {\infer[\ref{rule:hopill-cut}]
    {\judge{\Theta}{\res{xy}{P}}{\hyp{G} \pp \Gamma, \query \Delta}}
    {\deduce
      {\judge{\Theta}{P}{\hyp{G} \pp \Gamma, \cht{x}{\query A} \pp \query \Delta,\cht{y}{\bang \dual A}}}
      {\der{D}}}}
  {\tau}
  {\infer[\ref{rule:hopill-cut}]
    {\judge{\Theta}{\res{x_1y_1}{P'}}{\hyp{G} \pp \Gamma, \query \Delta}}
    {\deduce
      {\judge{\Theta}{P'}{\hyp{G} \pp \Gamma, \cht{x_1}{\query A\parr\query A} \pp \query \Delta,\cht{y_1}{\bang \dual A \tensor\bang \dual A}}}
      {\der{D'}}}}}
{\plto
  {\deduce
    {\judge{\Theta}{P}{\hyp{G} \pp \Gamma, \cht{x}{\query A} \pp \query \Delta,\cht{y}{\bang \dual A}}}
    {\der{D}}}
  {\lsync
    {\ldup{x}{x_1}{x_2}}
    {\lcodup{y}{y_1}{y_2}}}
  {\deduce
    {\judge{\Theta}{P'}{\hyp{G} \pp \Gamma, \cht{x_1}{\query A\parr\query A} \pp \query \Delta,\cht{y_1}{\bang \dual A \tensor\bang \dual A}}}
    {\der{D'}}}}
\\
\infer[\rlabel{\ref*{rule:pill-cut-weaken}}{rule:hopill-cut-weaken}]
{\plto
  {\infer[\ref{rule:hopill-cut}]
    {\judge{\Theta}{\res{xy}{P}}{\hyp{G} \pp \Gamma, \query \Delta}}
    {\deduce
      {\judge{\Theta}{P}{\hyp{G} \pp \Gamma, \cht{x}{\query A} \pp \query \Delta,\cht{y}{\bang \dual A}}}
      {\der{D}}}}
  {\tau}
  {\infer[\ref{rule:hopill-cut}]
    {\judge{\Theta}{\res{xy}{P'}}{\hyp{G} \pp \Gamma, \query \Delta}}
    {\deduce
      {\judge{\Theta}{P'}{\hyp{G} \pp \Gamma, \cht{x}{\bot} \pp \query \Delta,\cht{y}{\one}}}
      {\der{D'}}}}}
{\plto
  {\deduce
    {\judge{\Theta}{P}{\hyp{G} \pp \Gamma, \cht{x}{\query A} \pp \query \Delta,\cht{y}{\bang \dual A}}}
    {\der{D}}}
  {\lsync
    {\plbl{\ldisp{x}}{\lweaken{A}}}
    {\plbl{\lcodisp{y}}{\lcoweaken{\dual{A}}}}}
  {\deduce
    {\judge{\Theta}{P'}{\hyp{G} \pp \Gamma, \cht{x}{\bot} \pp \query \Delta,\cht{y}{\one}}}
    {\der{D'}}}}
\\
\infer[\ref*{rule:hopill-cut-provide}]
  {\plto
    {\infer[\ref{rule:hopill-cut}]
      {\judge{\Theta,\Pi}{\res{xy}{Q}}{\hyp{G} \pp \Gamma}}
      {\deduce
        {\judge{\Theta,\Pi}{Q}{\hyp{G}\pp \cht{x}{\provide{\hyp{H}}} \pp \Gamma, \cht{y}{\assume{\hyp{H}}}}}
        {\der{D}}}}
    {\tau}
    {\infer[\ref{rule:hopill-chop}]
      {\judge{\Theta,\Pi}{\chop{p}{\abstr \rho P}{Q'}}{\hyp{G} \pp \Gamma}}
      {\deduce
        {\judge{\Pi}{P}{\hyp{H}\rho}}
        {\der{E}}
      &\deduce
        {\judge{\Theta,\pvt{p}{\hyp{H}}}{Q'}{\hyp{G} \pp \Gamma}}
        {\der{D'}}}}}
  {\plto
    {\deduce
      {\judge{\Theta,\Pi}{Q}{\hyp{G}\pp \cht{x}{\provide{\hyp{H}}} \pp \Gamma, \cht{y}{\assume{\hyp{H}}}}}
      {\der{D}}}
    {\lsync
      {\lsendho{x}{\abstr{\rho}{\der{E}}}}
      {\lrecvho{y}{p}}}
    {\deduce
      {\judge{\Theta,\pvt{p}{\hyp{H}}}{Q'}{\hyp{G} \pp \Gamma}}
      {\der{D'}}}}
\\
\end{drules}

\subsection{SOS of processes}
\label{sec:hopill-full-sos-processes}

\begin{highlight}\small
\begin{spreadlines}{\termltsskipamount}
  \headertext{Actions}
  \begin{gather*}
    {\phl{\query x[].P} \lto{\query x[]} \phl{\wait xP}}
    \qquad
    {\phl{\ldup{x}{x_1}{x_2}.P} \lto{\ldup{x}{x_1}{x_2}} \phl{x_1(x_2).P}}
    \qquad
    {\phl{\recvtype{x}{X}{P}} \lto{\lrecvtype{x}{A}} \phl{P\{A/X\}}}
    \\
    \deduce
      {\phl{\choice xPQ} \lto{\lcoinr x} \phl{Q}}
      {\phl{\choice xPQ} \lto{\lcoinl x} \phl{P}}
    \qquad
    \infer
      {\phl{\pi.P} \lto{\vphantom{!(}\pi} \phl{P}}
      {\phl{\pi \neq \lrecvtype{x}{A},\ldisp{x},\ldup{x}{x_1}{x_2}}}
    \quad
    \infer
      {\phl{\server{x}{x'}P} \lto{\lcodisp{x}} \phl{\clientdisp{z_1}{\cdots\clientdisp{z_n}{\close{x}{\nil}}}}}
      {\phl{\fn(P)\setminus\{x'\} = \{z_1, \ldots, z_n\}}}
    \\
    \infer
      {\phl{\server{x}{x'}P} \lto{\lcodup{x}{x_1}{x_2}} \phl{\clientdup{z_1}{z_1\sigma_1}{z_1\sigma_2}{\dots\clientdup{z_n}{z_n\sigma_1}{z_n\sigma_2}{\send{x_1}{x_2}{(\server{x_1}{x'\sigma_1}{P_1} \pp \server{x_2}{x'\sigma_2}{P_2})}}}}}
      {\phl{P_1 = P\sigma_1}
      & \phl{P_2 = P\sigma_2}
      & \phl{\fn(P_1) \cap \fn(P_2) = \emptyset}
      & \phl{\fn(P)\setminus\{x'\} = \{z_1,\ldots,z_n\}}}
   \end{gather*}%
  \headertext{Structural}
  \begin{gather*}
    {\phl{\forward xy} \lto{\lforward yx} \phl{\nil}}
    \qquad
    {\phl{\forward xy} \lto{\lforward xy} \phl{\nil}}
    \qquad
    {\phl{\invoke[\hyp{G}]{p}{\sigma}} \lto{\lsubst[\hyp{G}]{p}{\abstr{\rho}{P}}} \phl{P\{\sigma \circ \rho^{-1}\}}}
    \\
    \infer[\rrelabel{rule:pill-cut-axiom}{rule:hopill-proc-cut-forward}]
      {\phl{\res{xy}{P}} \lto{\phl\tau} \phl{P'\{x/z\}}}
      {\phl{P} \lto{\lforward{y}{z}} \phl{P'}}
    \quad
    \infer[\rrelabel{rule:pill-proc-mix-1}{rule:hopill-proc-mix-1}]
      {\phl{P \pp Q} \lto{l} \phl{P' \pp Q}}
      {\phl{P} \lto{l} \phl{P'}
      &\phl{\bn(l) \cap \fn(Q) = \emptyset}}
    \quad
    \infer[\rrelabel{rule:pill-proc-mix-2}{rule:hopill-proc-mix-2}]
      {\phl{P \pp Q} \lto{l} \phl{P \pp Q'}}
      {\phl{Q} \lto{l} \phl{Q'}
      &\phl{\bn(l) \cap \fn(P) = \emptyset}}
    \\
    \infer[\rrelabel{rule:pill-proc-mix-sync}{rule:hopill-proc-mix-sync}]
      {\phl{P \pp Q} \lto{\lsync{l}{l'}} \phl{P' \pp Q'}}
      {\phl{P} \lto{l} \phl{P'}
      &\phl{Q} \lto{l'} \phl{Q'}
      &\phl{\bn(l) \cap \bn(l') = \emptyset}}
    \qquad
    \infer[\rrelabel{rule:pill-proc-cut-res}{rule:hopill-proc-cut-res}]
      {\phl{\res{xy}{P}} \lto{l} \phl{\res{xy}{P'}}}
      {\phl{P} \lto{l} \phl{P'}
      &\phl{x,y \notin \cn(l)}}
    \qquad
    \infer[\rrelabel{rule:pill-proc-alpha}{rule:hopill-proc-alpha}]
      {\phl{P} \lto{l} \phl{R}}
      {\phl{P \aleq Q}
      &\phl{Q} \lto{l} \phl{R}}
    \\
    \infer[\rlabel{\ref*{rule:hopill-chop-id}}{rule:hopill-proc-chop-id}]
      {\phl{\chop{q}{\abstr{\rho}{Q}}{P}} \lto{\tau} \phl{P'}}
      {\phl{P} \lto{\lsubst[\hyp{G}]{q}{\abstr{\rho}{Q}}} \phl{P'}}
    \quad
    \infer[\rlabel{\ref*{rule:hopill-chop-provide}}{rule:hopill-proc-chop-provide}]
      {\phl{\chop{q}{\abstr{\rho}{Q}}{P}} \lto{\lsendho{x}{\abstr{\sigma}{\chop{q}{\abstr{\rho}{Q}}{R}}}} \phl{P'}}
      {\phl{P} \lto{\lsendho{x}{\abstr{\sigma}{R}}} \phl{P'}} 
    \quad
    \infer[\rlabel{\ref*{rule:hopill-chop-res}}{rule:hopill-proc-chop-res}]
      {\phl{\chop{q}{\abstr{\rho}{Q}}{P}} \lto{l} \phl{\chop{q}{\abstr{\rho}{Q}}{P'}}}
      {\phl{P} \lto{l} \phl{P'}
      &\phl{q \notin \fpv(l)}} 
  \end{gather*}%
  \headertext{Communications}
  \begin{gather*}
    \infer[\rrelabel{rule:pill-proc-cut-tensor}{rule:hopill-proc-cut-tensor}]
      {\phl{\res{xy}{P}} \lto{\tau} \phl{\res{xy}{\res{x'y'}{P'}}}}
      {\phl{P} \lto{\lsync{\lsend{x}{x'}}{\lrecv{y}{y'}}} \phl{P'}}
    \qquad
    \infer[\rrelabel{rule:pill-proc-cut-one}{rule:hopill-proc-cut-one}]
      {\phl{\res{xy}{P}} \lto{\tau} \phl{P'}}
      {\phl{P} \lto{\lsync{\lclose{x}}{\lwait{y}}} \phl{P'}}
    \qquad
    \infer[\rrelabel{rule:pill-proc-cut-oplus_1}{rule:hopill-proc-cut-oplus_1}]
      {\phl{\res{xy}{P}} \lto{\tau} \phl{\res{xy}{P'}}}
      {\phl{P} \lto{\lsync{\linl{x}}{\lcoinl{y}}} \phl{P'}}
    \\
    \infer[\rrelabel{rule:pill-proc-cut-oplus_2}{rule:hopill-proc-cut-oplus_2}]
     {\phl{\res{xy}{P}} \lto{\tau} \phl{\res{xy}{P'}}}
     {\phl{P} \lto{\lsync{\linr{x}}{\lcoinr{y}}} \phl{P'}}
    \qquad
    \infer[\rrelabel{rule:pill-proc-cut-?}{rule:hopill-proc-cut-?}]
      {\phl{\res{xy}{P}} \lto{\tau} \phl{\res{x'y'}{P'}}}
      {\phl{P} \lto{\lsync{\luse{x}{x'}}{\lcouse{y}{y'}}} \phl{P'}}
    \qquad
    \infer[\rrelabel{rule:pill-proc-cut-weaken}{rule:hopill-proc-cut-weaken}]
      {\phl{\res{xy}{P}} \lto{\phl{\tau}} \phl{\res{xy}{P'}}}
      {\phl{P} \lto{\lsync{\ldisp{x}}{\lcodisp{y}}} \phl{P'}}
    \\
    \infer[\rrelabel{rule:pill-proc-cut-contract}{rule:hopill-proc-cut-contract}]
      {\phl{\res{xy}{P}} \lto{\tau} \phl{\phl{\res{x_1y_1}{P'}}}}
      {\phl{P} \lto{\lsync{\ldup{x}{x_1}{x_2}}{\lcodup{y}{y_1}{y_2}}} \phl{P'}}
    \qquad
    \infer[\rrelabel{rule:pill-proc-cut-exists}{rule:hopill-proc-cut-exists}]
      {\phl{\res{xy}{P}} \lto{\tau} \phl{\res{xy}{P'}}}
      {\phl{P} \lto{\lsync{\lsendtype{x}{A}}{\lrecvtype{y}{A}}} \phl{P'}}
    \qquad
    \infer[\rlabel{\ref*{rule:hopill-cut-provide}}{rule:hopill-proc-cut-provide}]
      {\phl{\res{xy}{P}} \lto{\tau} \phl{\chop{q}{\abstr{\rho}{Q}}{P'}}}
      {\phl{P} \lto{\lsync{\lsendho{x}{\abstr{\rho}{Q}}}{\lrecvho{y}{q}}} \phl{P'}}
  \end{gather*}
   \footer%
  \end{spreadlines}%
\end{highlight}

%% file: main.bbl

\begin{thebibliography}{65}


\ifx \showCODEN    \undefined \def \showCODEN     #1{\unskip}     \fi
\ifx \showDOI      \undefined \def \showDOI       #1{#1}\fi
\ifx \showISBNx    \undefined \def \showISBNx     #1{\unskip}     \fi
\ifx \showISBNxiii \undefined \def \showISBNxiii  #1{\unskip}     \fi
\ifx \showISSN     \undefined \def \showISSN      #1{\unskip}     \fi
\ifx \showLCCN     \undefined \def \showLCCN      #1{\unskip}     \fi
\ifx \shownote     \undefined \def \shownote      #1{#1}          \fi
\ifx \showarticletitle \undefined \def \showarticletitle #1{#1}   \fi
\ifx \showURL      \undefined \def \showURL       {\relax}        \fi
\providecommand\bibfield[2]{#2}
\providecommand\bibinfo[2]{#2}
\providecommand\natexlab[1]{#1}
\providecommand\showeprint[2][]{arXiv:#2}

\bibitem[\protect\citeauthoryear{Abadi, Cardelli, Curien, and L{\'{e}}vy}{Abadi
  et~al\mbox{.}}{1991}]%
        {ACCL91}
\bibfield{author}{\bibinfo{person}{Mart{\'{\i}}n Abadi}, \bibinfo{person}{Luca
  Cardelli}, \bibinfo{person}{Pierre{-}Louis Curien}, {and}
  \bibinfo{person}{Jean{-}Jacques L{\'{e}}vy}.}
  \bibinfo{year}{1991}\natexlab{}.
\newblock \showarticletitle{Explicit Substitutions}.
\newblock \bibinfo{journal}{\emph{J. Funct. Program.}} \bibinfo{volume}{1},
  \bibinfo{number}{4} (\bibinfo{year}{1991}), \bibinfo{pages}{375--416}.
\newblock


\bibitem[\protect\citeauthoryear{Abramsky}{Abramsky}{1994}]%
        {A94}
\bibfield{author}{\bibinfo{person}{Samson Abramsky}.}
  \bibinfo{year}{1994}\natexlab{}.
\newblock \showarticletitle{Proofs as Processes}.
\newblock \bibinfo{journal}{\emph{TCS}} \bibinfo{volume}{135},
  \bibinfo{number}{1} (\bibinfo{year}{1994}), \bibinfo{pages}{5--9}.
\newblock
\urldef\tempurl%
\url{https://doi.org/10.1016/0304-3975(94)00103-0}
\showDOI{\tempurl}


\bibitem[\protect\citeauthoryear{Ancona, Bono, Bravetti, Campos, Castagna,
  Deni{\'{e}}lou, Gay, Gesbert, Giachino, Hu, Johnsen, Martins, Mascardi,
  Montesi, Neykova, Ng, Padovani, Vasconcelos, and Yoshida}{Ancona
  et~al\mbox{.}}{2016}]%
        {Aetal16}
\bibfield{author}{\bibinfo{person}{Davide Ancona}, \bibinfo{person}{Viviana
  Bono}, \bibinfo{person}{Mario Bravetti}, \bibinfo{person}{Joana Campos},
  \bibinfo{person}{Giuseppe Castagna}, \bibinfo{person}{Pierre{-}Malo
  Deni{\'{e}}lou}, \bibinfo{person}{Simon~J. Gay}, \bibinfo{person}{Nils
  Gesbert}, \bibinfo{person}{Elena Giachino}, \bibinfo{person}{Raymond Hu},
  \bibinfo{person}{Einar~Broch Johnsen}, \bibinfo{person}{Francisco Martins},
  \bibinfo{person}{Viviana Mascardi}, \bibinfo{person}{Fabrizio Montesi},
  \bibinfo{person}{Rumyana Neykova}, \bibinfo{person}{Nicholas Ng},
  \bibinfo{person}{Luca Padovani}, \bibinfo{person}{Vasco~T. Vasconcelos},
  {and} \bibinfo{person}{Nobuko Yoshida}.} \bibinfo{year}{2016}\natexlab{}.
\newblock \showarticletitle{Behavioral Types in Programming Languages}.
\newblock \bibinfo{journal}{\emph{Foundations and Trends in Programming
  Languages}} \bibinfo{volume}{3}, \bibinfo{number}{2-3}
  (\bibinfo{year}{2016}), \bibinfo{pages}{95--230}.
\newblock
\urldef\tempurl%
\url{https://doi.org/10.1561/2500000031}
\showDOI{\tempurl}


\bibitem[\protect\citeauthoryear{Atkey, Lindley, and Morris}{Atkey
  et~al\mbox{.}}{2016}]%
        {ALM16}
\bibfield{author}{\bibinfo{person}{Robert Atkey}, \bibinfo{person}{Sam
  Lindley}, {and} \bibinfo{person}{J.~Garrett Morris}.}
  \bibinfo{year}{2016}\natexlab{}.
\newblock \showarticletitle{Conflation Confers Concurrency}. In
  \bibinfo{booktitle}{\emph{A List of Successes That Can Change the World -
  Essays Dedicated to Philip Wadler on the Occasion of His 60th Birthday}}
  \emph{(\bibinfo{series}{Lecture Notes in Computer Science})},
  \bibfield{editor}{\bibinfo{person}{Sam Lindley}, \bibinfo{person}{Conor
  McBride}, \bibinfo{person}{Philip~W. Trinder}, {and} \bibinfo{person}{Donald
  Sannella}} (Eds.), Vol.~\bibinfo{volume}{9600}.
  \bibinfo{publisher}{Springer}, \bibinfo{pages}{32--55}.
\newblock
\urldef\tempurl%
\url{https://doi.org/10.1007/978-3-319-30936-1\_2}
\showDOI{\tempurl}


\bibitem[\protect\citeauthoryear{Avron}{Avron}{1991}]%
        {A91}
\bibfield{author}{\bibinfo{person}{Arnon Avron}.}
  \bibinfo{year}{1991}\natexlab{}.
\newblock \showarticletitle{Hypersequents, logical consequence and intermediate
  logics for concurrency}.
\newblock \bibinfo{journal}{\emph{Annals of Mathematics and Artificial
  Intelligence}}  \bibinfo{volume}{4} (\bibinfo{year}{1991}),
  \bibinfo{pages}{225--248}.
\newblock


\bibitem[\protect\citeauthoryear{Baelde}{Baelde}{2012}]%
        {B12}
\bibfield{author}{\bibinfo{person}{David Baelde}.}
  \bibinfo{year}{2012}\natexlab{}.
\newblock \showarticletitle{Least and Greatest Fixed Points in Linear Logic}.
\newblock \bibinfo{journal}{\emph{{ACM} Trans. Comput. Log.}}
  \bibinfo{volume}{13}, \bibinfo{number}{1} (\bibinfo{year}{2012}),
  \bibinfo{pages}{2:1--2:44}.
\newblock
\urldef\tempurl%
\url{https://doi.org/10.1145/2071368.2071370}
\showDOI{\tempurl}


\bibitem[\protect\citeauthoryear{Balzer and Pfenning}{Balzer and
  Pfenning}{2017}]%
        {BP17}
\bibfield{author}{\bibinfo{person}{Stephanie Balzer} {and}
  \bibinfo{person}{Frank Pfenning}.} \bibinfo{year}{2017}\natexlab{}.
\newblock \showarticletitle{Manifest sharing with session types}.
\newblock \bibinfo{journal}{\emph{{PACMPL}}} \bibinfo{volume}{1},
  \bibinfo{number}{{ICFP}} (\bibinfo{year}{2017}),
  \bibinfo{pages}{37:1--37:29}.
\newblock
\urldef\tempurl%
\url{https://doi.org/10.1145/3110281}
\showDOI{\tempurl}


\bibitem[\protect\citeauthoryear{Balzer, Pfenning, and Toninho}{Balzer
  et~al\mbox{.}}{2018}]%
        {BPT18}
\bibfield{author}{\bibinfo{person}{Stephanie Balzer}, \bibinfo{person}{Frank
  Pfenning}, {and} \bibinfo{person}{Bernardo Toninho}.}
  \bibinfo{year}{2018}\natexlab{}.
\newblock \showarticletitle{A Universal Session Type for Untyped Asynchronous
  Communication}. In \bibinfo{booktitle}{\emph{29th International Conference on
  Concurrency Theory, {CONCUR} 2018, September 4-7, 2018, Beijing, China}}
  \emph{(\bibinfo{series}{LIPIcs})}, \bibfield{editor}{\bibinfo{person}{Sven
  Schewe} {and} \bibinfo{person}{Lijun Zhang}} (Eds.),
  Vol.~\bibinfo{volume}{118}. \bibinfo{publisher}{Schloss Dagstuhl -
  Leibniz-Zentrum fuer Informatik}, \bibinfo{pages}{30:1--30:18}.
\newblock
\urldef\tempurl%
\url{https://doi.org/10.4230/LIPIcs.CONCUR.2018.30}
\showDOI{\tempurl}


\bibitem[\protect\citeauthoryear{Balzer, Toninho, and Pfenning}{Balzer
  et~al\mbox{.}}{2019}]%
        {BTP19}
\bibfield{author}{\bibinfo{person}{Stephanie Balzer}, \bibinfo{person}{Bernardo
  Toninho}, {and} \bibinfo{person}{Frank Pfenning}.}
  \bibinfo{year}{2019}\natexlab{}.
\newblock \showarticletitle{Manifest Deadlock-Freedom for Shared Session
  Types}. In \bibinfo{booktitle}{\emph{Programming Languages and Systems - 28th
  European Symposium on Programming, {ESOP} 2019, Held as Part of the European
  Joint Conferences on Theory and Practice of Software, {ETAPS} 2019, Prague,
  Czech Republic, April 6-11, 2019, Proceedings}}
  \emph{(\bibinfo{series}{Lecture Notes in Computer Science})},
  \bibfield{editor}{\bibinfo{person}{Lu{\'{\i}}s Caires}} (Ed.),
  Vol.~\bibinfo{volume}{11423}. \bibinfo{publisher}{Springer},
  \bibinfo{pages}{611--639}.
\newblock
\urldef\tempurl%
\url{https://doi.org/10.1007/978-3-030-17184-1\_22}
\showDOI{\tempurl}


\bibitem[\protect\citeauthoryear{Bartoletti, Scalas, and Zunino}{Bartoletti
  et~al\mbox{.}}{2014}]%
        {BSZ14}
\bibfield{author}{\bibinfo{person}{Massimo Bartoletti},
  \bibinfo{person}{Alceste Scalas}, {and} \bibinfo{person}{Roberto Zunino}.}
  \bibinfo{year}{2014}\natexlab{}.
\newblock \showarticletitle{A Semantic Deconstruction of Session Types}. In
  \bibinfo{booktitle}{\emph{{CONCUR} 2014 - Concurrency Theory - 25th
  International Conference, {CONCUR} 2014, Rome, Italy, September 2-5, 2014.
  Proceedings}} \emph{(\bibinfo{series}{Lecture Notes in Computer Science})},
  \bibfield{editor}{\bibinfo{person}{Paolo Baldan} {and}
  \bibinfo{person}{Daniele Gorla}} (Eds.), Vol.~\bibinfo{volume}{8704}.
  \bibinfo{publisher}{Springer}, \bibinfo{pages}{402--418}.
\newblock
\urldef\tempurl%
\url{https://doi.org/10.1007/978-3-662-44584-6\_28}
\showDOI{\tempurl}


\bibitem[\protect\citeauthoryear{Bellin and Scott}{Bellin and Scott}{1994}]%
        {BS94}
\bibfield{author}{\bibinfo{person}{Gianluigi Bellin} {and}
  \bibinfo{person}{Philip~J. Scott}.} \bibinfo{year}{1994}\natexlab{}.
\newblock \showarticletitle{On the pi-Calculus and Linear Logic}.
\newblock \bibinfo{journal}{\emph{TCS}} \bibinfo{volume}{135},
  \bibinfo{number}{1} (\bibinfo{year}{1994}), \bibinfo{pages}{11--65}.
\newblock
\urldef\tempurl%
\url{https://doi.org/10.1016/0304-3975(94)00104-9}
\showDOI{\tempurl}


\bibitem[\protect\citeauthoryear{Bock and Sch{\"{u}}rmann}{Bock and
  Sch{\"{u}}rmann}{2015}]%
        {BS15}
\bibfield{author}{\bibinfo{person}{Peter~Brottveit Bock} {and}
  \bibinfo{person}{Carsten Sch{\"{u}}rmann}.} \bibinfo{year}{2015}\natexlab{}.
\newblock \showarticletitle{A Contextual Logical Framework}. In
  \bibinfo{booktitle}{\emph{{LPAR}}} \emph{(\bibinfo{series}{Lecture Notes in
  Computer Science})}, Vol.~\bibinfo{volume}{9450}.
  \bibinfo{publisher}{Springer}, \bibinfo{pages}{402--417}.
\newblock


\bibitem[\protect\citeauthoryear{Bono, Coppo, Dezani{-}Ciancaglini, and
  Venneri}{Bono et~al\mbox{.}}{2017}]%
        {BCDV17}
\bibfield{author}{\bibinfo{person}{Viviana Bono}, \bibinfo{person}{Mario
  Coppo}, \bibinfo{person}{Mariangiola Dezani{-}Ciancaglini}, {and}
  \bibinfo{person}{Betti Venneri}.} \bibinfo{year}{2017}\natexlab{}.
\newblock \showarticletitle{Data-driven adaptation for smart sessions}.
\newblock \bibinfo{journal}{\emph{J. Log. Algebr. Meth. Program.}}
  \bibinfo{volume}{90} (\bibinfo{year}{2017}), \bibinfo{pages}{31--49}.
\newblock


\bibitem[\protect\citeauthoryear{Brengos, Miculan, and Peressotti}{Brengos
  et~al\mbox{.}}{2015}]%
        {BMP15}
\bibfield{author}{\bibinfo{person}{Tomasz Brengos}, \bibinfo{person}{Marino
  Miculan}, {and} \bibinfo{person}{Marco Peressotti}.}
  \bibinfo{year}{2015}\natexlab{}.
\newblock \showarticletitle{Behavioural equivalences for coalgebras with
  unobservable moves}.
\newblock \bibinfo{journal}{\emph{Journal of Logical and Algebraic Methods in
  Programming}} \bibinfo{volume}{84}, \bibinfo{number}{6}
  (\bibinfo{year}{2015}), \bibinfo{pages}{826--852}.
\newblock
\urldef\tempurl%
\url{https://doi.org/10.1016/j.jlamp.2015.09.002}
\showDOI{\tempurl}


\bibitem[\protect\citeauthoryear{Caires and P{\'{e}}rez}{Caires and
  P{\'{e}}rez}{2016}]%
        {CP16}
\bibfield{author}{\bibinfo{person}{Lu{\'{\i}}s Caires} {and}
  \bibinfo{person}{Jorge~A. P{\'{e}}rez}.} \bibinfo{year}{2016}\natexlab{}.
\newblock \showarticletitle{Multiparty Session Types Within a Canonical Binary
  Theory, and Beyond}. In \bibinfo{booktitle}{\emph{Formal Techniques for
  Distributed Objects, Components, and Systems - 36th {IFIP} {WG} 6.1
  International Conference, {FORTE} 2016, Held as Part of the 11th
  International Federated Conference on Distributed Computing Techniques,
  DisCoTec 2016, Heraklion, Crete, Greece, June 6-9, 2016, Proceedings}}
  \emph{(\bibinfo{series}{Lecture Notes in Computer Science})},
  \bibfield{editor}{\bibinfo{person}{Elvira Albert} {and} \bibinfo{person}{Ivan
  Lanese}} (Eds.), Vol.~\bibinfo{volume}{9688}. \bibinfo{publisher}{Springer},
  \bibinfo{pages}{74--95}.
\newblock
\urldef\tempurl%
\url{https://doi.org/10.1007/978-3-319-39570-8\_6}
\showDOI{\tempurl}


\bibitem[\protect\citeauthoryear{Caires and Pfenning}{Caires and
  Pfenning}{2010}]%
        {CP10}
\bibfield{author}{\bibinfo{person}{Lu\'{\i}s Caires} {and}
  \bibinfo{person}{Frank Pfenning}.} \bibinfo{year}{2010}\natexlab{}.
\newblock \showarticletitle{Session Types as Intuitionistic Linear
  Propositions}. In \bibinfo{booktitle}{\emph{CONCUR}}.
  \bibinfo{pages}{222--236}.
\newblock


\bibitem[\protect\citeauthoryear{Caires, Pfenning, and Toninho}{Caires
  et~al\mbox{.}}{2016}]%
        {CPT16}
\bibfield{author}{\bibinfo{person}{Lu{\'{\i}}s Caires}, \bibinfo{person}{Frank
  Pfenning}, {and} \bibinfo{person}{Bernardo Toninho}.}
  \bibinfo{year}{2016}\natexlab{}.
\newblock \showarticletitle{Linear logic propositions as session types}.
\newblock \bibinfo{journal}{\emph{Mathematical Structures in Computer Science}}
  \bibinfo{volume}{26}, \bibinfo{number}{3} (\bibinfo{year}{2016}),
  \bibinfo{pages}{367--423}.
\newblock
\urldef\tempurl%
\url{https://doi.org/10.1017/S0960129514000218}
\showDOI{\tempurl}


\bibitem[\protect\citeauthoryear{Carbone, Cruz{-}Filipe, Montesi, and
  Murawska}{Carbone et~al\mbox{.}}{2018a}]%
        {CCMM18}
\bibfield{author}{\bibinfo{person}{Marco Carbone}, \bibinfo{person}{Lu{\'{\i}}s
  Cruz{-}Filipe}, \bibinfo{person}{Fabrizio Montesi}, {and}
  \bibinfo{person}{Agata Murawska}.} \bibinfo{year}{2018}\natexlab{a}.
\newblock \showarticletitle{Multiparty Classical Choreographies}. In
  \bibinfo{booktitle}{\emph{Logic-Based Program Synthesis and Transformation -
  28th International Symposium, {LOPSTR} 2018, Frankfurt/Main, Germany,
  September 4-6, 2018, Revised Selected Papers}}
  \emph{(\bibinfo{series}{Lecture Notes in Computer Science})},
  \bibfield{editor}{\bibinfo{person}{Fred Mesnard} {and}
  \bibinfo{person}{Peter~J. Stuckey}} (Eds.), Vol.~\bibinfo{volume}{11408}.
  \bibinfo{publisher}{Springer}, \bibinfo{pages}{59--76}.
\newblock
\urldef\tempurl%
\url{https://doi.org/10.1007/978-3-030-13838-7\_4}
\showDOI{\tempurl}


\bibitem[\protect\citeauthoryear{Carbone, Lindley, Montesi, Sch{\"{u}}rmann,
  and Wadler}{Carbone et~al\mbox{.}}{2016}]%
        {CLMSW16}
\bibfield{author}{\bibinfo{person}{Marco Carbone}, \bibinfo{person}{Sam
  Lindley}, \bibinfo{person}{Fabrizio Montesi}, \bibinfo{person}{Carsten
  Sch{\"{u}}rmann}, {and} \bibinfo{person}{Philip Wadler}.}
  \bibinfo{year}{2016}\natexlab{}.
\newblock \showarticletitle{Coherence Generalises Duality: {A} Logical
  Explanation of Multiparty Session Types}. In \bibinfo{booktitle}{\emph{27th
  International Conference on Concurrency Theory, {CONCUR} 2016, August 23-26,
  2016, Qu{\'{e}}bec City, Canada}} \emph{(\bibinfo{series}{LIPIcs})},
  \bibfield{editor}{\bibinfo{person}{Jos{\'{e}}e Desharnais} {and}
  \bibinfo{person}{Radha Jagadeesan}} (Eds.), Vol.~\bibinfo{volume}{59}.
  \bibinfo{publisher}{Schloss Dagstuhl - Leibniz-Zentrum f{\"{u}}r Informatik},
  \bibinfo{pages}{33:1--33:15}.
\newblock
\urldef\tempurl%
\url{https://doi.org/10.4230/LIPIcs.CONCUR.2016.33}
\showDOI{\tempurl}


\bibitem[\protect\citeauthoryear{Carbone, Montesi, and Sch{\"{u}}rmann}{Carbone
  et~al\mbox{.}}{2018b}]%
        {CMS18}
\bibfield{author}{\bibinfo{person}{Marco Carbone}, \bibinfo{person}{Fabrizio
  Montesi}, {and} \bibinfo{person}{Carsten Sch{\"{u}}rmann}.}
  \bibinfo{year}{2018}\natexlab{b}.
\newblock \showarticletitle{Choreographies, logically}.
\newblock \bibinfo{journal}{\emph{Distributed Computing}} \bibinfo{volume}{31},
  \bibinfo{number}{1} (\bibinfo{year}{2018}), \bibinfo{pages}{51--67}.
\newblock
\urldef\tempurl%
\url{https://doi.org/10.1007/s00446-017-0295-1}
\showDOI{\tempurl}
\newblock
\shownote{Also: CONCUR, pages 47--62, 2014.}


\bibitem[\protect\citeauthoryear{Carbone, Montesi, Sch{\"{u}}rmann, and
  Yoshida}{Carbone et~al\mbox{.}}{2017}]%
        {CMSY17}
\bibfield{author}{\bibinfo{person}{Marco Carbone}, \bibinfo{person}{Fabrizio
  Montesi}, \bibinfo{person}{Carsten Sch{\"{u}}rmann}, {and}
  \bibinfo{person}{Nobuko Yoshida}.} \bibinfo{year}{2017}\natexlab{}.
\newblock \showarticletitle{{Multiparty session types as coherence proofs}}.
\newblock \bibinfo{journal}{\emph{Acta Informatica}} (\bibinfo{year}{2017}).
\newblock
\urldef\tempurl%
\url{https://doi.org/10.1007/s00236-016-0285-y}
\showDOI{\tempurl}
\newblock
\shownote{Also: CONCUR 2015.}


\bibitem[\protect\citeauthoryear{Ciancia}{Ciancia}{2013}]%
        {CHM13}
\bibfield{author}{\bibinfo{person}{Vincenzo Ciancia}.}
  \bibinfo{year}{2013}\natexlab{}.
\newblock \showarticletitle{Interaction and Observation: Categorical Semantics
  of Reactive Systems Through Dialgebras}. In \bibinfo{booktitle}{\emph{Algebra
  and Coalgebra in Computer Science - 5th International Conference, {CALCO}
  2013, Warsaw, Poland, September 3-6, 2013. Proceedings}}
  \emph{(\bibinfo{series}{Lecture Notes in Computer Science})},
  \bibfield{editor}{\bibinfo{person}{Reiko Heckel} {and}
  \bibinfo{person}{Stefan Milius}} (Eds.), Vol.~\bibinfo{volume}{8089}.
  \bibinfo{publisher}{Springer}, \bibinfo{pages}{110--125}.
\newblock
\showISBNx{978-3-642-40206-7}
\urldef\tempurl%
\url{https://doi.org/10.1007/978-3-642-40206-7\_10}
\showDOI{\tempurl}


\bibitem[\protect\citeauthoryear{Ciobanu and Horne}{Ciobanu and Horne}{2015}]%
        {CH15}
\bibfield{author}{\bibinfo{person}{Gabriel Ciobanu} {and} \bibinfo{person}{Ross
  Horne}.} \bibinfo{year}{2015}\natexlab{}.
\newblock \showarticletitle{Behavioural Analysis of Sessions Using the Calculus
  of Structures}. In \bibinfo{booktitle}{\emph{Perspectives of System
  Informatics - 10th International Andrei Ershov Informatics Conference, {PSI}
  2015, in Memory of Helmut Veith, Kazan and Innopolis, Russia, August 24-27,
  2015, Revised Selected Papers}} \emph{(\bibinfo{series}{Lecture Notes in
  Computer Science})}, \bibfield{editor}{\bibinfo{person}{Manuel Mazzara} {and}
  \bibinfo{person}{Andrei Voronkov}} (Eds.), Vol.~\bibinfo{volume}{9609}.
  \bibinfo{publisher}{Springer}, \bibinfo{pages}{91--106}.
\newblock
\urldef\tempurl%
\url{https://doi.org/10.1007/978-3-319-41579-6\_8}
\showDOI{\tempurl}


\bibitem[\protect\citeauthoryear{Dalla~Preda, Gabbrielli, Giallorenzo, Lanese,
  and Mauro}{Dalla~Preda et~al\mbox{.}}{2017}]%
        {DGGLM16}
\bibfield{author}{\bibinfo{person}{Mila Dalla~Preda}, \bibinfo{person}{Maurizio
  Gabbrielli}, \bibinfo{person}{Saverio Giallorenzo}, \bibinfo{person}{Ivan
  Lanese}, {and} \bibinfo{person}{Jacopo Mauro}.}
  \bibinfo{year}{2017}\natexlab{}.
\newblock \showarticletitle{Dynamic Choreographies: Theory And Implementation}.
\newblock \bibinfo{journal}{\emph{Logical Methods in Computer Science}}
  \bibinfo{volume}{13}, \bibinfo{number}{2} (\bibinfo{year}{2017}).
\newblock


\bibitem[\protect\citeauthoryear{Dardha, Giachino, and Sangiorgi}{Dardha
  et~al\mbox{.}}{2017}]%
        {DGS17}
\bibfield{author}{\bibinfo{person}{Ornela Dardha}, \bibinfo{person}{Elena
  Giachino}, {and} \bibinfo{person}{Davide Sangiorgi}.}
  \bibinfo{year}{2017}\natexlab{}.
\newblock \showarticletitle{Session types revisited}.
\newblock \bibinfo{journal}{\emph{Inf. Comput.}}  \bibinfo{volume}{256}
  (\bibinfo{year}{2017}), \bibinfo{pages}{253--286}.
\newblock
\urldef\tempurl%
\url{https://doi.org/10.1016/j.ic.2017.06.002}
\showDOI{\tempurl}


\bibitem[\protect\citeauthoryear{Di~Giusto and P{\'{e}}rez}{Di~Giusto and
  P{\'{e}}rez}{2015}]%
        {DP15}
\bibfield{author}{\bibinfo{person}{Cinzia Di~Giusto} {and}
  \bibinfo{person}{Jorge~A. P{\'{e}}rez}.} \bibinfo{year}{2015}\natexlab{}.
\newblock \showarticletitle{Disciplined structured communications with
  disciplined runtime adaptation}.
\newblock \bibinfo{journal}{\emph{Sci. Comput. Program.}}  \bibinfo{volume}{97}
  (\bibinfo{year}{2015}), \bibinfo{pages}{235--265}.
\newblock


\bibitem[\protect\citeauthoryear{Dragoni, Giallorenzo, Lluch{-}Lafuente,
  Mazzara, Montesi, Mustafin, and Safina}{Dragoni et~al\mbox{.}}{2017}]%
        {DGLMMMS17}
\bibfield{author}{\bibinfo{person}{Nicola Dragoni}, \bibinfo{person}{Saverio
  Giallorenzo}, \bibinfo{person}{Alberto Lluch{-}Lafuente},
  \bibinfo{person}{Manuel Mazzara}, \bibinfo{person}{Fabrizio Montesi},
  \bibinfo{person}{Ruslan Mustafin}, {and} \bibinfo{person}{Larisa Safina}.}
  \bibinfo{year}{2017}\natexlab{}.
\newblock \showarticletitle{Microservices: Yesterday, Today, and Tomorrow}.
\newblock In \bibinfo{booktitle}{\emph{Present and Ulterior Software
  Engineering.}}, \bibfield{editor}{\bibinfo{person}{Manuel Mazzara} {and}
  \bibinfo{person}{Bertrand Meyer}} (Eds.). \bibinfo{publisher}{Springer},
  \bibinfo{pages}{195--216}.
\newblock
\urldef\tempurl%
\url{https://doi.org/10.1007/978-3-319-67425-4_12}
\showDOI{\tempurl}


\bibitem[\protect\citeauthoryear{Fu}{Fu}{2013}]%
        {F13}
\bibfield{author}{\bibinfo{person}{Yuxi Fu}.} \bibinfo{year}{2013}\natexlab{}.
\newblock \showarticletitle{The Value-Passing Calculus}. In
  \bibinfo{booktitle}{\emph{Theories of Programming and Formal Methods - Essays
  Dedicated to Jifeng He on the Occasion of His 70th Birthday}}
  \emph{(\bibinfo{series}{Lecture Notes in Computer Science})},
  \bibfield{editor}{\bibinfo{person}{Zhiming Liu}, \bibinfo{person}{Jim
  Woodcock}, {and} \bibinfo{person}{Huibiao Zhu}} (Eds.),
  Vol.~\bibinfo{volume}{8051}. \bibinfo{publisher}{Springer},
  \bibinfo{pages}{166--195}.
\newblock
\urldef\tempurl%
\url{https://doi.org/10.1007/978-3-642-39698-4\_11}
\showDOI{\tempurl}


\bibitem[\protect\citeauthoryear{Fu}{Fu}{2017}]%
        {F17}
\bibfield{author}{\bibinfo{person}{Yuxi Fu}.} \bibinfo{year}{2017}\natexlab{}.
\newblock \showarticletitle{On the Power of Name-Passing Communication}. In
  \bibinfo{booktitle}{\emph{28th International Conference on Concurrency
  Theory, {CONCUR} 2017, September 5-8, 2017, Berlin, Germany}}
  \emph{(\bibinfo{series}{LIPIcs})}, \bibfield{editor}{\bibinfo{person}{Roland
  Meyer} {and} \bibinfo{person}{Uwe Nestmann}} (Eds.),
  Vol.~\bibinfo{volume}{85}. \bibinfo{publisher}{Schloss Dagstuhl -
  Leibniz-Zentrum fuer Informatik}, \bibinfo{pages}{22:1--22:15}.
\newblock
\urldef\tempurl%
\url{https://doi.org/10.4230/LIPIcs.CONCUR.2017.22}
\showDOI{\tempurl}


\bibitem[\protect\citeauthoryear{Girard}{Girard}{1987}]%
        {G87}
\bibfield{author}{\bibinfo{person}{Jean{-}Yves Girard}.}
  \bibinfo{year}{1987}\natexlab{}.
\newblock \showarticletitle{Linear Logic}.
\newblock \bibinfo{journal}{\emph{Theor. Comput. Sci.}}  \bibinfo{volume}{50}
  (\bibinfo{year}{1987}), \bibinfo{pages}{1--102}.
\newblock
\urldef\tempurl%
\url{https://doi.org/10.1016/0304-3975(87)90045-4}
\showDOI{\tempurl}


\bibitem[\protect\citeauthoryear{Hasuo}{Hasuo}{2006}]%
        {H06}
\bibfield{author}{\bibinfo{person}{Ichiro Hasuo}.}
  \bibinfo{year}{2006}\natexlab{}.
\newblock \showarticletitle{Generic Forward and Backward Simulations}. In
  \bibinfo{booktitle}{\emph{{CONCUR} 2006 - Concurrency Theory, 17th
  International Conference, {CONCUR} 2006, Bonn, Germany, August 27-30, 2006,
  Proceedings}} \emph{(\bibinfo{series}{Lecture Notes in Computer Science})},
  \bibfield{editor}{\bibinfo{person}{Christel Baier} {and}
  \bibinfo{person}{Holger Hermanns}} (Eds.), Vol.~\bibinfo{volume}{4137}.
  \bibinfo{publisher}{Springer}, \bibinfo{pages}{406--420}.
\newblock
\urldef\tempurl%
\url{https://doi.org/10.1007/11817949\_27}
\showDOI{\tempurl}


\bibitem[\protect\citeauthoryear{Honda}{Honda}{1993}]%
        {H93}
\bibfield{author}{\bibinfo{person}{Kohei Honda}.}
  \bibinfo{year}{1993}\natexlab{}.
\newblock \showarticletitle{Types for Dyadic Interaction}. In
  \bibinfo{booktitle}{\emph{{CONCUR} '93, 4th International Conference on
  Concurrency Theory, Hildesheim, Germany, August 23-26, 1993, Proceedings}}
  \emph{(\bibinfo{series}{Lecture Notes in Computer Science})},
  \bibfield{editor}{\bibinfo{person}{Eike Best}} (Ed.),
  Vol.~\bibinfo{volume}{715}. \bibinfo{publisher}{Springer},
  \bibinfo{pages}{509--523}.
\newblock
\urldef\tempurl%
\url{https://doi.org/10.1007/3-540-57208-2\_35}
\showDOI{\tempurl}


\bibitem[\protect\citeauthoryear{Honda, Vasconcelos, and Kubo}{Honda
  et~al\mbox{.}}{1998}]%
        {HVK98}
\bibfield{author}{\bibinfo{person}{Kohei Honda}, \bibinfo{person}{Vasco
  Vasconcelos}, {and} \bibinfo{person}{Makoto Kubo}.}
  \bibinfo{year}{1998}\natexlab{}.
\newblock \showarticletitle{Language primitives and type disciplines for
  structured communication-based programming}. In
  \bibinfo{booktitle}{\emph{ESOP}}. \bibinfo{pages}{22--138}.
\newblock


\bibitem[\protect\citeauthoryear{Honda, Yoshida, and Carbone}{Honda
  et~al\mbox{.}}{2016}]%
        {HYC16}
\bibfield{author}{\bibinfo{person}{Kohei Honda}, \bibinfo{person}{Nobuko
  Yoshida}, {and} \bibinfo{person}{Marco Carbone}.}
  \bibinfo{year}{2016}\natexlab{}.
\newblock \showarticletitle{Multiparty Asynchronous Session Types}.
\newblock \bibinfo{journal}{\emph{JACM}} \bibinfo{volume}{63},
  \bibinfo{number}{1} (\bibinfo{year}{2016}), \bibinfo{pages}{9}.
\newblock
\newblock
\shownote{Also: POPL, 2008, pages 273--284.}


\bibitem[\protect\citeauthoryear{H{\"{u}}ttel, Lanese, Vasconcelos, Caires,
  Carbone, Deni{\'{e}}lou, Mostrous, Padovani, Ravara, Tuosto, Vieira, and
  Zavattaro}{H{\"{u}}ttel et~al\mbox{.}}{2016}]%
        {Hetal16}
\bibfield{author}{\bibinfo{person}{Hans H{\"{u}}ttel}, \bibinfo{person}{Ivan
  Lanese}, \bibinfo{person}{Vasco~T. Vasconcelos}, \bibinfo{person}{Lu{\'{\i}}s
  Caires}, \bibinfo{person}{Marco Carbone}, \bibinfo{person}{Pierre{-}Malo
  Deni{\'{e}}lou}, \bibinfo{person}{Dimitris Mostrous}, \bibinfo{person}{Luca
  Padovani}, \bibinfo{person}{Ant{\'{o}}nio Ravara}, \bibinfo{person}{Emilio
  Tuosto}, \bibinfo{person}{Hugo~Torres Vieira}, {and}
  \bibinfo{person}{Gianluigi Zavattaro}.} \bibinfo{year}{2016}\natexlab{}.
\newblock \showarticletitle{Foundations of Session Types and Behavioural
  Contracts}.
\newblock \bibinfo{journal}{\emph{{ACM} Comput. Surv.}} \bibinfo{volume}{49},
  \bibinfo{number}{1} (\bibinfo{year}{2016}), \bibinfo{pages}{3:1--3:36}.
\newblock
\urldef\tempurl%
\url{https://doi.org/10.1145/2873052}
\showDOI{\tempurl}


\bibitem[\protect\citeauthoryear{Kobayashi, Pierce, and Turner}{Kobayashi
  et~al\mbox{.}}{1999}]%
        {KPT99}
\bibfield{author}{\bibinfo{person}{Naoki Kobayashi},
  \bibinfo{person}{Benjamin~C. Pierce}, {and} \bibinfo{person}{David~N.
  Turner}.} \bibinfo{year}{1999}\natexlab{}.
\newblock \showarticletitle{Linearity and the pi-calculus}.
\newblock \bibinfo{journal}{\emph{{ACM} Trans. Program. Lang. Syst.}}
  \bibinfo{volume}{21}, \bibinfo{number}{5} (\bibinfo{year}{1999}),
  \bibinfo{pages}{914--947}.
\newblock
\urldef\tempurl%
\url{https://doi.org/10.1145/330249.330251}
\showDOI{\tempurl}


\bibitem[\protect\citeauthoryear{Kokke, Montesi, and Peressotti}{Kokke
  et~al\mbox{.}}{2018}]%
        {KMP18}
\bibfield{author}{\bibinfo{person}{Wen Kokke}, \bibinfo{person}{Fabrizio
  Montesi}, {and} \bibinfo{person}{Marco Peressotti}.}
  \bibinfo{year}{2018}\natexlab{}.
\newblock \showarticletitle{Taking Linear Logic Apart}. In
  \bibinfo{booktitle}{\emph{Proceedings Joint International Workshop on
  Linearity {\&} Trends in Linear Logic and Applications, Linearity-TLLA@FLoC
  2018, Oxford, UK, 7-8 July 2018}} \emph{(\bibinfo{series}{{EPTCS}})},
  \bibfield{editor}{\bibinfo{person}{Thomas Ehrhard}, \bibinfo{person}{Maribel
  Fern{\'{a}}ndez}, \bibinfo{person}{Valeria de~Paiva}, {and}
  \bibinfo{person}{Lorenzo~Tortora de~Falco}} (Eds.),
  Vol.~\bibinfo{volume}{292}. \bibinfo{pages}{90--103}.
\newblock
\urldef\tempurl%
\url{https://doi.org/10.4204/EPTCS.292.5}
\showDOI{\tempurl}


\bibitem[\protect\citeauthoryear{Kokke, Montesi, and Peressotti}{Kokke
  et~al\mbox{.}}{2019}]%
        {KMP19}
\bibfield{author}{\bibinfo{person}{Wen Kokke}, \bibinfo{person}{Fabrizio
  Montesi}, {and} \bibinfo{person}{Marco Peressotti}.}
  \bibinfo{year}{2019}\natexlab{}.
\newblock \showarticletitle{Better late than never: a fully-abstract semantics
  for classical processes}.
\newblock \bibinfo{journal}{\emph{{PACMPL}}} \bibinfo{volume}{3},
  \bibinfo{number}{{POPL}} (\bibinfo{year}{2019}),
  \bibinfo{pages}{24:1--24:29}.
\newblock
\urldef\tempurl%
\url{https://doi.org/10.1145/3290337}
\showDOI{\tempurl}


\bibitem[\protect\citeauthoryear{Kouzapas, P{\'{e}}rez, and Yoshida}{Kouzapas
  et~al\mbox{.}}{2016}]%
        {KPY16}
\bibfield{author}{\bibinfo{person}{Dimitrios Kouzapas},
  \bibinfo{person}{Jorge~A. P{\'{e}}rez}, {and} \bibinfo{person}{Nobuko
  Yoshida}.} \bibinfo{year}{2016}\natexlab{}.
\newblock \showarticletitle{On the Relative Expressiveness of Higher-Order
  Session Processes}. In \bibinfo{booktitle}{\emph{Programming Languages and
  Systems - 25th European Symposium on Programming, {ESOP} 2016, Held as Part
  of the European Joint Conferences on Theory and Practice of Software, {ETAPS}
  2016, Eindhoven, The Netherlands, April 2-8, 2016, Proceedings}}
  \emph{(\bibinfo{series}{Lecture Notes in Computer Science})},
  \bibfield{editor}{\bibinfo{person}{Peter Thiemann}} (Ed.),
  Vol.~\bibinfo{volume}{9632}. \bibinfo{publisher}{Springer},
  \bibinfo{pages}{446--475}.
\newblock
\urldef\tempurl%
\url{https://doi.org/10.1007/978-3-662-49498-1\_18}
\showDOI{\tempurl}


\bibitem[\protect\citeauthoryear{Kouzapas and Yoshida}{Kouzapas and
  Yoshida}{2014}]%
        {KY14}
\bibfield{author}{\bibinfo{person}{Dimitrios Kouzapas} {and}
  \bibinfo{person}{Nobuko Yoshida}.} \bibinfo{year}{2014}\natexlab{}.
\newblock \showarticletitle{Globally Governed Session Semantics}.
\newblock \bibinfo{journal}{\emph{Logical Methods in Computer Science}}
  \bibinfo{volume}{10}, \bibinfo{number}{4} (\bibinfo{year}{2014}).
\newblock


\bibitem[\protect\citeauthoryear{Lanese, P{\'{e}}rez, Sangiorgi, and
  Schmitt}{Lanese et~al\mbox{.}}{2008}]%
        {LPSS08}
\bibfield{author}{\bibinfo{person}{Ivan Lanese}, \bibinfo{person}{Jorge~A.
  P{\'{e}}rez}, \bibinfo{person}{Davide Sangiorgi}, {and} \bibinfo{person}{Alan
  Schmitt}.} \bibinfo{year}{2008}\natexlab{}.
\newblock \showarticletitle{On the Expressiveness and Decidability of
  Higher-Order Process Calculi}. In \bibinfo{booktitle}{\emph{Proceedings of
  the Twenty-Third Annual {IEEE} Symposium on Logic in Computer Science, {LICS}
  2008, 24-27 June 2008, Pittsburgh, PA, {USA}}}. \bibinfo{publisher}{{IEEE}
  Computer Society}, \bibinfo{pages}{145--155}.
\newblock
\urldef\tempurl%
\url{https://doi.org/10.1109/LICS.2008.8}
\showDOI{\tempurl}


\bibitem[\protect\citeauthoryear{Lanese, P{\'{e}}rez, Sangiorgi, and
  Schmitt}{Lanese et~al\mbox{.}}{2011}]%
        {LPSS11}
\bibfield{author}{\bibinfo{person}{Ivan Lanese}, \bibinfo{person}{Jorge~A.
  P{\'{e}}rez}, \bibinfo{person}{Davide Sangiorgi}, {and} \bibinfo{person}{Alan
  Schmitt}.} \bibinfo{year}{2011}\natexlab{}.
\newblock \showarticletitle{On the expressiveness and decidability of
  higher-order process calculi}.
\newblock \bibinfo{journal}{\emph{Inf. Comput.}} \bibinfo{volume}{209},
  \bibinfo{number}{2} (\bibinfo{year}{2011}), \bibinfo{pages}{198--226}.
\newblock
\urldef\tempurl%
\url{https://doi.org/10.1016/j.ic.2010.10.001}
\showDOI{\tempurl}


\bibitem[\protect\citeauthoryear{Lindley and Morris}{Lindley and
  Morris}{2016}]%
        {LM16}
\bibfield{author}{\bibinfo{person}{Sam Lindley} {and}
  \bibinfo{person}{J.~Garrett Morris}.} \bibinfo{year}{2016}\natexlab{}.
\newblock \showarticletitle{Talking bananas: structural recursion for session
  types}. In \bibinfo{booktitle}{\emph{Proceedings of the 21st {ACM} {SIGPLAN}
  International Conference on Functional Programming, {ICFP} 2016, Nara, Japan,
  September 18-22, 2016}}, \bibfield{editor}{\bibinfo{person}{Jacques
  Garrigue}, \bibinfo{person}{Gabriele Keller}, {and} \bibinfo{person}{Eijiro
  Sumii}} (Eds.). \bibinfo{publisher}{{ACM}}, \bibinfo{pages}{434--447}.
\newblock
\urldef\tempurl%
\url{https://doi.org/10.1145/2951913.2951921}
\showDOI{\tempurl}


\bibitem[\protect\citeauthoryear{Merro and Sangiorgi}{Merro and
  Sangiorgi}{2004}]%
        {MS04}
\bibfield{author}{\bibinfo{person}{Massimo Merro} {and} \bibinfo{person}{Davide
  Sangiorgi}.} \bibinfo{year}{2004}\natexlab{}.
\newblock \showarticletitle{On asynchrony in name-passing calculi}.
\newblock \bibinfo{journal}{\emph{Mathematical Structures in Computer Science}}
  \bibinfo{volume}{14}, \bibinfo{number}{5} (\bibinfo{year}{2004}),
  \bibinfo{pages}{715--767}.
\newblock


\bibitem[\protect\citeauthoryear{Mezzina and P{\'{e}}rez}{Mezzina and
  P{\'{e}}rez}{2017}]%
        {MP17}
\bibfield{author}{\bibinfo{person}{Claudio~Antares Mezzina} {and}
  \bibinfo{person}{Jorge~A. P{\'{e}}rez}.} \bibinfo{year}{2017}\natexlab{}.
\newblock \showarticletitle{Causally consistent reversible choreographies: a
  monitors-as-memories approach}. In \bibinfo{booktitle}{\emph{Proceedings of
  the 19th International Symposium on Principles and Practice of Declarative
  Programming, Namur, Belgium, October 09 - 11, 2017}},
  \bibfield{editor}{\bibinfo{person}{Wim Vanhoof} {and}
  \bibinfo{person}{Brigitte Pientka}} (Eds.). \bibinfo{publisher}{{ACM}},
  \bibinfo{pages}{127--138}.
\newblock
\urldef\tempurl%
\url{https://doi.org/10.1145/3131851.3131864}
\showDOI{\tempurl}


\bibitem[\protect\citeauthoryear{Milner}{Milner}{1989}]%
        {M89}
\bibfield{author}{\bibinfo{person}{Robin Milner}.}
  \bibinfo{year}{1989}\natexlab{}.
\newblock \bibinfo{booktitle}{\emph{Communication and Concurrency}}.
\newblock \bibinfo{publisher}{Prentice-Hall}.
\newblock


\bibitem[\protect\citeauthoryear{Milner, Parrow, and Walker}{Milner
  et~al\mbox{.}}{1992}]%
        {MPW92}
\bibfield{author}{\bibinfo{person}{Robin Milner}, \bibinfo{person}{Joachim
  Parrow}, {and} \bibinfo{person}{David Walker}.}
  \bibinfo{year}{1992}\natexlab{}.
\newblock \showarticletitle{A Calculus of Mobile Processes, {I}}.
\newblock \bibinfo{journal}{\emph{Inf. Comput.}} \bibinfo{volume}{100},
  \bibinfo{number}{1} (\bibinfo{year}{1992}), \bibinfo{pages}{1--40}.
\newblock


\bibitem[\protect\citeauthoryear{Montesi}{Montesi}{2013}]%
        {M13:phd}
\bibfield{author}{\bibinfo{person}{Fabrizio Montesi}.}
  \bibinfo{year}{2013}\natexlab{}.
\newblock \emph{\bibinfo{title}{Choreographic Programming}}.
\newblock Ph.{D}. Thesis. \bibinfo{school}{IT University of Copenhagen}.
\newblock
\newblock
\shownote{\url{https://www.fabriziomontesi.com/files/choreographic_programming.pdf}.}


\bibitem[\protect\citeauthoryear{Montesi}{Montesi}{2018}]%
        {M18}
\bibfield{author}{\bibinfo{person}{Fabrizio Montesi}.}
  \bibinfo{year}{2018}\natexlab{}.
\newblock \showarticletitle{Classical Higher-Order Processes}.
\newblock \bibinfo{journal}{\emph{CoRR}}  \bibinfo{volume}{abs/1802.02917}
  (\bibinfo{year}{2018}).
\newblock
\showeprint[arxiv]{1802.02917}
\urldef\tempurl%
\url{http://arxiv.org/abs/1802.02917}
\showURL{%
\tempurl}


\bibitem[\protect\citeauthoryear{Montesi and Peressotti}{Montesi and
  Peressotti}{2018}]%
        {MP18}
\bibfield{author}{\bibinfo{person}{Fabrizio Montesi} {and}
  \bibinfo{person}{Marco Peressotti}.} \bibinfo{year}{2018}\natexlab{}.
\newblock \showarticletitle{Classical Transitions}.
\newblock \bibinfo{journal}{\emph{CoRR}}  \bibinfo{volume}{abs/1803.01049}
  (\bibinfo{year}{2018}).
\newblock
\showeprint[arxiv]{1803.01049}
\urldef\tempurl%
\url{http://arxiv.org/abs/1803.01049}
\showURL{%
\tempurl}


\bibitem[\protect\citeauthoryear{Montesi and Weber}{Montesi and Weber}{2016}]%
        {MW16}
\bibfield{author}{\bibinfo{person}{Fabrizio Montesi} {and}
  \bibinfo{person}{Janine Weber}.} \bibinfo{year}{2016}\natexlab{}.
\newblock \showarticletitle{Circuit Breakers, Discovery, and {API} Gateways in
  Microservices}.
\newblock \bibinfo{journal}{\emph{CoRR}}  \bibinfo{volume}{abs/1609.05830}
  (\bibinfo{year}{2016}).
\newblock
\showeprint[arxiv]{1609.05830}
\urldef\tempurl%
\url{http://arxiv.org/abs/1609.05830}
\showURL{%
\tempurl}


\bibitem[\protect\citeauthoryear{Montesi and Yoshida}{Montesi and
  Yoshida}{2013}]%
        {MY13}
\bibfield{author}{\bibinfo{person}{Fabrizio Montesi} {and}
  \bibinfo{person}{Nobuko Yoshida}.} \bibinfo{year}{2013}\natexlab{}.
\newblock \showarticletitle{Compositional Choreographies}. In
  \bibinfo{booktitle}{\emph{CONCUR}}. \bibinfo{pages}{425--439}.
\newblock


\bibitem[\protect\citeauthoryear{Mostrous and Yoshida}{Mostrous and
  Yoshida}{2015}]%
        {MY15}
\bibfield{author}{\bibinfo{person}{Dimitris Mostrous} {and}
  \bibinfo{person}{Nobuko Yoshida}.} \bibinfo{year}{2015}\natexlab{}.
\newblock \showarticletitle{Session typing and asynchronous subtyping for the
  higher-order {\(\pi\)}-calculus}.
\newblock \bibinfo{journal}{\emph{Inf. Comput.}}  \bibinfo{volume}{241}
  (\bibinfo{year}{2015}), \bibinfo{pages}{227--263}.
\newblock


\bibitem[\protect\citeauthoryear{Nanevski, Pfenning, and Pientka}{Nanevski
  et~al\mbox{.}}{2008}]%
        {NPP08}
\bibfield{author}{\bibinfo{person}{Aleksandar Nanevski}, \bibinfo{person}{Frank
  Pfenning}, {and} \bibinfo{person}{Brigitte Pientka}.}
  \bibinfo{year}{2008}\natexlab{}.
\newblock \showarticletitle{Contextual modal type theory}.
\newblock \bibinfo{journal}{\emph{{ACM} Trans. Comput. Log.}}
  \bibinfo{volume}{9}, \bibinfo{number}{3} (\bibinfo{year}{2008}),
  \bibinfo{pages}{23:1--23:49}.
\newblock


\bibitem[\protect\citeauthoryear{P{\'{e}}rez}{P{\'{e}}rez}{2016}]%
        {P16}
\bibfield{author}{\bibinfo{person}{Jorge~A. P{\'{e}}rez}.}
  \bibinfo{year}{2016}\natexlab{}.
\newblock \showarticletitle{The Challenge of Typed Expressiveness in
  Concurrency}. In \bibinfo{booktitle}{\emph{{FORTE}}}
  \emph{(\bibinfo{series}{Lecture Notes in Computer Science})},
  Vol.~\bibinfo{volume}{9688}. \bibinfo{publisher}{Springer},
  \bibinfo{pages}{239--247}.
\newblock


\bibitem[\protect\citeauthoryear{Plotkin}{Plotkin}{2004}]%
        {P04}
\bibfield{author}{\bibinfo{person}{Gordon~D. Plotkin}.}
  \bibinfo{year}{2004}\natexlab{}.
\newblock \showarticletitle{A structural approach to operational semantics}.
\newblock \bibinfo{journal}{\emph{J. Log. Algebr. Program.}}
  \bibinfo{volume}{60-61} (\bibinfo{year}{2004}), \bibinfo{pages}{17--139}.
\newblock


\bibitem[\protect\citeauthoryear{Sangiorgi}{Sangiorgi}{1993}]%
        {S93}
\bibfield{author}{\bibinfo{person}{Davide Sangiorgi}.}
  \bibinfo{year}{1993}\natexlab{}.
\newblock \showarticletitle{From pi-Calculus to Higher-Order pi-Calculus - and
  Back}. In \bibinfo{booktitle}{\emph{{TAPSOFT}}}
  \emph{(\bibinfo{series}{Lecture Notes in Computer Science})},
  Vol.~\bibinfo{volume}{668}. \bibinfo{publisher}{Springer},
  \bibinfo{pages}{151--166}.
\newblock


\bibitem[\protect\citeauthoryear{Sangiorgi}{Sangiorgi}{1996}]%
        {S96}
\bibfield{author}{\bibinfo{person}{Davide Sangiorgi}.}
  \bibinfo{year}{1996}\natexlab{}.
\newblock \showarticletitle{Pi-Calculus, Internal Mobility, and Agent-Passing
  Calculi}.
\newblock \bibinfo{journal}{\emph{TCS}} \bibinfo{volume}{167},
  \bibinfo{number}{1{\&}2} (\bibinfo{year}{1996}), \bibinfo{pages}{235--274}.
\newblock


\bibitem[\protect\citeauthoryear{Sangiorgi}{Sangiorgi}{2011}]%
        {S11}
\bibfield{author}{\bibinfo{person}{Davide Sangiorgi}.}
  \bibinfo{year}{2011}\natexlab{}.
\newblock \bibinfo{booktitle}{\emph{Introduction to Bisimulation and
  Coinduction}}.
\newblock \bibinfo{publisher}{Cambridge University Press}.
\newblock
\urldef\tempurl%
\url{https://doi.org/10.1017/CBO9780511777110}
\showDOI{\tempurl}


\bibitem[\protect\citeauthoryear{Sangiorgi and Walker}{Sangiorgi and
  Walker}{2001}]%
        {SW01}
\bibfield{author}{\bibinfo{person}{Davide Sangiorgi} {and}
  \bibinfo{person}{David Walker}.} \bibinfo{year}{2001}\natexlab{}.
\newblock \bibinfo{booktitle}{\emph{The Pi-Calculus - a theory of mobile
  processes}}.
\newblock \bibinfo{publisher}{Cambridge University Press}.
\newblock


\bibitem[\protect\citeauthoryear{Toninho, Caires, and Pfenning}{Toninho
  et~al\mbox{.}}{2013}]%
        {TCP13}
\bibfield{author}{\bibinfo{person}{Bernardo Toninho},
  \bibinfo{person}{Lu{\'{\i}}s Caires}, {and} \bibinfo{person}{Frank
  Pfenning}.} \bibinfo{year}{2013}\natexlab{}.
\newblock \showarticletitle{Higher-Order Processes, Functions, and Sessions:
  {A} Monadic Integration}. In \bibinfo{booktitle}{\emph{{ESOP}}}
  \emph{(\bibinfo{series}{Lecture Notes in Computer Science})},
  Vol.~\bibinfo{volume}{7792}. \bibinfo{publisher}{Springer},
  \bibinfo{pages}{350--369}.
\newblock


\bibitem[\protect\citeauthoryear{Toninho and Yoshida}{Toninho and
  Yoshida}{2018}]%
        {TY18}
\bibfield{author}{\bibinfo{person}{Bernardo Toninho} {and}
  \bibinfo{person}{Nobuko Yoshida}.} \bibinfo{year}{2018}\natexlab{}.
\newblock \showarticletitle{On Polymorphic Sessions and Functions - {A} Tale of
  Two (Fully Abstract) Encodings}. In \bibinfo{booktitle}{\emph{Programming
  Languages and Systems - 27th European Symposium on Programming, {ESOP} 2018,
  Held as Part of the European Joint Conferences on Theory and Practice of
  Software, {ETAPS} 2018, Thessaloniki, Greece, April 14-20, 2018,
  Proceedings}} \emph{(\bibinfo{series}{Lecture Notes in Computer Science})},
  \bibfield{editor}{\bibinfo{person}{Amal Ahmed}} (Ed.),
  Vol.~\bibinfo{volume}{10801}. \bibinfo{publisher}{Springer},
  \bibinfo{pages}{827--855}.
\newblock
\urldef\tempurl%
\url{https://doi.org/10.1007/978-3-319-89884-1\_29}
\showDOI{\tempurl}


\bibitem[\protect\citeauthoryear{Vasconcelos}{Vasconcelos}{2012}]%
        {V12}
\bibfield{author}{\bibinfo{person}{Vasco~T. Vasconcelos}.}
  \bibinfo{year}{2012}\natexlab{}.
\newblock \showarticletitle{Fundamentals of session types}.
\newblock \bibinfo{journal}{\emph{Inf. Comput.}}  \bibinfo{volume}{217}
  (\bibinfo{year}{2012}), \bibinfo{pages}{52--70}.
\newblock


\bibitem[\protect\citeauthoryear{Wadler}{Wadler}{2014}]%
        {W14}
\bibfield{author}{\bibinfo{person}{Philip Wadler}.}
  \bibinfo{year}{2014}\natexlab{}.
\newblock \showarticletitle{Propositions as sessions}.
\newblock \bibinfo{journal}{\emph{JFP}} \bibinfo{volume}{24},
  \bibinfo{number}{2--3} (\bibinfo{year}{2014}), \bibinfo{pages}{384--418}.
\newblock
\newblock
\shownote{Also: ICFP, pages 273--286, 2012.}


\bibitem[\protect\citeauthoryear{Wadler}{Wadler}{2015}]%
        {W15}
\bibfield{author}{\bibinfo{person}{Philip Wadler}.}
  \bibinfo{year}{2015}\natexlab{}.
\newblock \showarticletitle{Propositions as types}.
\newblock \bibinfo{journal}{\emph{Commun. {ACM}}} \bibinfo{volume}{58},
  \bibinfo{number}{12} (\bibinfo{year}{2015}), \bibinfo{pages}{75--84}.
\newblock
\urldef\tempurl%
\url{https://doi.org/10.1145/2699407}
\showDOI{\tempurl}


\end{thebibliography}
